\documentclass[a4paper,12pt]{article}
\pdfoutput=1
\usepackage{geometry}
\usepackage{amsmath,amssymb,amsfonts}
\usepackage{xcolor,graphicx,cite,soul}
\usepackage{array,booktabs,longtable}
\usepackage{caption,microtype}
\usepackage{hyperref}

\newcommand{\lsim}
{\;\raisebox{-.3em}{$\stackrel{\displaystyle <}{\sim}$}\;}
\newcommand{\gsim}
{\;\raisebox{-.3em}{$\stackrel{\displaystyle >}{\sim}$}\;}
\newcommand{\id}{{\rm 1\kern-.12em
\rule{0.3pt}{1.5ex}\raisebox{0.0ex}{\rule{0.1em}{0.3pt}}}}

\newcommand{\SLASH}[2]{\makebox[#2ex][l]{$#1$}/}
\newcommand{\pslash}{\SLASH{p}{.2}}

\newcommand\tb{\tan\beta}

\newcommand{\sinb}{\sin \beta\,}
\newcommand{\cosb}{\cos \beta\,}

\newcommand{\Sbe}{\sin \beta}
\newcommand{\Cb}{\cos \beta}

\newcommand{\OP}{\omega_+}
\newcommand{\OM}{\omega_-}

\newcommand\LP{\left(}
\newcommand\RP{\right)}
\newcommand\LB{\left[}
\newcommand\RB{\right]}
\newcommand\LV{\left\{}
\newcommand\RV{\right\}}

\renewcommand\Re{\mathop{\mathrm{Re}}}

\newcommand\ReDiag{\mathop{%
  \raise .5pt\hbox{[}%
  \widetilde{\mathrm{Re}}%
  \raise .5pt\hbox{]}}}
\newcommand\ReOffDiag{\mathop{%
  \raise .5pt\hbox{$\llbracket$}%
  \widetilde{\mathrm{Re}}%
  \raise .5pt\hbox{$\rrbracket$}}}
\newcommand\SE[1]{\Sigma_{#1}}
\newcommand\rSE[1]{\hat{\Sigma}_{#1}}
\def\Si{\Sigma}

\def\hSi{\hat{\Sigma}}
\newcommand\OS{\mathrm{OS}}
\newcommand\Os{\mathrm{os}}
\newcommand\DRbar{\ensuremath{\smash{\overline{\mathrm{DR}}}}}

\newcommand\matr[1]{\mathbf{#1}}

\newcommand\cL{{\cal L}}
\newcommand{\cZ}{{\cal Z}}
\newcommand{\bcZ}{\bar{\cal Z}}

\newcommand\SW{s_\mathrm{w}}
\newcommand\CW{c_\mathrm{w}}
\newcommand\sw{\SW}
\newcommand\cw{\CW}
\newcommand\MW{M_W}
\newcommand\MZ{M_Z}

\newcommand{\dcZ}[1]{\delta \cZ_{#1}}
\newcommand{\dbcZ}[1]{\delta \bcZ_{#1}}
\newcommand\dZm[1]{\delta\matr{Z}_{#1}}
\newcommand\dbZm[1]{\delta\matr{\breve Z}_{#1}}
\newcommand\dBZm[1]{\delta\matr{\breve{\bar Z}}{}_{#1}}

\newcommand{\dZZm}[1]{\bigl[\dZm{#1}\bigr]}

\newcommand\ino[1]{\tilde\chi_{#1}}

\newcommand\chapm[1]{\ino{#1}^\pm}

\newcommand\chap[1]{\ino{#1}^+}
\newcommand\cham[1]{\ino{#1}^-}
\newcommand\cha{\chapm}
\newcommand\mcha[1]{m_{\chapm{#1}}}

\newcommand\neu[1]{\ino{#1}^0}
\newcommand\mneu[1]{m_{\neu{#1}}}
\newcommand{\mneuOS}[1]{\hat{m}_{\tilde \chi^0_{#1}}}

\newcommand\refeq[1]{Eq.~(\ref{#1})}
\newcommand\refeqs[1]{Eqs.~(\ref{#1})}
\newcommand\refta[1]{Tab.~\ref{#1}}
\newcommand\refse[1]{Sect.~\ref{#1}}

\newcommand\citere[1]{Ref.~\cite{#1}}
\newcommand\citeres[1]{Refs.~\cite{#1}}

\newcommand{\CP}{{\cal CP}}
\newcommand{\cp}{{\CP}}
\newcommand{\os}{\mathrm{os}}
\newcommand{\re}{\mathop{\mathrm{Re}}}
\newcommand{\im}{\mathop{\mathrm{Im}}}
\newcommand{\wtre}{\widetilde\Re}

\newcommand{\tev}{\,\, \mathrm{TeV}}
\newcommand{\gev}{\,\, \mathrm{GeV}}
\newcommand{\mev}{\,\, \mathrm{MeV}}

\newcommand\edz{\tfrac{1}{2}}

\def\ed#1{\frac{1}{#1}}
\newcommand\FA{\texttt{FeynArts}}
\newcommand\FC{\texttt{FormCalc}}
\newcommand\LT{\texttt{LoopTools}}

\newcommand{\sig}{\sigma}

\newcommand{\ccn}[1]{\ensuremath{\mbox{CCN}_{#1}}}
\newcommand{\cnn}[3]{\ensuremath{\mbox{CNN}_{#1 #2 #3}}}
\newcommand{\nnn}[3]{\ensuremath{\mbox{NNN}_{#1 #2 #3}}}
\newcommand{\rs}[1]{\ensuremath{\mbox{RS}_{#1}}}

\def\order#1{\ensuremath{{\cal O}(#1)}}
\def\reffi#1{\mbox{Fig.~\ref{#1}}}
\def\reffis#1{\mbox{Figs.~\ref{#1}}}
\newcommand{\non}{\nonumber}

\def\Ga{\Gamma}

\def\de{\delta}

\newcommand{\MOne}{\ensuremath{M_1}}
\newcommand{\MTwo}{\ensuremath{M_2}}

\definecolor{Orange}{named}{orange}
\definecolor{Purple}{named}{purple}
\definecolor{Lightblue}{cmyk}{0.9,0.1,0.1,0.3}
\definecolor{dgelborange}{cmyk}{0.,0.3,0.5, 0.}
\definecolor{Lila}{rgb}{0.5,0.,1}

\graphicspath{{figs/}}
\allowdisplaybreaks
\begin{document}
\thispagestyle{empty}

\def\thefootnote{\fnsymbol{footnote}}

\begin{flushright}
IFT--UAM/CSIC--23-012 \\
\end{flushright}

\vspace{0.5cm}

\begin{center}

{\large\sc 
{\bf Automated Choice for the Best Renormalization Scheme}} 

\vspace{0.4cm}

{\large\sc {\bf in BSM Models}}

\vspace{1cm}

{\sc
S.~Heinemeyer$^{1}$%
\footnote{email: Sven.Heinemeyer@cern.ch}%
~and F.~von der Pahlen$^{2}$%
\footnote{email: federico.vonderpahlen@udea.edu.co}%
}

\vspace*{.7cm}

{\sl
$^1$Instituto de F\'isica Te\'orica (UAM/CSIC), 
Universidad Aut\'onoma de Madrid, \\ 
Cantoblanco, 28049, Madrid, Spain

\vspace*{0.1cm}

$^2$Instituto de F\'isica, Universidad de Antioquia,
Calle 70 No. 52-21, Medell\'in, Colombia

}

\end{center}

\vspace*{0.1cm}

\begin{abstract}
\noindent
The explorations of models beyond the Standard Model (BSM) naturally
involve scans over the unknown BSM parameters. On the other
hand, high precision 
predictions require calculations at the loop-level and thus a
renormalization of (some of) the BSM parameters. Often many choices are
possible for the renormalization scheme (RS). This concerns the choice
of the set of to-be-renormalized parameters out of a larger set of BSM
parameters, but can also concern the type of renormalization condition
which is chosen for a specific parameter. A given RS can be well suited
to yield ``stable'' and ``well behaved'' higher-order corrections in one
part of the BSM parameter space, but can fail completely in other
parts, which may not even be noticed numerically if an isolated
parameter point is investigated, or when the higher-order BSM
calculations are performed in an automated, not supervised set-up.
Consequently, the (automated) exploration of BSM models requires
a choice of  
a good RS {\em before} the calculation is performed. We propose 
a new method with which such a choice can be performed.
We demonstrate the
feasibility of our new method in the chargino/neutralino sector of the
Minimal Supersymmetric Standard Model (MSSM), but stress the general
applicability of our method to all types of BSM models.
\end{abstract}


\def\thefootnote{\arabic{footnote}}
\setcounter{page}{0}
\setcounter{footnote}{0}

\newpage


\section{Introduction}

The Standard Model (SM) provides a good description of nearly all
experimental data in high-energy physics. On the other hand, there is
clear evidence that the SM cannot be the ultimate theory. Gravity is not
included, it does not contain a suitable Dark Matter candidate, it does
not explain the Baryon asymmetry in the universe, neutrinos are massless, etc.
Consequently, the exploration
 of models beyond the Standard Model (BSM)
is one of the main tasks of the upcoming LHC Runs, including the
high-luminosity LHC (HL-LHC), as well as possible future
collider projects. Reliable investigations, going beyond the very
initial exploration of a BSM model, require
 the inclusion of
higher-order corrections to, e.g., the production cross sections of BSM
particles at the HL-LHC. This in turn requires the renormalization of
the BSM model.

The renormalization of BSM models is much less explored than the
renormalization of the SM~\cite{Bohm:1986rj,Hollik:1988ii}.
Examples for ``full one-loop
renormalizations'' can be found for the Two Higgs Doublet Model
(2HDM)~\cite{Altenkamp:2017ldc,Kanemura:2017wtm,Krause:2019qwe}
(see also \citere{Denner:2018opp}), the Minimal Supersymmetric Standard Model 
(MSSM)~\cite{Baro:2009gn,Baro:2008bg,MSSMCT,Stop2decay},
and the Next-to-MSSM
(NMSSM)~\cite{Belanger:2016tqb,Belanger:2017rgu,Ender:2011qh,Graf:2012hh,Drechsel:2016jdg}.
These analyses showed that many different choices of renormalization schemes
(RS) are possible. This can concern the choice
of the set of to-be-renormalized parameters out of a larger set of BSM
parameters, but can also concern the type of renormalization condition
that is chosen for a specific parameter. 

BSM models naturally possess several new BSM parameters. The number of
new parameters can vary from \order{1} to \order{10}, or even
higher. Often multi-dimensional parameter scans are employed, or methods
such as Markow-Chain Monte-Carlo (MCMC)
analyses~\cite{Bagnaschi:2017tru,GAMBIT:2017zdo}
to find the phenomenological best-appealing parameters in the
multi-dimensional BSM parameter space. The above mentioned BSM analyses
also demonstrated that a given RS can be well suited
to yield ``stable'' and ``well behaved'' higher-order corrections (more
details will be given below) in one
part of the BSM parameter space, but can fail completely in other
parts. The latter may not even be noticed numerically if only isolated
parameter points are investigated, which is natural in a scan, or MCMC
analyses.
Consequently, the exploration of BSM
models requires a choice of 
a good RS {\em before}  the higher-order calculation of the physical
  observable is performed (counterterm calculations in various schemes,
  however, will be required).
An RS ``fails'' if 
the set of mass matrix counterterms
for which we have set on-shell conditions
is (approximately) independent of
one of the parameter counterterms (or a combination of them). 
This failure can manifest itself in 
\begin{itemize}
\item ``unnaturally'' large higher-order corrections,
\item large (numerical) differences between \DRbar\ and OS masses,
\item large (numerical) differences between \DRbar\ and OS parameters.
\end{itemize}
In this work we propose a new method 
with which
such a situation can be avoided,
 i.e.\ how a ``good'' RS can be chosen. This method is based on
the properties of the transformation matrix that connects the various
mass counterterms with the parameter counterterms.
This allows a
point-by-point test of all ``available'' or ``possible'' RS, and the
``best'' one can be chosen to perform the calculation.
(The term ``best'' will be defined more clearly below in
\refse{sec:general}, where we describe our general idea.)
Only this type of selection of a good RS before the calculation
will allow a fully automated set-up of BSM higher-order calculations.

Our idea is designed to work in all cases of RS choices (in BSM
models). In the detailed definition we will, however, tackle a slightly
more specific question: in many BSM models one can be faced with the
situation that one has $m$ underlying Lagrangian parameters
and $n > m$ particles or particle masses that can be renormalized
on-shell (OS). The calculation of the production and/or decay of BSM
particles naturally requires OS renormalizations of the particles
involved. Each choice of $m$ particles renormalized OS defines an
\rs{l}, of which we have $N$ in total. We will demonstrate how out of
these $N$ \rs{l} one can choose the ``best'' \rs{L}.

The numerical examples will be performed within the MSSM, concretely in
the sector of charginos and neutralinos, the supersymmetric (SUSY)
partners of the SM gauge bosons and the 2HDM-like Higgs sector. 
While this constitutes a very specific example, we would like to stress 
that we expect a much more general
applicability of our method to many types of BSM models and
types of RS choices.

\medskip
The paper is organized as follows. In \refse{sec:general} we present 
our general idea, i.e.\ how
to choose $n$ particles to be renormalized OS, when
only $m < n$ free parameters are available. The concrete implementation
for the chargino/neutralino sector will be given in
\refse{sec:concrete}. Here we will define this sector in all details, 
discuss which different schemes are available and explain the treatment
of the $m-n$ particles that cannot be renormalized OS. In
\refse{sec:numerical} we present concrete numerical examples and
demonstrate that our method selects a stable and well behaved
renormalization over the full parameter space. Our conclusions are given
in \refse{sec:conclusions}.


\section{General idea}
\label{sec:general}

As discussed above, the idea of how to choose a stable and well behaved
RS is generally applicable. However, here we will outline it focusing a
more concrete problem:
in our theory we have $m$ underlying Lagrangian parameters
and $n > m$ particles or particle masses that can be renormalized OS. 
Each choice of $m$ particles renormalized OS defines an
\rs{l}, of which we have $N$ in total. How can one choose the
``best'' \rs{L}?%
\footnote{%
The word ``best'' does not have a real scientific meaning. 
We use it synonymously for a RS which maximizes our numerical criterion,
and which is expected to be stable
(for the parameter point under investigation).
}%

\bigskip
\noindent
There are two possible starting points for input parameters in our analysis:
\begin{itemize}
\item[\DRbar:]
  The masses of the BSM particles under investigation have not (yet)
  been measured. Then we start with $\DRbar$ parameters.%
\item[OS:]
  All the masses (or a subset) have been determined experimentally. Then
  one starts with OS masses (or a subset of them). 
\end{itemize}

Currently, we are clearly in the \DRbar\ case. Consequently, we will 
focus on the \DRbar\ choice, and leave the OS case for future work.

\bigskip
\noindent
The general idea for the automated choice of the \rs{L} in the
\DRbar\ case can be outlined for two possible levels of refinement. The
first one is called ``semi-OS scheme'', and the second one ``full-OS
scheme'' (where in our numerical examples we will focus on the
latter). The two cases are defined as follows.


\subsubsection*{Semi-OS scheme:}
\begin{enumerate}
\item
  We start with $m$ \DRbar\ parameters, $P_i^{\DRbar}$, from the Lagrangian. 
\item
  We have $N$ \rs{l}. 
\item
  For each \rs{l}, i.e.\ each different choice of $m$ particles
  renormalized OS, we evaluate the corresponding OS parameters
  \begin{align}
    P_{i,l}^{\Os} = P_i^{\DRbar} - \de P_{i,l|{\rm fin}}^{\Os}
    \label{Pilos}
  \end{align}
  with the matrix $\matr{A}^{\DRbar}_l$, defining the transformation matrix 
  from the set of parameter counterterms to mass counterterms
 (more details will be given below, see \refeq{ADRbar} and Appendix~\ref{sec:matrA}). 
With the label \mbox{}$^{\Os}$ we denote
    the OS-like counterterms in the semi-OS scheme (whereas below in the
    full-OS scheme the label \mbox{}$^{\OS}$ will be used).
\item
  It will be argued that a ``bad'' scheme \rs{l}\ has a small or even
  vanishing $|\det\matr{A}^{\DRbar}_l|$.
\item
  Comparing the various $|\det\matr{A}^{\DRbar}_l|$ and choosing
    the scheme with the largest $|\det\matr{A}^{\DRbar}_l|$ yields \rs{L}.
  While there is no formal proof that this procedure yields the
    ``best'' RS, 
a vanishing determinant yields an ill defined RS because the
transformation cannot be inverted. 
A small determinant, while
technically allowed, may yield numerically unstable RS,  
as is indeed the case with all our numerical tests described below.
Conversely, going away from zero implies that none of the
coefficients of the inverted matrix are exceptionally large,  
leading us to expect a stable RS (as will be found in the numerical
  analysis below).
Furthermore, the choice of
    the largest determinant yields a clear numerical recipe.
\item
  Inserting $P_{i,L}^{\Os}$ into the Lagrangian yields $n$ particle
  masses out of which $m$ are by definition given as their OS values.
  The remaining OS masses have to be determined calculating $n-m$ finite
  shifts.
\item
  The counterterms for the $P_{i,L}^{\Os}$ are already known from
  \refeq{Pilos} as $\de P_{i,L}^{\Os}$ and can be inserted as
  counterterms in a loop calculation.
\end{enumerate}

\noindent
This procedure yields all ingredients for an OS scheme. However, the
OS counterterms $\de P_{i,L}^{\Os}$ and thus also the OS parameters
themselves, $P_{i,L}^{\Os}$, are calculated in terms of
\DRbar\ parameters, i.e.\ one has $\de P_{i,L}^{\Os}(P_i^{\DRbar})$ and
$P_{i,L}^{\Os}(P_i^{\DRbar})$.
This is unsatisfactory for a ``true'' OS scheme, i.e.\ one would like to have
$\de P_{i,L}^{\OS}(P_{i,L}^{\OS})$.
Furthermore, when a \rs{l}\ ``starts to turn bad'' as a function of a
\DRbar\ parameter, large differences between the $P_{i,l}^{\Os}$ and
$P_i^{\DRbar}$ occur, shedding doubt on the above outlined procedure.
These problems can be circumvented by extending the above scheme to an
evaluation of the counterterms in terms of OS parameters. The general
idea starts as above, but deviates from step~4 on.

Before we proceed, it should be noted that a different type of OS
  renormalization 
could be applied. In principle it is also possible to renormalize all
particles OS, yielding higher-order corrections to the respective (now
broken) tree-level relations. However, we deem this method more
complicated for an automated procedure, and it is not clear to us whether
this general choice avoids numerical instabilities.


\subsubsection*{Full-OS scheme:}
\begin{enumerate}
\item
  We start with $m$ \DRbar\ parameters, $P_i^{\DRbar}$, from the Lagrangian. 
\item
  We have $N$ \rs{l}. 
\item
  For each \rs{l}, i.e.\ each different choice of $m$ particles
  renormalized OS, we evaluate the corresponding OS parameters
  \begin{align}
    P_{i,l}^{\Os} = P_i^{\DRbar} - \de P_{i,l|{\rm fin}}^{\Os}
    \label{PilOs}
  \end{align}
  with the matrix matrix $\matr{A}^{\DRbar}_l$, defining the transformation matrix 
    from the set of parameter counterterms to mass counterterms
 (more details will be given below, see \refeq{AOS} and Appendix~\ref{sec:matrA}).
\item
  Inserting $P_{i,l}^{\Os}$ into the Lagrangian yields $n$ particle
  masses out of which $m$ are by definition given as their $\Os_l$ values.
  The remaining $\Os_l$ masses have to be determined calculating $n-m$ finite
  shifts.
\item
  \rs{l} is applied again on the OS$_l$ Lagrangian.
\item
  This yields now OS counterterms in terms of $\Os_l$ parameters,
  \begin{align}
    \de P_{i,l}^{\OS}(P_{i,l}^{\Os})
    \label{PilOS}
  \end{align}
  with the matrix $\matr{A}^{\OS}_l$, defining the
    transformation matrix from the set of 
    OS~parameter counterterms to OS~mass counterterms
  (more details will be given below).
\item
  It will be argued that a ``bad'' scheme \rs{l}\ has a small or even
  vanishing $|\det\matr{A}^{\DRbar}_l|$ and/or $|\det\matr{A}^{\OS}_l|$.
\item
  Comparing the various
  \begin{align}
      \min \LV |\det\matr{A}^{\DRbar}_l|, |\det\matr{A}^{\OS}_l| \RV
\label{eq.detA}
  \end{align}
  and choosing the RS with the largest value yields \rs{L}.
  Here the same remark as for the semi-OS scheme holds.
Furthermore, the recipe is later corroborated in our numerical analysis. 

\item
  The counterterms for the $P_{i,L}^{\OS}$ are already known from
  \refeq{PilOS} as $\de P_{i,L}^{\OS}$ and can be inserted as
  counterterms in a loop calculation.
\end{enumerate}
Steps 4-6 could be iterated until convergence is reached. We will not do
this. 

\bigskip
\noindent
In the following sections we will explain and demonstrate the concrete
implementation of this procedure for the case of the chargino/neutralino
sector in the MSSM.


\section{Concrete implementation}
\label{sec:concrete}

The concrete implementation concerns the calculation of physics
processes with (external) charginos and/or 
neutralinos, $\cha{c} (c = 1,2)$ and $\neu{n} (n = 1, 2, 3, 4)$ at the loop
level. This requires the choice of a (numerically well behaved) RS.
In an OS RS 
the three Lagrangian parameters  $M_1$ , $M_2$ and $\mu$ of the MSSM 
 are renormalized through the OS conditions on three chargino/neutralino
 fields.
The possible scheme choices are ($n'' > n' > n$)
\begin{align}
\ccn{n}, \quad \cnn{c}{n}{n'}, \quad \nnn{n}{n'}{n''} 
\quad c = 1,2; \; n,n',n'' = 1,2,3,4~.
\end{align}
Here \ccn{n} denotes a scheme where the two charginos and the
neutralino~$n$, $\neu{n}$, are renormalized OS. \cnn{c}{n}{n'} denotes a
scheme were chargino~$c$, $\cha{c}$, as well as neutralinos~$n,n'$,
$\neu{n}, \neu{n'}$, are renormalized OS. Finally \nnn{n}{n'}{n''} denotes
a scheme with three neutralinos renormalized OS
(see also \citere{Chatterjee:2011wc}).
For the sake of simplicity, in the following we neglect the \nnn{n}{n'}{n''}
schemes. 

In the following we will describe the concrete implementation of the
general scheme described in \refse{sec:general} in the case of the
chargino/neutralino sector of the MSSM. The starting point will be
\DRbar\ input parameters. The case of an observation of (parts of) the
spectrum is beyond the scope of this publication.


\subsection{Notation}
\label{sec:notation}

In this section we briefly review our notation of the
chargino/neutralino sector of the MSSM, as well as the relevant
(derived) quantities, such as (renormalized) self-energies and mass
shifts (more details and applications can be found in
\citeres{LHCxC1,LHCxC2,LHCxN,LHCxNprod,eeIno}).

The starting point for the renormalization of the
chargino/neutralino sector is the part of the Fourier transformed MSSM
Lagrangian which is bilinear in the chargino and neutralino fields,
\begin{align}
\cL^{\text{bil.}}_{\cham{},\tilde{\chi}^0} &= 
  \overline{\cham{i}}\, \pslash\, \OM \cham{i} 
+ \overline{\cham{i}}\, \pslash\, \OP \cham{i} 
- \overline{\cham{i}}\, [\matr{V}^* \matr{X}^\top \matr{U}^\dagger]_{ij} \,
  \OM \cham{j} 
- \overline{\cham{i}}\, [\matr{U} \matr{X}^* \matr{V}^{\top}]_{ij} \,
  \OP \cham{j} \non \\
&\quad + \frac{1}{2} \LP
  \overline{\neu{k}}\, \pslash\, \OM \neu{k}, 
+ \overline{\neu{k}}\, \pslash\, \OP \neu{k} 
- \overline{\neu{k}}\, [\matr{N}^*\matr{Y} \matr{N}^\dagger]_{kl} \,
  \OM \neu{l} 
- \overline{\neu{k}}\, [\matr{N} \matr{Y}^* \matr{N}^{\top}]_{kl} \,
  \OP \neu{l} \RP~, 
\end{align}
already expressed in terms of the chargino and neutralino mass eigenstates
$\cham{i}$ and $\neu{k}$, respectively, 
and $i,j = 1,2$ and $k,l = 1,2,3,4$.
Here, ${\omega}_{\pm} = \frac{1}{2}(\id \pm \gamma_5)$
are the right- and left-handed projectors, respectively.
The mass eigenstates can be determined via unitary 
transformations where the corresponding matrices diagonalize the chargino and
neutralino mass matrix, $\matr{X}$ and $\matr{Y}$, respectively. 

In the chargino case, two $2 \times 2$ matrices $\matr{U}$ and
$\matr{V}$ are necessary for the diagonalization of the chargino mass
matrix~$\matr{X}$, 
\begin{align}
\matr{M}_{\cham{}} = \matr{V}^* \, \matr{X}^\top \, \matr{U}^{\dagger} =
  \begin{pmatrix} m_{\tilde{\chi}^\pm_1} & 0 \\ 
                  0 & m_{\tilde{\chi}^\pm_2} \end{pmatrix}  \quad
\text{with} \quad
  \matr{X} =
  \begin{pmatrix}
    \MTwo & \sqrt{2} \sinb \MW \\
    \sqrt{2} \cosb \MW & \mu
\label{eq:X}
  \end{pmatrix}~,
\end{align}
where $\matr{M}_{\cham{}}$ is the diagonal mass matrix with the chargino
masses $\mcha{1}, \mcha{2}$ as entries, which are determined as the
(real and positive) singular values of $\matr{X}$. 
The singular value decomposition of $\matr{X}$ also yields results for 
$\matr{U}$ and~$\matr{V}$. 
Using the transformation matrices $\matr{U}$ and $\matr{V}$, the interaction
Higgsino and wino spinors $\tilde{H}^-_1$, $\tilde{H}^+_2$ and
$\tilde{W}^\pm$, which are two component Weyl spinors, can be transformed into
the mass eigenstates 
\begin{align}
\cham{i} = 
\begin{pmatrix} \psi^L_i
   \\ \overline{\psi^R_i} \end{pmatrix}
\quad \text{with} \quad \psi^L_i = U_{ij} \begin{pmatrix} \tilde{W}^-
  \\ \tilde{H}^-_1 \end{pmatrix}_j \quad \text{and} \quad
 \psi^R_i = V_{ij} \begin{pmatrix} \tilde{W}^+
  \\ \tilde{H}^+_2 \end{pmatrix}_j
\end{align}
where the $i$th mass eigenstate can be expressed in terms of either 
the Weyl spinors $\psi^L_i$ and $\psi^R_i$  or the Dirac spinor $\cham{i}$.

In the neutralino case, as the neutralino mass matrix $\matr{Y}$ is
symmetric, one $4 \times 4$~matrix is sufficient for the diagonalization
\begin{align}
\matr{M}_{\neu{}} = \matr{N}^* \, \matr{Y} \, \matr{N}^{\dagger} =
\text{\bf diag}(m_{\neu{1}}, m_{\neu{2}}, m_{\neu{3}}, m_{\neu{4}})
\end{align}
with
\begin{align}
\matr{Y} &=
  \begin{pmatrix}
    \MOne                  & 0                & -\MZ \, \sw \cosb
    & \MZ \, \sw \sinb \\ 
    0                      & \MTwo            & \quad \MZ \, \cw \cosb
    & -\MZ \, \cw \sinb \\ 
    -\MZ \, \sw \cosb      & \MZ \, \cw \cosb & 0
    & -\mu             \\ 
    \quad \MZ \, \sw \sinb & -\MZ \, \cw \sinb & -\mu              & 0
  \end{pmatrix}~.
\label{eq:Y}
\end{align}
$\MZ$ and $\MW$ are the masses of the $Z$~and $W$~boson, 
$\cw = \MW/\MZ$ and $\sw = \sqrt{1 - \cw^2}$. 
The unitary 4$\times$4 matrix $\matr{N}$ and the physical neutralino
(tree-level) masses $\mneu{k}$ ($k = 1,2,3,4$) result from a numerical Takagi 
factorization \cite{Takagi} of $\matr{Y}$. 
Starting from the original bino/wi\-no/higg\-si\-no basis, the mass
eigenstates can be determined with the help of the transformation matrix~$\matr{N}$, 
\begin{align}
\neu{k} = \begin{pmatrix} \psi^0_k \\[.2em] \overline{\psi^0_k} 
\end{pmatrix} 
\qquad \text{with} \qquad 
\psi^0_k = N_{kl}
{\LP \tilde{B}^0, \tilde{W}^0, \tilde{H}^0_1, \tilde{H}^0_2 \RP^{\top}_l}
\end{align}
where $\psi^0_k$ denotes the two component Weyl spinor and $\neu{k}$
the four component Majorana spinor of the $k$th neutralino field.

\medskip
Concerning the renormalization of this sector, the
following replacements of the parameters and the
fields are performed according to the multiplicative renormalization
procedure, which is formally identical for the two set-ups:
\begin{align}
M_1 \; &\to \; M_1 + \de M_1 ~, \\
M_2 \; &\to \; M_2 + \de M_2 ~, \\
\mu \; &\to \; \mu + \de \mu ~, \\
\OM \chapm{i} \; &\to \; \LB \id + \edz \dZm{\chapm{}}^L \RB_{ij}
                         \OM \chapm{j} \qquad (i,j = 1,2)~, \\
\OP \chapm{i} \; &\to \; \LB \id + \edz \dZm{\chapm{}}^R \RB_{ij}
                         \OP \chapm{j} \qquad (i,j = 1,2)~, \\
\OM \neu{k} \; &\to \; \LB \id + \edz \dZm{\neu{}}^{} \RB_{kl}
                       \OM \neu{l} \qquad (k,l = 1,2,3,4)~, \\
\OP \neu{k} \; &\to \; \LB \id + \edz \dZm{\neu{}}^* \RB_{kl}
                       \OP \neu{l} \qquad (k,l = 1,2,3,4)~.
\label{dZNeuR}
\end{align}
It should be noted that the parameter counterterms are complex
counterterms which each need two renormalization conditions to be fixed.
The transformation matrices are not renormalized (a choice that
simplifies the procedure), so that, using the notation 
of replacing a matrix by its renormalized matrix and a counterterm matrix 
\begin{align}
\label{deX}
\matr{X} &\to \matr{X} + \de\matr{X} ~, \\
\matr{Y} &\to \matr{Y} + \de\matr{Y} ~
\end{align}
with
\begin{align}\label{deltaX}
\de\matr{X} &= 
  \begin{pmatrix} \de M_2 & \sqrt{2}\, \de(\MW \Sbe) \\
                  \sqrt{2}\, \de(\MW \Cb) & \de \mu
  \end{pmatrix}~, \\[.4em]
\de\matr{Y} &= 
  \begin{pmatrix} 
      \de M_1 & 0 & -\de(\MZ\sw\Cb) & \de(\MZ\sw\Sbe) \\
      0 & \de M_2 & \de(\MZ\cw\Cb) & -\de(\MZ\cw\Sbe) \\
      -\de(\MZ\sw\Cb) & \de(\MZ\cw\Cb) & 0 & -\de\mu  \\
      \de(\MZ\sw\Sbe) & -\de(\MZ\cw\Sbe) & -\de\mu & 0
  \end{pmatrix}~,
\end{align}
the replacements of the matrices $\matr{M}_{\cham{}}$ and $\matr{M}_{\neu{}}$
can be expressed as
\begin{align}
\matr{M}_{\cham{}} &\to \matr{M}_{\cham{}} + \de\matr{M}_{\cham{}}
   = \matr{M}_{\cham{}} + \matr{V}^* \de\matr{X}^\top \matr{U}^\dagger \\
\label{Mneu}
\matr{M}_{\neu{}} &\to \matr{M}_{\neu{}} + \de\matr{M}_{\neu{}}
   = \matr{M}_{\neu{}} + \matr{N}^* \de\matr{Y} \matr{N}^\dagger~. 
\end{align}

\noindent
For convenience, we decompose the self-energies into left- and right-handed
vector and scalar coefficients via
\begin{equation}
\LB \Si_{\tilde \chi}^{}(p^2)\RB_{nm} = \displaystyle{\not}p\, {\omega_-}
\LB \Si_{\tilde \chi}^L(p^2) \RB_{nm} 
+ \displaystyle{\not}p\, {\omega_+} \LB \Si_{\tilde \chi}^R(p^2) \RB_{nm}
+ {\omega_-} \LB \Si_{\tilde \chi}^{SL}(p^2) \RB_{nm} 
+ {\omega_+} \LB \Si_{\tilde \chi}^{SR}(p^2) \RB_{nm}~,
\label{decomposition}
\end{equation}
Now the coefficients of the renormalized self-energies are given by $(i,j = 1,2; k,l = 1,2,3,4)$
\begin{align}\label{renSEcha_L}
\LB \hSi_{\chapm{}}^L(p^2) \RB_{ij} &=
  \LB \Si_{\chapm{}}^L(p^2) \RB_{ij} 
+ \edz \LB \dZm{\chapm{}}^L + \dZm{\chapm{}}^{L\dagger} \RB_{ij}~, \\
\LB \hSi_{\chapm{}}^R(p^2) \RB_{ij} &=
  \LB \Si_{\chapm{}}^R(p^2) \RB_{ij} 
+ \edz \LB \dZm{\chapm{}}^R + \dZm{\chapm{}}^{R\dagger} \RB_{ij}~, \\
\LB \hSi_{\chapm{}}^{SL}(p^2) \RB_{ij} &=
  \LB \Si_{\chapm{}}^{SL}(p^2) \RB_{ij}
  - \LB \edz \dZm{\chapm{}}^{R\dagger} \matr{M}_{\cham{}}
        + \edz \matr{M}_{\cham{}} \dZm{\chapm{}}^L
        + \de\matr{M}_{\cham{}} \RB_{ij}~, \\
\LB \hSi_{\chapm{}}^{SR}(p^2) \RB_{ij} &=
  \LB \Si_{\chapm{}}^{SR}(p^2) \RB_{ij}
  - \LB \edz \dZm{\chapm{}}^{L\dagger} \matr{M}_{\cham{}}^\dagger
       + \edz \matr{M}_{\cham{}}^\dagger \dZm{\chapm{}}^R
       + \de\matr{M}_{\cham{}}^\dagger \RB_{ij}~, \\[.4em]
\LB \hSi_{\neu{}}^L(p^2) \RB_{kl} &=
  \LB \Si_{\neu{}}^L(p^2) \RB_{kl} 
+ \edz \LB \dZm{\neu{}} + \dZm{\neu{}}^{\dagger} \RB_{kl}~, \\
\LB \hSi_{\neu{}}^R(p^2) \RB_{kl} &=
  \LB \Si_{\neu{}}^R(p^2) \RB_{kl} 
+ \edz \LB \dZm{\neu{}}^* + \dZm{\neu{}}^{\top} \RB_{kl}~, \\
\LB \hSi_{\neu{}}^{SL}(p^2) \RB_{kl} &=
  \LB \Si_{\neu{}}^{SL}(p^2) \RB_{kl}
  - \LB \edz \dZm{\neu{}}^{\top} \matr{M}_{\neu{}}
       + \edz \matr{M}_{\neu{}} \dZm{\neu{}}
        + \de\matr{M}_{\neu{}} \RB_{kl}~, \\
\label{renSEneu_SR}
\LB \hSi_{\neu{}}^{SR}(p^2) \RB_{kl} &=
  \LB \Si_{\neu{}}^{SR}(p^2) \RB_{kl}
  - \LB \edz \dZm{\neu{}}^{\dagger} \matr{M}_{\neu{}}^{\dagger} 
       + \edz \matr{M}_{\neu{}}^{\dagger} \dZm{\neu{}}^*
       + \de\matr{M}_{\neu{}}^\dagger \RB_{kl}~.
\end{align}

\medskip
In the following we will give some general expressions for a chargino or
neutralino to be renormalized OS.
The OS conditions read:
\begin{align}
\label{mcha-OS_org}
\Bigl(\LB \wtre \hSi_{\cham{}} (p)\RB_{ii} 
     \cham{i}(p)\Bigr)\Big|_{p^2 = \mcha{i}^2} &= 0 \qquad  (i = 1,2)~, \\
\label{mneu-OS_org}
\quad \Bigl(\LB\wtre \hSi_{\neu{}} (p)\RB_{kk} 
      \neu{k}(p)\Bigr)\Big|_{p^2 = \mneu{k}^2} &= 0 \qquad (k = 1,2,3,4)~.
\end{align}
These conditions can be rewritten in terms of the expressions for the
renormalized self-energies, 
\begin{align}
\label{mcha-OS}
\wtre \LB \mcha{i} \LP \hSi_{\cham{}}^{L}(\mcha{i}^2)
                       +\hSi_{\cham{}}^{R}(\mcha{i}^2) \RP 
                       +\hSi_{\cham{}}^{SL}(\mcha{i}^2)
                       +\hSi_{\cham{}}^{SR}(\mcha{i}^2) 
      \RB_{ii} &= 0~, \\
\label{mcha-OS-2}
\wtre \LB \mcha{i} \LP \hSi_{\cham{}}^{L}(\mcha{i}^2)
                       -\hSi_{\cham{}}^{R}(\mcha{i}^2) \RP 
                       -\hSi_{\cham{}}^{SL}(\mcha{i}^2)
                       +\hSi_{\cham{}}^{SR}(\mcha{i}^2) 
      \RB_{ii} &= 0~, \\
\label{mneu-OS}
\wtre \LB \mneu{k} \LP \hSi_{\neu{}}^{L}(\mneu{k}^2)
                       +\hSi_{\neu{}}^{R}(\mneu{k}^2) \RP
                       +\hSi_{\neu{}}^{SL}(\mneu{k}^2)
                       +\hSi_{\neu{}}^{SR}(\mneu{k}^2) 
      \RB_{kk} &= 0~, \\
\label{mneu-OS-2}
\wtre \LB \mneu{k} \LP \hSi_{\neu{}}^{L}(\mneu{k}^2)
                       -\hSi_{\neu{}}^{R}(\mneu{k}^2) \RP
                       -\hSi_{\neu{}}^{SL}(\mneu{k}^2)
                       +\hSi_{\neu{}}^{SR}(\mneu{k}^2) 
      \RB_{kk} &= 0~.
\end{align}
Equations~(\ref{mcha-OS-2}) and (\ref{mneu-OS-2}) are related to the
axial and axial-vector component of the renormalized self energy and therefore
the l.h.s.\ vanishes in the case of real couplings.
Therefore, in the rMSSM only Eqs.~(\ref{mcha-OS}) and (\ref{mneu-OS}) remain.
It should be noted that since the lightest neutralino is stable there
are no absorptive 
contributions from its self energy and $\wtre$ can be dropped from 
Eqs.~(\ref{mneu-OS_org},\ref{mneu-OS},\ref{mneu-OS-2}) for $k = 1$. 

For the further determination of the field renormalization constants,
we also impose
\begin{align}
\lim_{p^2 \to \mcha{i}^2} 
\Bigl(\frac{(\pslash\, + \mcha{i}) \bigl[ \wtre \hSi_{\cham{}}(p)\bigr]_{ii}}
           {p^2 - \mcha{i}^2} \cham{i}(p)\Bigr) &= 0 \qquad (i = 1,2)~, 
\label{fieldRCcha}\\ 
 \lim_{p^2 \to {\mneu{k}^2}} 
\Bigl(\frac{(\pslash\, + {\mneu{k}})\bigl[ \wtre \hSi_{\neu{}}(p)\bigr]_{{kk}}}
           {p^2 - {\mneu{k}^2}}{\neu{k}}(p)\Bigr) &= 0 \qquad (k = 1,2,3,4)~,
\label{fieldRCneu}
\end{align} 
{
which, together with \refeqs{mcha-OS-2} and (\ref{mneu-OS-2}),
lead to the following set of equations
}
\begin{align}
\non
\wtre\Bigl[\edz \LP \hSi_{\cham{}}^{L}(\mcha{i}^2)
                  + \hSi_{\cham{}}^{R}(\mcha{i}^2) \RP
                  + \mcha{i}^2 \LP \hSi_{\cham{}}^{L'}(\mcha{i}^2)
                  + \hSi_{\cham{}}^{R'}(\mcha{i}^2) \RP \quad\ &\\
                  + \mcha{i} \LP \hSi_{\cham{}}^{SL'}(\mcha{i}^2)
                  + \hSi_{\cham{}}^{SR'}(\mcha{i}^2) \RP 
      \Bigr]_{ii} &= 0~, \label{Zchadiag-OS}\\
\wtre \LB \hSi_{\cham{}}^{L}(\mcha{i}^2)
                       -\hSi_{\cham{}}^{R}(\mcha{i}^2) 
      \RB_{ii} &= 0~, \label{Zchadiag-OS-2}\\
\non
\wtre \Bigl[\edz \LP \hSi_{\neu{}}^{L}(\mneu{{k}}^2)
                   + \hSi_{\neu{}}^{R}(\mneu{{k}}^2) \RP 
                   + \mneu{{k}}^2 \LP \hSi_{\neu{}}^{L'}(\mneu{{k}}^2) 
                   + \hSi_{\neu{}}^{R'}(\mneu{{k}}^2) \RP \quad\ &\\
                   + \mneu{{k}} \LP \hSi_{\neu{}}^{SL'}(\mneu{{k}}^2)
                   + \hSi_{\neu{}}^{SR'}(\mneu{{k}}^2) \RP
      \Bigr]_{{kk}} 
                   &= 0~, \label{Zneudiag-OS}\\
\wtre \LB \hSi_{\neu{}}^{L}(\mneu{{k}}^2)
                       -\hSi_{\neu{}}^{R}(\mneu{{k}}^2)
      \RB_{{kk}} 
                 &= 0~,
\label{Zneudiag-OS-2}
\end{align}
where we have used the short-hand 
$\Si'(m^2) \equiv (\partial \Si/\partial p^2)|_{p^2 = m^2}$. 
It should be noted that \refeq{Zneudiag-OS-2} is already fulfilled due to the 
Majorana nature of the neutralinos. 

\newpage

\smallskip
Inserting \refeqs{renSEcha_L}~--~(\ref{renSEneu_SR}) for the
renormalized self-energies in \refeqs{mcha-OS}~--~(\ref{mneu-OS-2})
and solving for 
$\LB \de\matr{M}_{\cham{}}\RB_{ii}$ and $\LB \de\matr{M}_{\neu{}} \RB_{kk}$
results in
\begin{align}
\label{eq:redZcha}
\re\LB \de\matr{M}_{\cham{}} \RB_{ii} &= \frac{1}{2}
 \wtre \LB \mcha{i} \LP \Si_{\chapm{}}^L(\mcha{i}^2) 
                        + \Si_{\chapm{}}^R(\mcha{i}^2) \RP
                        + \Si_{\chapm{}}^{SL}(\mcha{i}^2)
                        + \Si_{\chapm{}}^{SR}(\mcha{i}^2) \RB_{ii}~, \\
\im\LB\de\matr{M}_{\cham{}}\RB_{ii} &= \frac{i}{2}
   \wtre \LB \Si_{\chapm{}}^{SR}(\mcha{i}^2) 
            - \Si_{\chapm{}}^{SL}(\mcha{i}^2) \RB_{ii}
 - \frac{1}{2} \mcha{i} \im \LB \dZm{\cham{}}^L
           - \dZm{\cham{}}^R \RB_{ii}~, \label{eq:imdZcha}\\[.4em]
\re\LB\de\matr{M}_{\neu{}}\RB_{kk} &= \frac{1}{2}
   \wtre \LB \mneu{k} \LP \Si_{\neu{}}^L(\mneu{k}^2)
                         + \Si_{\neu{}}^R(\mneu{k}^2) \RP
                         + \Si_{\neu{}}^{SL}(\mneu{k}^2)
                         + \Si_{\neu{}}^{SR}(\mneu{k}^2) \RB_{kk}~,
\label{eq:redZneu}\\
\im\LB\de\matr{M}_{\neu{}}\RB_{kk} &= \frac{i}{2}
   \wtre \LB \Si_{\neu{}}^{SR}(\mneu{k}^2)
             -\Si_{\neu{}}^{SL}(\mneu{k}^2) \RB_{kk}
   -\mneu{k} \im \dZZm{\neu{}}_{kk}~,\label{eq:imdZneu}
\end{align}
where we have used the 
relations~(\ref{Zchadiag-OS-2})~and~(\ref{Zneudiag-OS-2}).

\smallskip
\refeqs{Zchadiag-OS} and (\ref{Zneudiag-OS}) define the real part of
the diagonal field renormalization constants of the chargino  and of
the neutralino fields. 
The imaginary parts of the diagonal field renormalization 
constants are still undefined. 
However, they can be obtained using
\refeqs{eq:imdZcha} and (\ref{eq:imdZneu}).
For the charginos and neutralinos \refeqs{eq:imdZcha} and
(\ref{eq:imdZneu}) define
the imaginary parts of $\LB\de\matr{M}_{\cha{}}\RB_{ii}$
($i = 1,2$), and $\LB\de\matr{M}_{\neu{}}\RB_{kk}$ ($k = 1,2,3,4$)
in terms of the imaginary part of the field
renormalization constants. Therefore these are simply set to zero (see
  below \refeqs{eq:dZcha} and \eqref{eq:dZneu}),
 which is possible as all
divergences are absorbed by other counterterms. 

\smallskip
The off-diagonal field renormalization constants are fixed by the
condition that 
\begin{align}
\Bigl(\LB \wtre \hSi_{\cham{}} (p)\RB_{ij} 
    \cham{j}(p)\Bigr)\Big|_{p^2 = \mcha{j}^2} &= 0 \qquad (i,j = 1,2)~,
\\
\Bigl(\LB\wtre \hSi_{\neu{}} (p)\RB_{kl} 
    \neu{l}(p)\Bigr)\Big|_{p^2 = \mneu{l}^2} &= 0 \qquad (k,l = 1,2,3,4)~.
\end{align}
Finally, this yields for the field renormalization
constants~\cite{dissTF} (where we now make the correct dependence on
tree-level and on-shell masses explicit), 
\begin{align}
\re \LB \dZm{\chapm{}}^{L/R} \RB_{ii} &=
        - \wtre \Big[ \Si_{\chapm{}}^{L/R}(\mcha{i}^2) 
\label{dZcha_iiRe}\\
&\qquad + \mcha{i}^2 \LP \Si_{\chapm{}}^{L'}(\mcha{i}^2)
                       + \Si_{\chapm{}}^{R'}(\mcha{i}^2) \RP
        + \mcha{i} \LP \Si_{\chapm{}}^{SL'}(\mcha{i}^2)
                    +  \Si_{\chapm{}}^{SR'}(\mcha{i}^2) \RP
      \Big]_{ii}~, \non \\
\im \LB \dZm{\chapm{}}^{L/R} \RB_{ii} &= 
       \pm \frac{1}{\mcha{i}} \LB \frac{i}{2} 
    \wtre \LV \Si_{\chapm{}}^{SR}(\mcha{i}^2) 
              - \Si_{\chapm{}}^{SL}(\mcha{i}^2) \RV
    - \im \de\matr{M}_{\cham{}} \RB_{ii} =0~, 
\label{eq:dZcha} 
\\
\LB \dZm{\chapm{}}^{L/R} \RB_{ij} &= \frac{2}{\mcha{i}^2 - \mcha{j}^2} 
  \wtre \Big[ \mcha{j}^2 \Si_{\chapm{}}^{L/R}(\mcha{j}^2) 
             +\mcha{i} \mcha{j} \Si_{\chapm{}}^{R/L}(\mcha{j}^2)
\label{dZcha_ij} \\
&\qquad + \mcha{i} \Si_{\chapm{}}^{SL/SR}(\mcha{j}^2)
        + \mcha{j} \Si_{\chapm{}}^{SR/SL}(\mcha{j}^2)
        - \mcha{i/j} \de\matr{M}_{\cham{}}
        - \mcha{j/i} \de\matr{M}_{\cham{}}^{\dagger} \Big]_{ij}~, \non \\
\re \LB \dZm{\neu{}}^{} \RB_{kk} &= 
  -\wtre \Big[  \Si_{\neu{}}^L(\mneuOS{k}^2) \\
&\qquad  + \mneu{k}^2 \LP \Si_{\neu{}}^{L'}(\mneuOS{k}^2)
                         +\Si_{\neu{}}^{R'}(\mneuOS{k}^2) \RP
         + \mneu{k}   \LP \Si_{\neu{}}^{SL'}(\mneuOS{k}^2)
                         +\Si_{\neu{}}^{SR'}(\mneuOS{k}^2) \RP
       \Big]_{kk}~, \non \\
\im \LB \dZm{\neu{}}^{} \RB_{kk} &= 
         \frac{1}{\mneu{k}} \LB \frac{i}{2} 
    \wtre \LV \Si_{\neu{}}^{SR}(\mneuOS{k}^2) \label{eq:dZneu}
              - \Si_{\neu{}}^{SL}(\mneuOS{k}^2) \RV
    - \im \de\matr{M}_{\neu{}} \RB_{kk} =0~,\\ 
\LB \dZm{\neu{}}^{} \RB_{kl} &= \frac{2}{\mneu{k}^2 - \mneu{l}^2}
 \wtre \Big[ \mneu{l}^2 \Si_{\neu{}}^L(\mneuOS{l}^2) 
            +\mneu{k}\mneu{l} \Si_{\neu{}}^R(\mneuOS{l}^2) \non \\
&\qquad + \mneu{k} \Si_{\neu{}}^{SL}(\mneuOS{l}^2)
        + \mneu{l} \Si_{\neu{}}^{SR}(\mneuOS{l}^2)
        - \mneu{k} \de\matr{M}_{\neu{}} 
        - \mneu{l} \de\matr{M}_{\neu{}}^\dagger \Big]_{kl}~.
\label{eq:dZneu_ij}
\end{align}

\medskip
Contributions to the partial
decay widths can arise from the product of the imaginary parts of the
loop-functions (absorptive contributions) of the self-energy type
contributions in the external legs and the imaginary parts of
complex couplings entering the decay vertex or the self-energies. 
It is possible to combine these additional contributions 
with the field renormalization constants in a single ``$Z$~factor'',
$\mathcal Z$, see
e.g.~\citere{Stop2decay,LHCxC1}
and references therein. In our notation they read (unbarred for an
incoming neutralino or an incoming negative chargino, 
barred for an outgoing neutralino or negative chargino), 
\begin{align}
\LB \dcZ{\chapm{}}^{L/R} \RB_{ii} &=
        - \Big[ \Si_{\chapm{}}^{L/R}(\mcha{i}^2) \\
&\qquad + \mcha{i}^2 \LP \Si_{\chapm{}}^{L'}(\mcha{i}^2)
                       + \Si_{\chapm{}}^{R'}(\mcha{i}^2) \RP
        + \mcha{i} \LP \Si_{\chapm{}}^{SL'}(\mcha{i}^2)
                    +  \Si_{\chapm{}}^{SR'}(\mcha{i}^2) \RP \Big]_{ii}~\non \\
&\qquad \pm \ed{2 \mcha{i}} \LB 
                      \Si_{\chapm{}}^{SL}(\mcha{i}^2)
                    - \Si_{\chapm{}}^{SR}(\mcha{i}^2)
                    - \de\matr{M}_{\cham{}}
                    + \de\matr{M}_{\cham{}}^{*} \RB_{ii}~, \non \\
\LB \dcZ{\chapm{}}^{L/R} \RB_{ij} &= \frac{2}{\mcha{i}^2 - \mcha{j}^2} 
        \Big[ \mcha{j}^2 \Si_{\chapm{}}^{L/R}(\mcha{j}^2) 
             +\mcha{i} \mcha{j} \Si_{\chapm{}}^{R/L}(\mcha{j}^2) \\
&\qquad + \mcha{i} \Si_{\chapm{}}^{SL/SR}(\mcha{j}^2)
        + \mcha{j} \Si_{\chapm{}}^{SR/SL}(\mcha{j}^2)
        - \mcha{i/j} \de\matr{M}_{\cham{}}
        - \mcha{j/i} \de\matr{M}_{\cham{}}^\dagger \Big]_{ij}~, \non \\
\LB \dcZ{\neu{}}^{L/R} \RB_{kk} &= 
  - \Big[ \Si_{\neu{}}^{L/R}(\mneuOS{k}^2) \\
&\qquad     + \mneu{k}^2 \LP \Si_{\neu{}}^{L'}(\mneuOS{k}^2)
                            +\Si_{\neu{}}^{R'}(\mneuOS{k}^2) \RP
            + \mneu{k} \LP \Si_{\neu{}}^{SL'}(\mneuOS{k}^2)
                           +\Si_{\neu{}}^{SR'}(\mneuOS{k}^2) \RP
       \Big]_{kk} \non \\
&\qquad \pm \ed{2 \mneu{k}} \LB \Si_{\neu{}}^{SL}(\mneuOS{k}^2)
            - \Si_{\neu{}}^{SR}(\mneuOS{k}^2)
            - \de\matr{M}_{\neu{}} 
            + \de\matr{M}_{\neu{}}^{*} \RB_{kk}~, \non \\
\LB \dcZ{\neu{}}^{L/R} \RB_{kl} &= \frac{2}{\mneu{k}^2 - \mneu{l}^2}
  \Big[ \mneu{l}^2 \Si_{\neu{}}^{L/R}(\mneuOS{l}^2) 
            + \mneu{k}\mneu{l} \Si_{\neu{}}^{R/L}(\mneuOS{l}^2) \\
&\qquad + \mneu{k} \Si_{\neu{}}^{SL/SR}(\mneuOS{l}^2)
        + \mneu{l} \Si_{\neu{}}^{SR/SL}(\mneuOS{l}^2)
        - \mneu{k/l} \de\matr{M}_{\neu{}} 
        - \mneu{l/k} \de\matr{M}_{\neu{}}^{\dagger} \Big]_{kl}~, \non
\end{align}
\begin{align}
\LB \dbcZ{\chapm{}}^{L/R} \RB_{ii} &=
        - \Big[ \Si_{\chapm{}}^{L/R}(\mcha{i}^2) \\
&\qquad + \mcha{i}^2 \LP \Si_{\chapm{}}^{L'}(\mcha{i}^2)
                       + \Si_{\chapm{}}^{R'}(\mcha{i}^2) \RP
        + \mcha{i} \LP \Si_{\chapm{}}^{SL'}(\mcha{i}^2)
                    +  \Si_{\chapm{}}^{SR'}(\mcha{i}^2) \RP \Big]_{ii}~\non \\
&\qquad \mp \ed{2 \mcha{i}} \LB 
                      \Si_{\chapm{}}^{SL}(\mcha{i}^2)
                    - \Si_{\chapm{}}^{SR}(\mcha{i}^2)
                      - \de\matr{M}_{\cham{}}
                      + \de\matr{M}_{\cham{}}^{*} \RB_{ii}~, \non \\
\LB \dbcZ{\chapm{}}^{L/R} \RB_{ij} &= \frac{2}{\mcha{j}^2 - \mcha{i}^2} 
        \Big[ \mcha{i}^2 \Si_{\chapm{}}^{L/R}(\mcha{i}^2) 
             +\mcha{i} \mcha{j} \Si_{\chapm{}}^{R/L}(\mcha{i}^2) \\
&\qquad + \mcha{i} \Si_{\chapm{}}^{SL/SR}(\mcha{i}^2)
       + \mcha{j} \Si_{\chapm{}}^{SR/SL}(\mcha{i}^2)
       - \mcha{i/j} \de\matr{M}_{\cham{}}
       - \mcha{j/i} \de\matr{M}_{\cham{}}^\dagger \Big]_{ij}~, \non \\
\label{diagneu}
\LB \dbcZ{\neu{}}^{L/R} \RB_{kk} &= 
 \LB \dcZ{\neu{}}^{R/L} \RB_{kk}~,  \\
\label{offdiagneu}
\LB \dbcZ{\neu{}}^{L/R} \RB_{kl} &= 
 \LB \dcZ{\neu{}}^{R/L} \RB_{lk}~.
\end{align}
The chargino/neutralino $\cZ$ factors obey
$\wtre\,\dbcZ{\tilde{\chi}}^{L/R} = 
[\wtre\,\dcZ{\tilde{\chi}}^{L/R}]^\dagger =
[\dZm{\tilde{\chi}}^{L/R}]^\dagger$, 
which is exactly the case without absorptive contributions. 
The Eqs.~\eqref{diagneu} and \eqref{offdiagneu} hold due to the Majorana 
character of the neutralinos.

\smallskip
After an OS renormalization (as will be discussed in the following
subsections) only the masses corresponding to the subset
of OS renormalized particles are OS masses. The other three masses then
require a finite shift to reach their OS value. 
The one-loop masses of the remaining charginos/neutralinos are 
obtained from the tree-level ones via the shifts~\cite{dissAF}:
\begin{alignat}{2}
\label{eq:Deltamcha}
\Delta \mcha{i} 
&= -\Re \bigg\{\mcha{i} \LP \SE{\cha{i}}^L(\mcha{i}^2) 
         + \frac{1}{2} \LB 
           \dZm{\cha{}}^L + \dBZm{\cha{}}^L 
                       \RB_{ii} \RP \notag \\
&\hspace{1.6cm} + \SE{\cha{i}}^{SL}(\mcha{i}^2) 
    - \frac{1}{2} \mcha{i} \LB 
      \dZm{\cha{}}^R + \dBZm{\cha{}}^R \RB_{ii}
    - \LB \delta \matr{M}_{\cha{}} \RB_{ii} \bigg\}\,, \\
\Delta \mneu{n} 
&= -\Re \bigg\{\mneu{k} \LP \SE{\neu{k}}^L(\mneu{k}^2) 
         + \frac{1}{2} \LB 
           \dZm{\neu{}}^R + \dbZm{\neu{}}^R 
                       \RB_{kk} \RP \notag \\
&\hspace{1.6cm} + \SE{\neu{k}}^{SL}(\mneu{k}^2) 
    - \frac{1}{2} \mneu{k} \LB 
      \dZm{\neu{}}^L + \dbZm{\neu{}}^L \RB_{kk}
    - \LB \delta \matr{M}_{\neu{}} \RB_{kk} \bigg\}
\label{eq:Deltamneu}
\end{alignat}
with $i = 1,2;\, k = 1,2,3,4$.
The (shifted) ``on-shell'' masses are obtained as
\begin{align}
\mcha{i}^\os = \mcha{i} + \Delta \mcha{i}\,, \qquad
\mneu{k}^\os = \mneu{k} + \Delta \mneu{k}\,.
\label{eq:minoOS}
\end{align}
These masses are used for external, on-shell particles. For
internal masses, i.e.\ of particles in the loops, the tree-level values
should be used. However, this can lead to IR divergences in the case of
charginos, when the divergence of a photon exchange has to cancel with
the real radiation. Therefore, also for charginos in the loops we use
shifted masses, leading to an IR- and UV-finite result for the cases
shown.
This issue is further discussed in \refse{sec:m2-var}; see also
the discussion in \citere{Stop2decay}.


\subsection{Semi-OS scheme}
\label{sec:semi-os}

\subsubsection{Concrete renormalization}
\label{sec:semi-os-ren}

We start with \DRbar\ mass matrices for charginos and neutralinos, collectively
denoted as $\matr{X}^{\DRbar}(P_i^{\DRbar})$, depending on the three input
parameters, 
\begin{align}
P_i^{\DRbar} &= M_1^{\DRbar}, M_2^{\DRbar}, \mu^{\DRbar} = \LV p_i^{\DRbar} \RV~%
, 
\end{align}
in addition to the (elsewhere renormalized) parameters $\tb, \MW, \MZ$, 
see eqs.~\refeqs{eq:X} and (\ref{eq:Y}).
The mass matrices can be diagonalized,\footnote{For the sake of simplicity 
we use the notation with a single diagonalization matrix as in the case of neutralinos. 
However, this can trivially be extended to the case of two diagonalization matrices as in the case of the charginos.
} 
\begin{align}
\matr{X}^{\DRbar} \to \matr{M}^{\DRbar} := 
(\matr{N}^{\DRbar})^\dagger \matr{X}^{\DRbar} \matr{N}^{\DRbar}~,
\end{align}
containing on the diagonal two charginos and four neutralino masses, $m_j$.\\

The $\matr{X}^{\DRbar}$ can be renormalized, 
\begin{align}
  \matr{X}^{\DRbar} &\to \matr{X}^{\DRbar} + \de\matr{X}^{\DRbar}(\de P_i^{\DRbar})~\\
\matr{M}^{\DRbar} &\to \matr{M}^{\DRbar} + \de\matr{M}^{\DRbar}(\de P_i^{\DRbar})
  = \matr{M}^{\DRbar} + (\matr{N}^{\DRbar})^\dagger 
                        \de\matr{X}^{\DRbar}(\de P_i^{\DRbar}) \matr{N}^{\DRbar}~.
                        \label{eq.XDRren}
\end{align}
So far, the $\de P_i^{\DRbar}$ are unknown.\\

The self-energies of the charginos and neutralinos can be written down as
\begin{align}
\SE{j}(P_i^{\DRbar}, \matr{X}^{\DRbar})~.
\end{align}

Now the RS is chosen: \ccn{n}\ or \cnn{c}{n}{n'}.
For each of these $N = 16$ 
schemes we perform the following. The scheme
is denoted as \rs{l} ($l = 1 \ldots 16$). 
Three renormalized self-energies are chosen to be zero, 
\begin{align}
\rSE{k,l}(P_i^{\DRbar}, \matr{X}^{\DRbar}) = 0~(k = 1,2,3)~,
\end{align}
corresponding to three $\Os$ masses, $m_k^\Os$.
The three renormalized
self-energies yield, for each RS$_l$, three conditions on
the counterterms $\de\matr{M}^{\DRbar}_k$
of the diagonal elements of $\matr{M}^{\DRbar}$,

\begin{align}
\de\matr{M}^{\DRbar}_{k,l} &= f^{\DRbar}_{k,l}(m_{k',l}^{\DRbar}, \SE{k'',l}) 
         + F^{\DRbar}_{k,l}(\de\tb, \de\MZ^2, \ldots) \label{eq.deltaMDR}\\[.5em]
\label{ADRbar}
&\downarrow \matr{A}_l^{\DRbar}\\[.5em]
\de P_{i,l}^\Os &= g^{\DRbar}_{i,l}(m_{k',l}^{\DRbar}, \SE{k'',l})
+ G^{\DRbar}_{i,l}(\de\tb, \de\MZ^2, \ldots)~,
\label{dePilOs}
\end{align}
yielding the $\Os$ values
\begin{align}
P_i^{\DRbar} &\to P_i^{\DRbar} - \de P_{i,l|{\rm fin}}^\Os \; = \; P_{i,l}^\Os~.
\label{PiOs}
\end{align}
It is worth noticing that in the r.h.s.\ of Eq.~(\ref{eq.deltaMDR})
$f_{k,l}$ is linear in $\de P_{i,l}^\Os$, while $F_{k,l}$ only depends on the
counterterm of the remaining model parameters. 
These relations define $\matr{A}_l^{\DRbar}$, 
the transformation matrix
from the set of parameter counterterms to mass counterterms 
(the explicit form of $\matr{A}_l^{\DRbar}$ can be found in
appendix~\ref{sec:matrA}), 
leading to
\begin{align}
\de P_{i,l}^\Os &= (\matr{A}_l^{\DRbar})^{-1}_{ik} 
        \LP \de \matr{M}^{\DRbar}_{k,l}
         - F_{k,l}(\de\tb, \de\MZ^2, \ldots) \RP~.
\label{eq.deltaPOsDR}
\end{align}
The $\Os$ masses $m_{k,l}^\Os$ are derived from 
\begin{align}
\matr{X}_l^\Os(P_{i,l}^{\Os}) \to \matr{M}_l^\Os := 
  (\matr{N}_l^\Os)^\dagger \matr{X}_l^\Os(P_{i,l}^{\Os}) \matr{N}^\Os~.
\label{MOs}
\end{align}
The three masses that are not obtained as $\Os$ masses so far can be
evaluated by adding finite shifts to them, see \refeq{eq:minoOS}.


\subsubsection{Selection of the best RS}
\label{sec:semi-os-sel}

As discussed above, an RS ``fails''
if  the set of mass matrix counterterms
for which we have set on-shell conditions
is (approximately) independent of
one of the parameter counterterms (or a combination of them).
This is exactly given in our ansatz if the matrix $\matr{A}_l^{\DRbar}$
does not provide a numerically ``well behaved'' transition 
\begin{align}
\de \matr{M}_{k.l}^{\DRbar} \stackrel{\matr{A}_l^{\DRbar}}{\to} \de P^\Os_{i,l}~,
\end{align}
see \refeqs{ADRbar}, suppressing terms involving other
counterterms ($\de\tb$, $\de\MZ^2$, \ldots).
Following the argument of the ``well behaved'' transition, \rs{l} fails if 
$\matr{A}^{\DRbar}_l$ becomes (approximately)
singular, or the normalized determinant, 
\begin{align}
\matr{D}^{\DRbar}_l := \frac{|\det\matr{A}^{\DRbar}_l|}
                           {||\matr{A}^{\DRbar}_l||} \ll 1~.
\end{align}
Conversely, the ``best'' scheme \rs{L} can be chosen 
as the one with the highest value of the normalized determinant,
\begin{align}
  \rs{L}^\Os \quad \Leftrightarrow \quad \matr{D}_L^{\DRbar} =
  \max_l \LV \matr{D}_l^{\DRbar} \RV~.
\end{align}
\noindent
Now all ingredients for physics calculations are at hand:
\begin{itemize}
\item
  The physical parameters $P_{i,L}^{\Os}$ are given via \refeq{PiOs}. 
\item
  The counterterms for the $P_{i,L}^{\Os}$ are known from
\refeq{dePilOs} as $\de P_{i,L}^{\Os}$ and can be inserted as
counterterms in a loop calculation.
\item
  Inserting $P_{i,L}^{\Os}$ into the Lagrangian yields six particle
masses out of which three are by definition given as their $\Os$ values.
The remaining $\Os$ masses have to be determined calculating three finite
shifts, see \refeq{eq:minoOS}.

\end{itemize}


\subsection{The full-OS scheme}
\label{sec:full-os}

\subsubsection{Full OS renormalization}
\label{sec:full-os-ren}

For each \rs{l} as evaluated in \refse{sec:semi-os-ren}
we now have $\Os$ mass matrices for charginos and neutralinos, collectively
denoted as $\matr{X}^\Os(P_{i,l}^{\Os})$ following \refeq{MOs}. We also
have $\Os$ parameters $P_{i,l}^\Os(P_i^{\DRbar})$ following \refeq{PiOs}
and $\de P_{i,l}^\Os(P_i^{\DRbar})$ following \refeq{dePilOs}.
This is unsatisfactory for a ``true'' OS scheme, i.e.\ one would like to have
$\de P_{i,l}^{\OS}(P_{i,l}^{\OS})$.
Furthermore, when a \rs{l}\ ``starts to turn bad'' as a function of a
\DRbar\ parameter, large differences between the $P_{i,l}^{\Os}$ and
$P_i^{\DRbar}$ occur, shedding doubt on the above outlined procedure.
These problems can be circumvented by extending the above scheme to an
evaluation of the counterterms in terms of OS parameters.

\medskip
We start with the $\Os$ parameters obtained in \refse{sec:semi-os-ren},
$P_{i,l}^\Os$. The mass matrices depend on these three input parameters,
\begin{align}
P_{i,l}^\Os &= M_{1,l}^\Os, M_{2,l}^\Os, \mu_l^\Os = \LV p_{i,l}^\Os \RV~.
\label{PilOs-start}
\end{align}
The mass matrices can be diagonalized as in \refeq{MOs}, 
\begin{align}
  \matr{X}_l^\Os \to \matr{M}_l^\Os :=
  (\matr{N}_l^\Os)^\dagger \matr{X}_l^\Os \matr{N}_l^\Os~,
\end{align}
containing on the diagonal two charginos and four neutralino masses,
$m_{j,}$, out of which three are by definition $\Os$ masses.
The remaining $\Os$ masses have to be determined calculating three finite
shifts, see \refeq{eq:minoOS}.

Now the renormalization process in \rs{l} is applied again, starting
from the above $\Os$ values.
The $\matr{X}_l^\Os$ can be renormalized, 
\begin{align}
\matr{X}_l^\Os &\to \matr{X}_l^\Os + \de\matr{X}_l^\Os(\de P_{i,l}^\Os)~\\
\matr{M}_l^\Os &\to \matr{M}_l^\Os + \de\matr{M}_l^\Os(\de P_{i,l}^\Os)
  = \matr{M}_l^\Os 
  + (\matr{N}_l^\Os)^\dagger \de\matr{X_l}^\Os(\de P_{i,l}^\Os) \matr{N}_l^\Os~.
\end{align}
So far the $\de P_{i,l}^\Os$ are unknown.\\

The self-energies of the charginos and neutralinos can be written down as
\begin{align}
\SE{j}(P_{i,l}^\Os, \matr{X}_l^\Os)~.
\end{align}

According to \rs{l} three
renormalized self-energies are chosen to be zero, 
\begin{align}
\rSE{k,l}(P_{i,l}^\Os, \matr{X}_l^\Os) = 0~(k = 1,2,3)~,
\end{align}
corresponding to three OS masses, $m_k^\OS$. The three renormalized
self-energies yield three conditions on $\de\matr{M}^\Os_k$,
\begin{align}
\label{eq.deltaMOs}
\de\matr{M}^\Os_{k,l} &= f^\Os_{k,l}(m_{k',l}^\Os, \SE{k'',l}) 
                   + F^\Os_{k,l}(\de\tb, \de\MZ^2, \ldots)~\\[.5em]
\label{AOS}
&\downarrow \matr{A}_l^\Os\\[.5em]
\de P_{i,l}^\OS &= g^\Os_{i,l}(m_{k',l}^\Os, \SE{k'',l})
   + G^\Os_{i,l}(\de\tb, \de\MZ^2, \ldots)~.
\label{dePilOS}
\end{align}
yielding the $\OS$ values
\begin{align}
  P_{i,l}^{\Os} &\to P_{i,l}^{\Os} + \de P_{i,l}^\Os
                         - \de P_{i,l}^\OS \; = \; P_{i,l}^\OS~.
\label{PilOS-full}
\end{align}
It is worth noticing that in the r.h.s.\ of Eq.~(\ref{eq.deltaMOs})
$f^\Os_{k,l}$ is linear in $\de P_{i,l}^\Os$ and $\de P_{i,l}^\OS$,
while $F^\Os_{k,l}$ only depends on the
counterterm of the remaining model parameters. 
These relations define
$\matr{A}_l^\Os$ as the transformation matrix 
from the set of parameter counterterms to mass counterterms 
(the explicit for of $\matr{A}_l^\Os$ can be obtained from the
formulas in appendix~\ref{sec:matrA} by replacing $\DRbar$ quantities
by $\Os$ quantities), leading to
\begin{align}
\de P_{i,l}^\OS &= (\matr{A}_l^\Os)^{-1}_{ik} 
    \LP \de \matr{M}^\Os_{k,l} - F_{k,l}^\Os(\de\tb, \de\MZ^2, \dots) \RP~.
\label{eq.deltaPOSOS}
\end{align}
$\OS$ masses $m_{k,l}^\OS$ are derived from 
\begin{align}
\matr{X}_l^\OS(P_{i,l}^{\OS}) \to \matr{M}_l^\OS := 
  (\matr{N}_l^\OS)^\dagger \matr{X}_l^\OS(P_{i,l}^{\OS}) \matr{N}^\OS~.
\label{MOS}
\end{align}
The three masses that are not obtained as $\OS$ masses so far can be
evaluated by adding finite shifts to them, see \refeq{eq:minoOS}. 


\subsubsection{Selection of the best RS}
\label{sec:full-os-sel}

As in the case of the semi-OS scheme, a bad \rs{l} is indicated if
in our ansatz the matrix $\matr{A}_l^\OS$ does not
provide a numerically ``well behaved'' transition 
\begin{align}
\de \matr{M}_{k,l}^\Os \stackrel{\matr{A}_l^\Os}{\to} \de P_{i,l}^\OS~,
\end{align}
see \refeq{AOS}, and suppressing terms involving other
counterterms ($\de\tb$, $\de\MZ^2$, \ldots).
Following the \refse{sec:semi-os-ren} (i.e.\ starting with \DRbar\ input
parameters), for each set of $P_i^{\DRbar}$ two such matrices exist for each
\rs{l}, 
\begin{align}
\rs{l} \quad \Rightarrow \quad \matr{A}^{\DRbar}_l, \matr{A}^\Os_l~.
\end{align}
For each $P_i^{\DRbar}$ one has two sets of $\matr{A}$ matrices,
\begin{align}
\LV \matr{A}^{\DRbar}_l \RV, \quad 
\LV \matr{A}^\Os_l \RV~.
\end{align}
Following the argument of the ``well behaved'' transition, \rs{l} fails
if either $\matr{A}^{\DRbar}_l$ or $\matr{A}^\Os_l$ become (approximately)
singular, or their normalized determinant, 
\begin{align}
\label{eq:matrA}
\matr{D}^{\DRbar}_l := \frac{|\det\matr{A}^{\DRbar}_l|}
                                 {||\matr{A}^{\DRbar}_l||} \ll 1
\quad\mbox{or}\quad
\matr{D}^{\Os}_l := \frac{|\det\matr{A}^{\Os}_l|}
                                 {||\matr{A}^{\Os}_l||} \ll 1~,
\end{align}
equivalent to
\begin{align}
  \label{eq:matrAOS}
  \matr{D}_l^\OS := 
  \min\LV \matr{D}^{\DRbar}_l, \matr{D}^{\Os}_l \RV \ll 1~.
\end{align}
Conversely, the ``best'' scheme \rs{L} can be chosen 
as the one with the highest value of the normalized determinant,
\begin{align}
  \rs{L}^\OS \quad \Leftrightarrow
      \quad \matr{D}^\OS_L = \max_l \LV \matr{D}^\OS_l \RV~.
\end{align}
\noindent
Now all ingredients for physics calculations are at hand:
\begin{itemize}
\item
  The physical parameters $P_{i,L}^{\OS}$ are given via \refeq{PilOS-full}. 
\item
  The counterterms for the $P_{i,L}^{\OS}$ are known from
\refeq{dePilOS} as $\de P_{i,L}^{\OS}$ and can be inserted as
counterterms in a loop calculation.
\item
  Inserting $P_{i,L}^{\OS}$ into the Lagrangian yields six particle
masses out of which three are by definition given as their $\OS$ values.
The remaining $\OS$ masses have to be determined calculating three finite
shifts, see \refeq{eq:minoOS}. 
\end{itemize}

\subsection{Possible iteration}
\label{sec:iteration}

For each \rs{l} as evaluated in \refse{sec:full-os-ren}
we now have $\OS$ mass matrices for charginos and neutralinos, collectively
denoted as $\matr{X}^\OS(P_{i,l}^{\OS})$ following \refeq{MOS}. We also
have $\OS$ parameters $P_{i,l}^\OS(P_i^{\Os})$ following \refeq{PilOS-full}
and $\de P_{i,l}^\OS(P_i^{\Os})$ following \refeq{dePilOS}.
This might still be unsatisfactory for a ``true'' OS scheme,
i.e.\ one would like to have $\de P_{i,l}^{\OS}(P_{i,l}^{\OS})$.
These problems can be circumvented by extending the above scheme to an
{\em iterative} evaluation of the counterterms in terms of OS parameters.

The procedure in the previous subsection can be repeated (starting from
\refeq{PilOs-start} on) using the previously obtained OS parameters as
$\Os$ parameters, i.e.
\begin{align}
  P_{i,l}^\OS &\to P_{i,l}^\Os~,\\
  \matr{X}_l^\OS &\to \matr{X}_l^\Os~,\\
  \matr{M}_l^\OS &\to \matr{M}_l^\Os~.
\end{align}
These parameters are used as indicated starting from
\refeq{PilOs-start}, until ``new'' OS parameters are derived,
$P_{i,l}^\OS$ in \refeq{PilOS-full}, $\matr{X}_l^\OS$ and $\matr{M}_l^\OS$
as in \refeq{MOS}, as well as the three finite mass shifts to OS values,
see \refeq{eq:minoOS}. 

The iteration can be stopped, when convergence is reached, i.e.\ the
comparison of the $\Os$ (or ``old'' OS) values and the (``new'') OS
values in one iteration are sufficiently close, 
\begin{align}
  \frac{|P_{i,l}^\OS - P_{i,l}^\Os|}{|P_{i,l}^\OS|} \le \epsilon ~\forall~i~,
\end{align}
with $\epsilon$ chosen appropriately.
If no convergence is observed, this indicates a ``bad'' \rs{l}, which
can be discarded at this point.

\smallskip
We will not use this iteration in our numerical evaluation, but remark
on the size of
\begin{align}
  \frac{|P_{i,l}^\OS - P_{i,l}^\Os|}{|P_{i,l}^\OS|} ~\forall~i
\end{align}
in our ``one step'' full-OS scheme.
  



\section{Numerical investigation}
\label{sec:numerical}

In our numerical investigation we will concentrate on the full-OS
scheme. The theoretical set-up presented in \refse{sec:full-os} will be
applied to concrete numerical scenarios, covering all relevant mass
hierarchies in the chargino/neutralino sector of the MSSM.
As physical observable we choose a decay width of a chargino or
neutralino, as will be specified below. We will
demonstrate that 
\begin{itemize}
\item[(i)] one can identify an \rs{L} for each $P_i^{\DRbar}$,
\item[(ii)] the calculated physics process, $\sig_L$ 
(in our case a decay width), employing \rs{L}, is
  numerically well behaved, i.e.\ the higher-order corrections are relatively
  small for all $P_i^{\DRbar}$, 
\item[(iii)] the full cross section (tree + one-loop corrections) is
  effectively smooth.
\end{itemize}


\subsection{Benchmark scenarios}
\label{sec:bench}

Our numerical investigation is performed in eight benchmark scenarios,
given in \refta{tab:bench1}. We specify the three mass parameters
governing the chargino/neutralino sector, $M_1$, $M_2$ and $\mu$, as
well as $\tb$. The other parameters that enter via higher-order
corrections are chosen as follows. 
    \begin{itemize}
    \item SUSY parameters:\\
      The diagonal soft SUSY-breaking parameters are chosen as
      $M_{\rm SUSY}=1.5\tev$ for all sleptons and squarks,\\
      The trilinear couplings are fixed as
    $|A_t|=|A_b|=2\tev$, $|A_\tau|=1.5\tev$ and are set to zero for the
      first and second generation.
    \item 3rd generation fermion masses:
      \begin{align}
    m_\tau =1776.86\mev~,\quad  
    m_t =173.1\gev~,\quad
    m_b(m_b)=4.18\gev\,. \non
      \end{align}
    \item The CKM matrix has been set to unity.
    \item Gauge boson masses:
      \begin{align}
        M_W & =80.385\gev, \quad M_Z=91.1876\gev~.
      \end{align}
    \item Coupling constants:
      \begin{align}
        \alpha &=\frac{e^2}{4\pi}=1/137.035999139\,, \non \\
        \alpha_s &:
        \mbox{computed at the process scale from~}
        \alpha_s(M_Z)=0.1182\,. \non \\
      \end{align}
    \item Higgs sector: $M_{H^\pm}=500\gev$,
The Higgs sector quantities have been evaluated using
{\tt FeynHiggs} version
2.13.0~\cite{feynhiggs,mhiggslong,mhiggsAEC,Mh-logresum,feynhiggs-new,mh-unc}
using the above SUSY parameters. 
    \end{itemize}

In each scenario given in \refta{tab:bench1} one mass parameter in the
chargino/neutralino sector is varied, while the others are kept
fixed. In this way we cover all relevant mass hierarchies and in
particular smoothly switch from one hierarchy to another. 
It should be noted here, that we are not aiming at a 
phenomenological analysis with parameter points that pass all existing
experimental and theoretical constraints. Our benchmark scenarios are
chosen to demonstrate how our algorithm for the automated choice of a
good RS works, while not being completely unrealistic. As an example,
the choice of soft SUSY-breaking parameters yields a light $\cp$-even
Higgs-boson mass of roughly $125 \gev$, but the parameters were not
tuned to yield this value precisely, since this is irrelevant for our
work (similarly, using an older version of {\tt FeynHiggs} does not play
a relevant role). 

\begin{table}[h]
\begin{center}
\caption{\label{tab:bench1} SUSY parameters in the chargino/neutralino
  sector, defining the benchmark scenarios for our numerical
  investigation. }
\begin{tabular}{|l||c|c|c|c|c}
  \hline
  Benchmark & $M_1[\gev]$ & $M_2[\gev]$  & $\mu[\gev]$ & $\tb$  \\
  \hline\hline
  B\_$\mu\, a$ & 200 & 500     & 150-700 & 10 \\
  \hline
  B\_$\mu\, b $ & -200 & 500     & 150-700 & 10 \\
  \hline
  B\_$\mu\, c $ & 500 & 300     & 150-700 & 10 \\
  \hline\hline
  B\_$M_1\, a $ & 100-700 & 200     & 500 & 10 \\
  \hline
  B\_$M_1\, b $ & 100-700 & 500     & 200 & 10 \\
  \hline\hline
  B\_$M_2\, a$ & 200 & 150-700     & 500 & 10 \\
  \hline
  B\_$M_2\, b $ & -200 & 150-700     & 500 & 10 \\
  \hline
  B\_$M_2\, c $ & 500 & 150-700     & 300 & 10 \\
  \hline
\end{tabular}
\end{center}
\end{table}

All numerical results are obtained using the \FA/\FC/\LT\
set-up~\cite{feynarts1,feynarts2,feynarts3,formcalc1,formcalc2} with the 
MSSM model file as defined in \citere{MSSMCT}
(\FA\ version 3.9, \FC\ version 9.5, \LT\ version 2.14).
We have modified the public \FA/\FC\ versions to include our new
method for the RS choice. We plan to make these modifications
public. For further illustration of our scenarios, we show the masses
of the charginos and neutralinos in \reffi{fig:chimasses} in 
appendix~\ref{sec:plots}. 



\subsection{Variation of \boldmath{$\mu$}}
\label{sec:mu-var}

\begin{figure}[h!]
\begin{center}
  \includegraphics[width=0.60\textwidth]{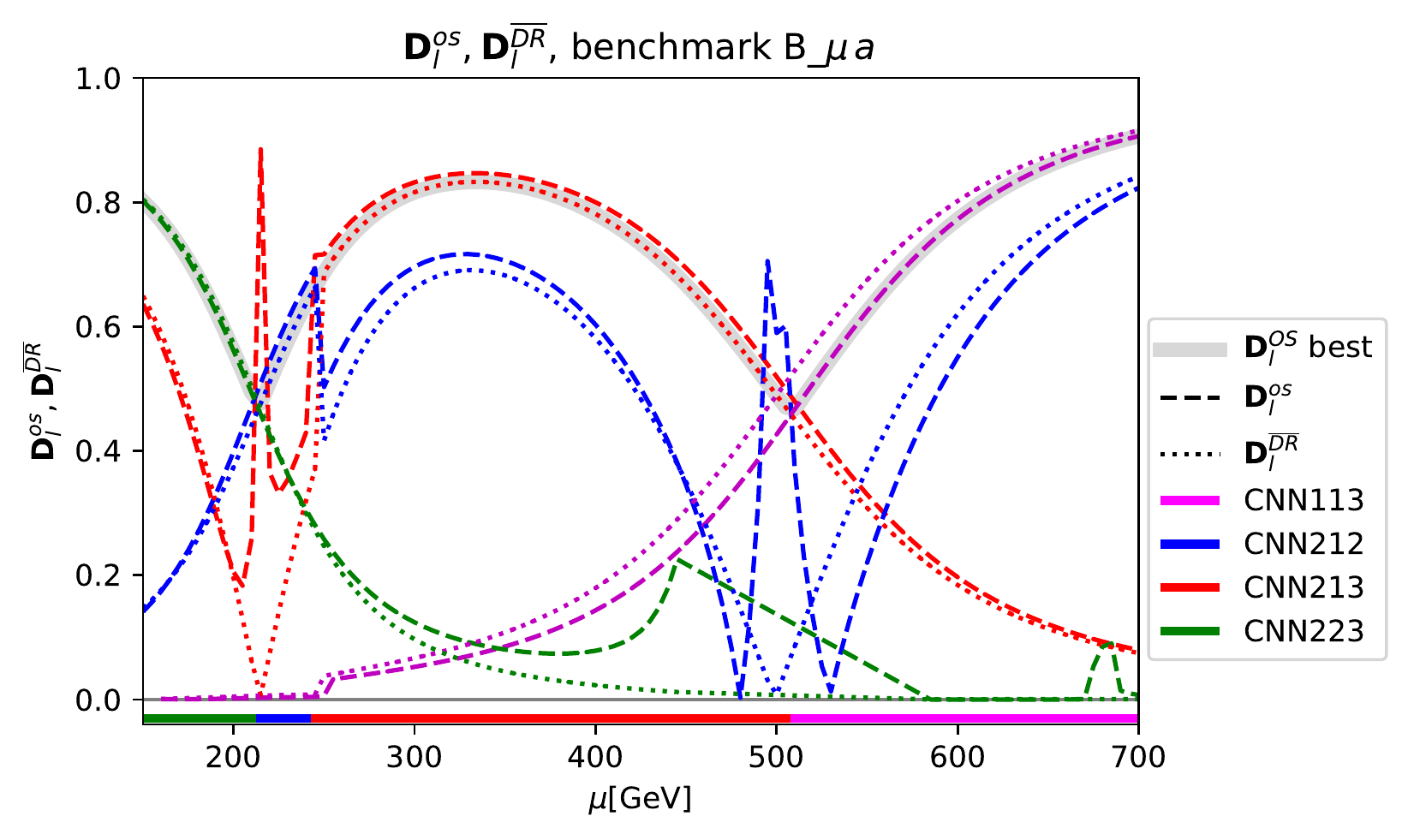}
  \includegraphics[width=0.45\textwidth]{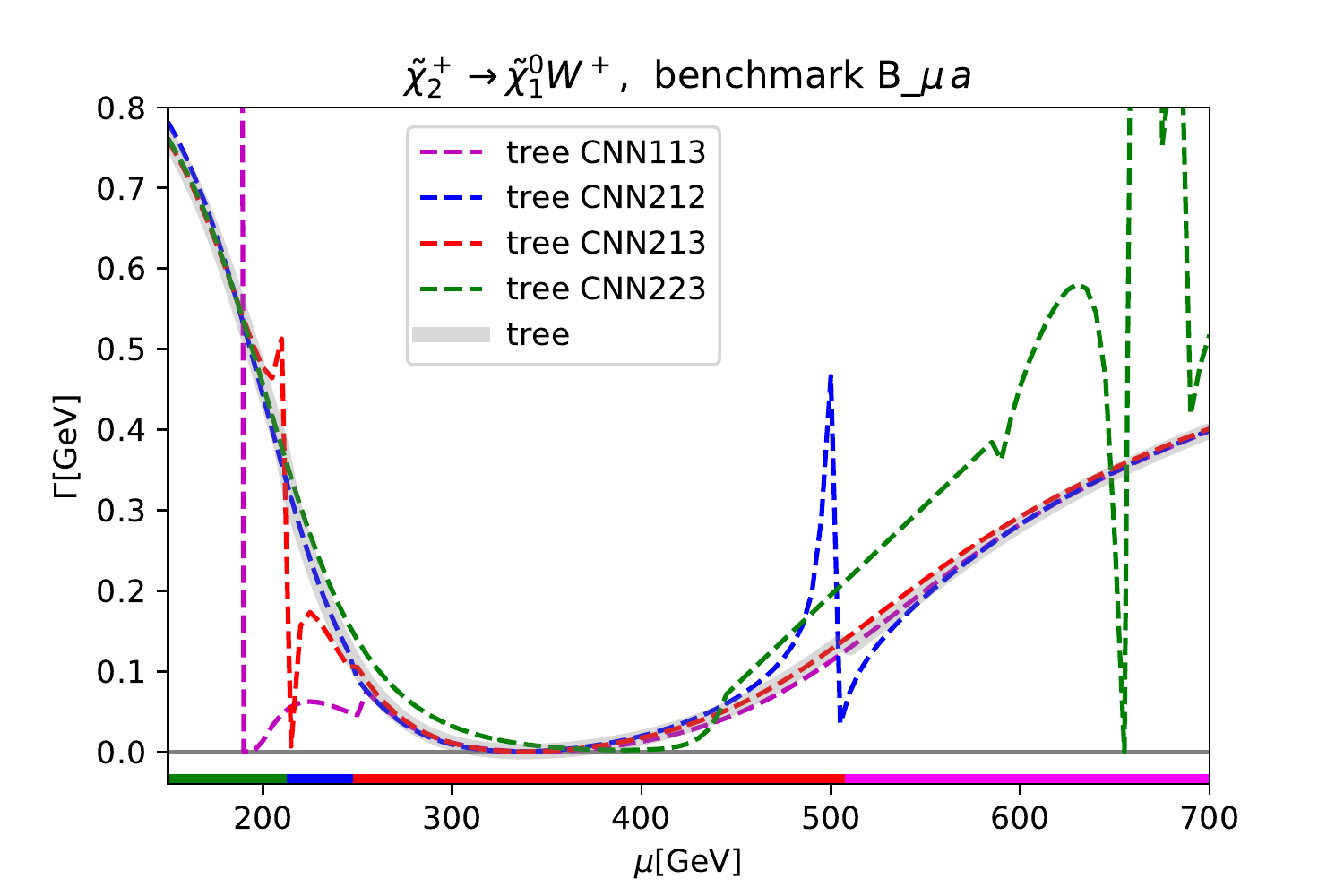}
  \includegraphics[width=0.45\textwidth]{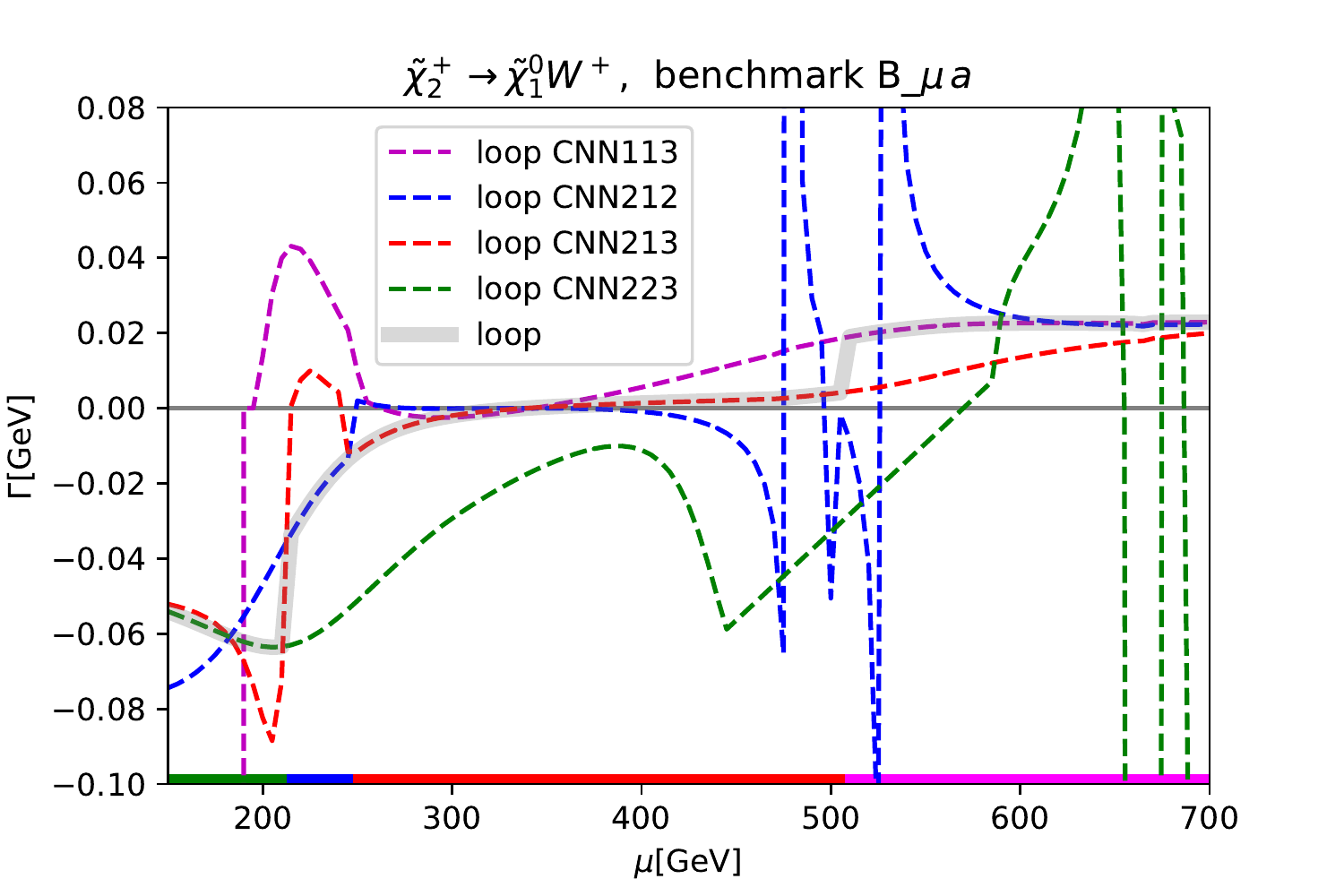}
  \includegraphics[width=0.45\textwidth]{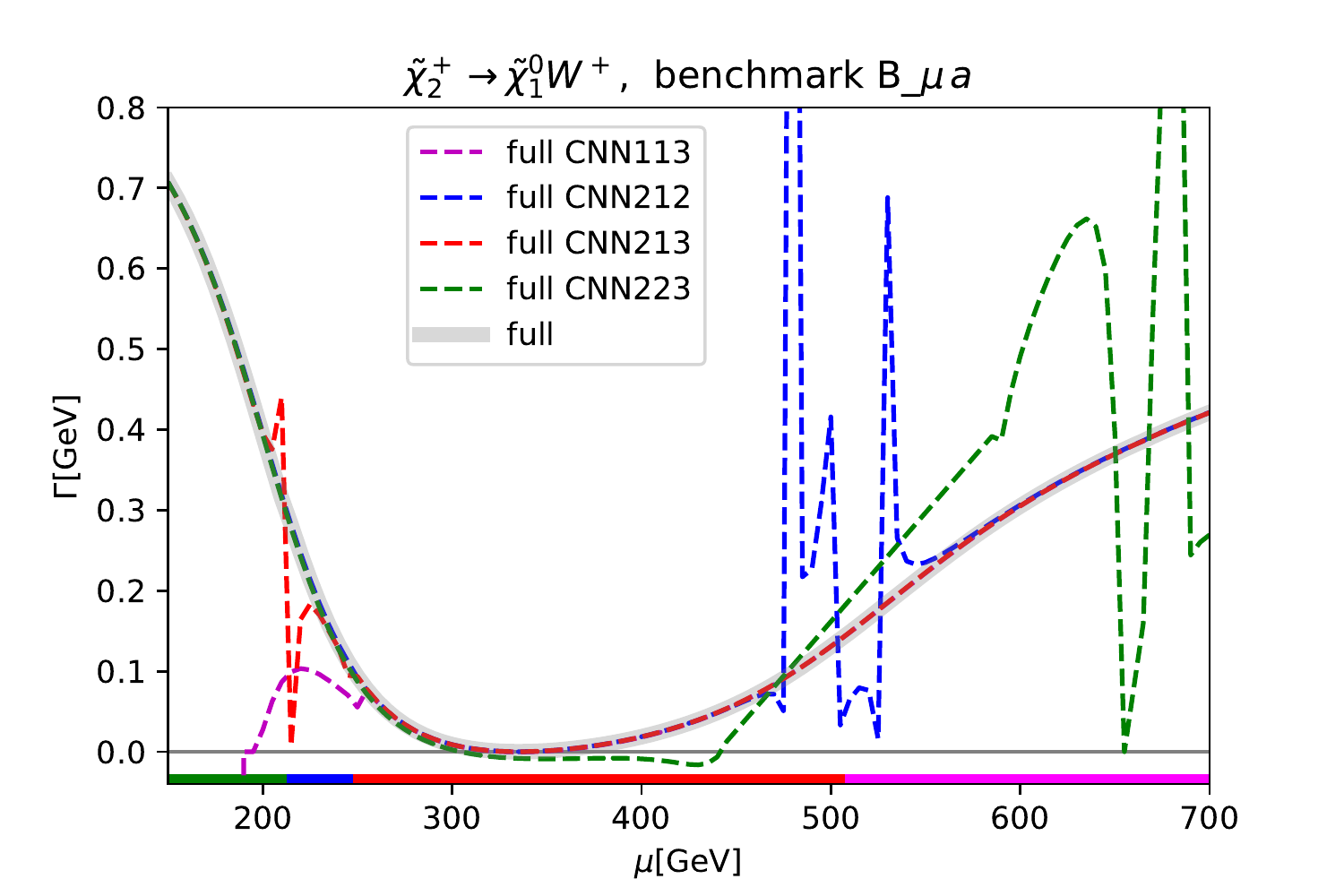}
  \includegraphics[width=0.45\textwidth]{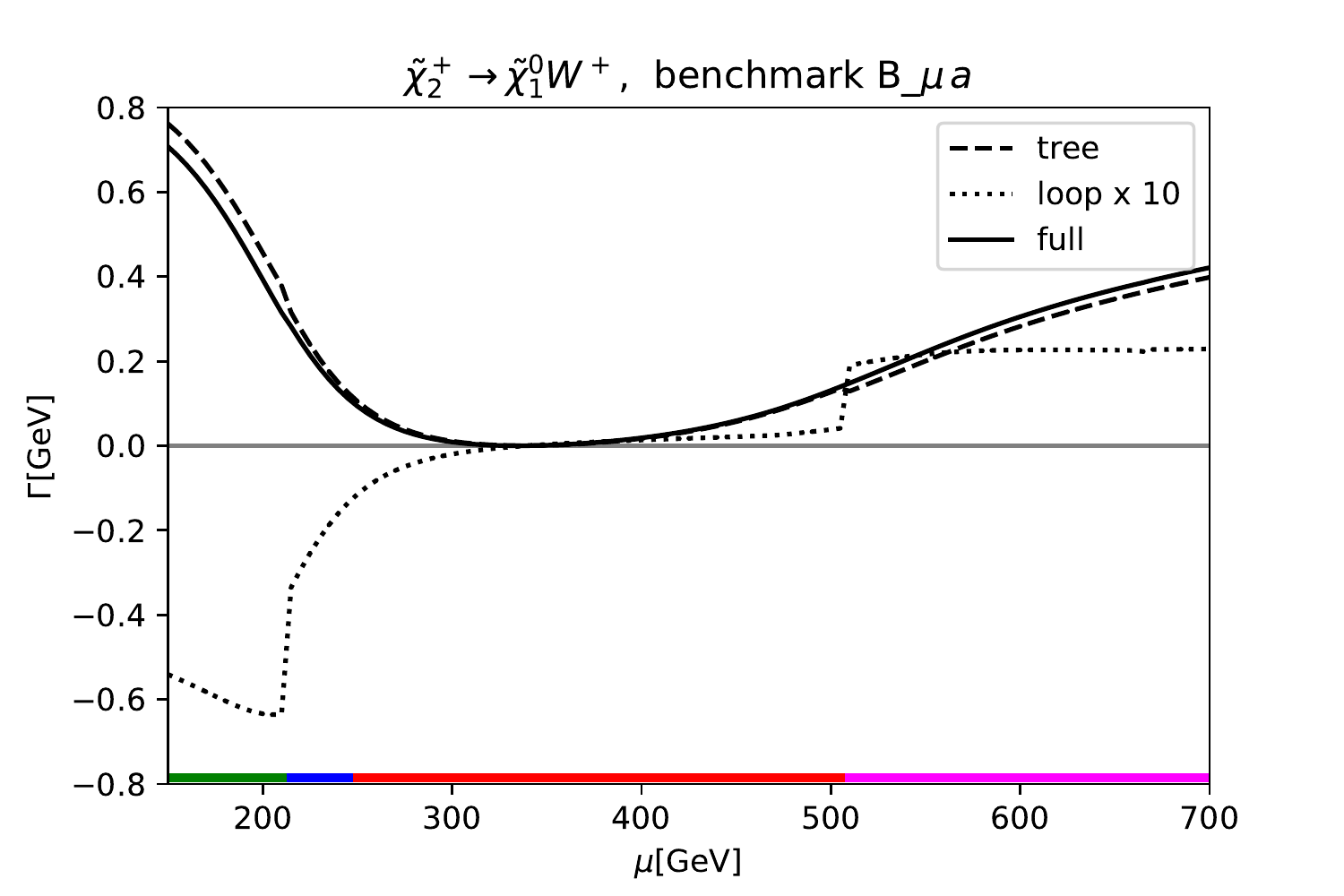}
\vspace{2em}
  \caption{Decay width for $\chap{2} \to \neu{1} W^+$ as a function of $\mu$
    in benchmark scenario B\_$\mu\,a$, see \refta{tab:bench1},
    with $M_1= 200\gev$, $M_2=500\gev$, $\tb= 10$. 
    Shown are the four ``best RS'' in this range of parameters (see text).
    The upper plot shows the (``naturally normalized'') determinants
    $\matr{D}^{\DRbar}_l $ (dotted) and $\matr{D}^{\Os}_l $ (dashed),
    see \refeq{eq:matrA}.
    The middle left (right) figure shows the tree (the loop plus real
    photon emission) results. 
    The bottom left figure shows the full NLO decay widths obtained
    summing results shown in the middle figures. 
    The bottom right figure shows the tree, loop and full results in the
    ``best'' RS for every parameter point. 
   The horizontal colored bar shows the best RS for the corresponding
   value of  $\mu$, 
   following  the same color coding as the curves:
   CNN$_{223}$ for $\mu\le 210\gev$ (green), 
   CNN$_{212}$ for $215 \gev\le\mu\le 245\gev$ (blue), 
   CNN$_{213}$ for $250 \gev\le\mu\le 505\gev$ (red), 
   CNN$_{113}$ for $510\gev\le\mu$ (pink). 
}
\label{fig:2022_C2N1W_mu_a}
\end{center}
\end{figure}

We start our numerical investigation with the benchmark scenario B\_$\mu\,a$.
In the scenario we have $200 \gev = M_1 < M_2 = 500 \gev$, while $\mu$
varies from $150 \gev$ to $700 \gev$, i.e.\ all hierarchies with $M_1 < M_2$
(with $M_1$ positive) are covered. In \reffi{fig:2022_C2N1W_mu_a} we
show the results for the
decay width for $\Ga(\chap{2} \to \neu{1} W^+)$ as a function of $\mu$
(for $M_1 = 200 \gev$, $M_2 = 500 \gev$ and $\tb = 10$), see also
\citere{ll-procs}. 
The upper plot shows the (``naturally normalized''%
) determinants
$\matr{D}^{\DRbar}_l $ (dotted) and $\matr{D}^{\Os}_l $ (dashed),
see \refeq{eq:matrA} in four colors for the four ``best RS''.
The results of the ``selected best RS'' are overlaid with a gray band.
The horizontal colored bar indicates this best RS for the corresponding
value of  $\mu$,  
following  the same color coding as the curves:%
~CNN$_{223}$ for $\mu \lsim 210\gev$ (green), 
CNN$_{212}$ for $215 \gev \lsim \mu \lsim 245\gev$ (blue), 
CNN$_{213}$ for $250 \gev \lsim \mu \lsim 505\gev$ (red), 
CNN$_{113}$ for $510 \gev \lsim \mu$ (pink). 
In this example the selected best scheme has determinants larger than
$\sim 0.5$, indicating that the 
counterterms can be determined reliably.

The middle left figure shows the tree results for the same four selected
RS as colored
dashed lines, and the results of the ``selected best RS'' are again overlaid
with a gray band.
One can observe that where a scheme is chosen, the tree level width
behaves ``well'' and smoothly. It reaches zero at $\mu \sim 330 \gev$
because the involved tree-level coupling has an (accidental) zero
crossing. On the other hand, outside the selected interval the
tree-level result behaves highly irregularly, induced by the shifts in the
mass matrices to obtain OS masses. 

The middle right plot shows the ``loop plus real photon emission''
results with the same color coding as in the middle left plot.
As for the tree-level result one can see that where a scheme is chosen the
loop corrections behave smoothly and the overall size stays at the level
of $\sim 10\%$ or less compared to the tree-level result. As above,
outside the chosen interval the loop corrections take irregular values,
which sometimes even diverge, owing to a vanishing determinant.

The lower left plot, using again the same color coding, shows the sum of
tree and higher-order corrections, i.e.\ of the two previous plots. The
same pattern of numerical behavior can be observed. The chosen scheme
yields a reliable higher-order corrected result, whereas other schemes
result in highly irregular and clearly unreliable results. This is
summarized in the lower right plot, where we show the selected tree-level
result as dashed line, the loop result as dotted (multiplied
by~10 for better visibility), and the full result as  
solid line. The overall behavior is completely well-behaved and
smooth. A remarkable feature can be observed at $\mu \sim 500
\gev$. Here the selected tree-level result has a kink, because of a
change in the shift in the OS values of the involved chargino/neutralino
masses, caused by the change from switching from \cnn{2}{1}{3} to
\cnn{1}{1}{3}. However, the loop corrections contain also a
corresponding kink, leading to a completely smooth full one-loop result.

\begin{figure}[h!]
\vspace{2em}
\begin{center}
  \includegraphics[width=0.60\textwidth]{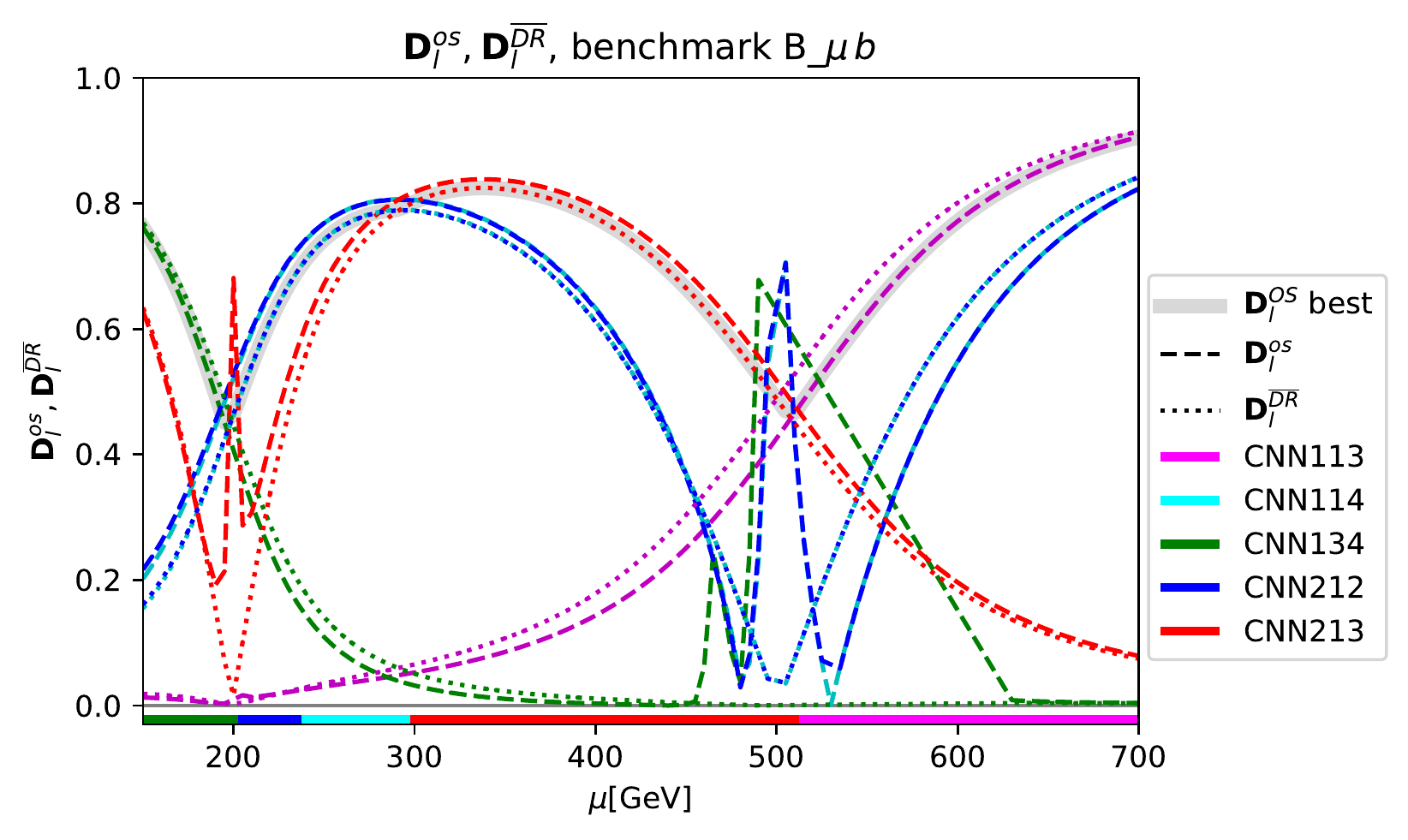}\\[1em]
  \includegraphics[width=0.45\textwidth]{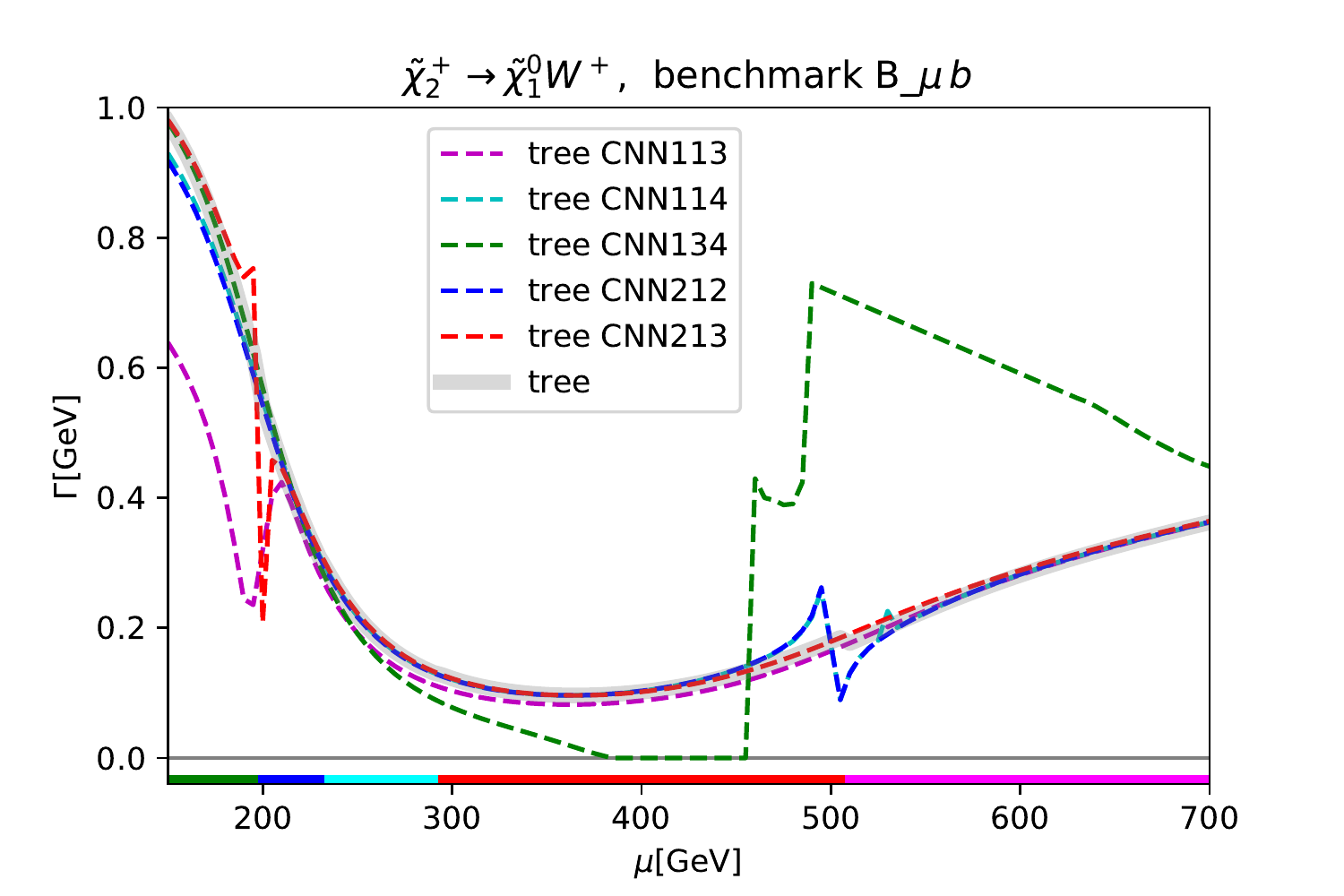}
  \includegraphics[width=0.45\textwidth]{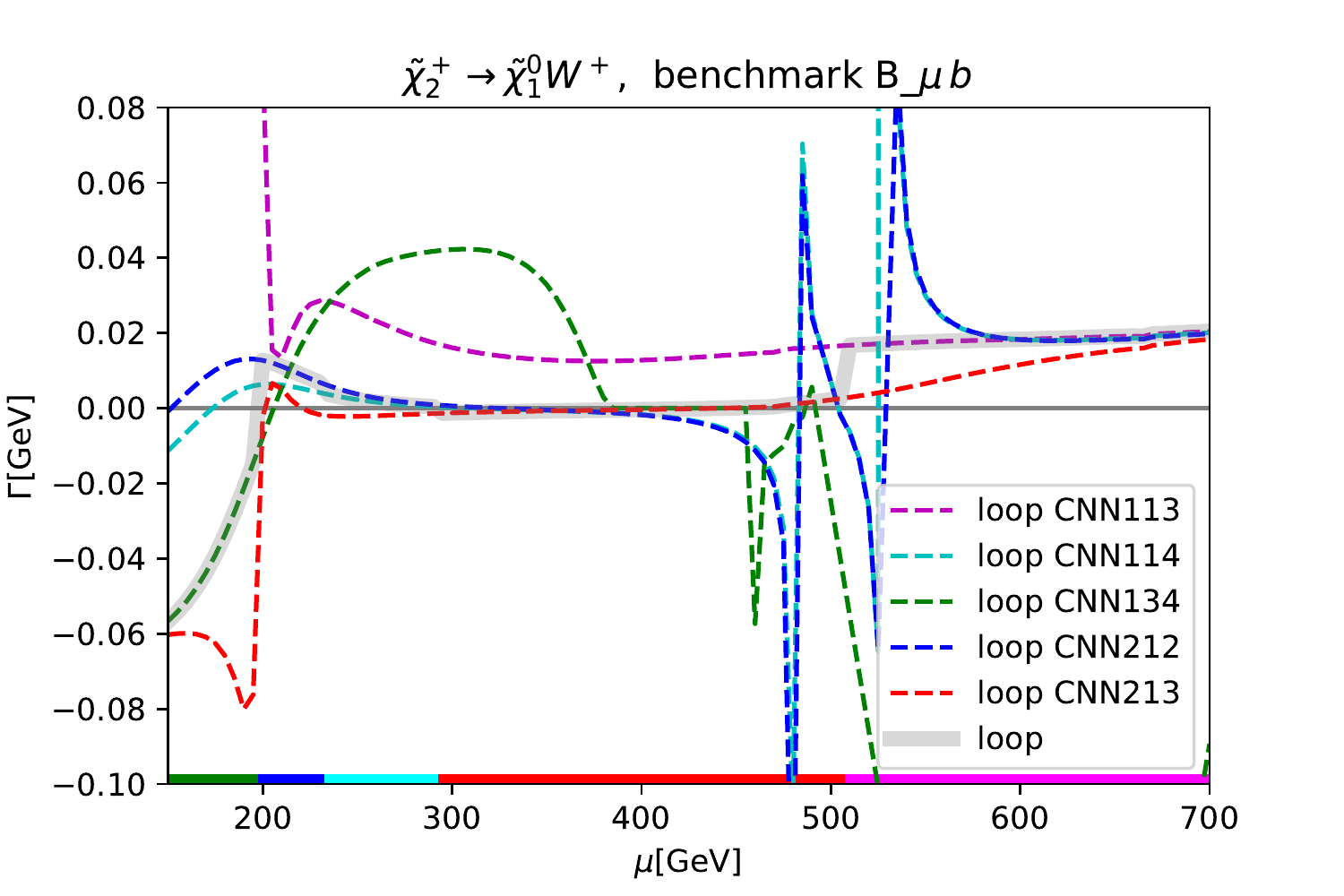}\\[1em]
  \includegraphics[width=0.45\textwidth]{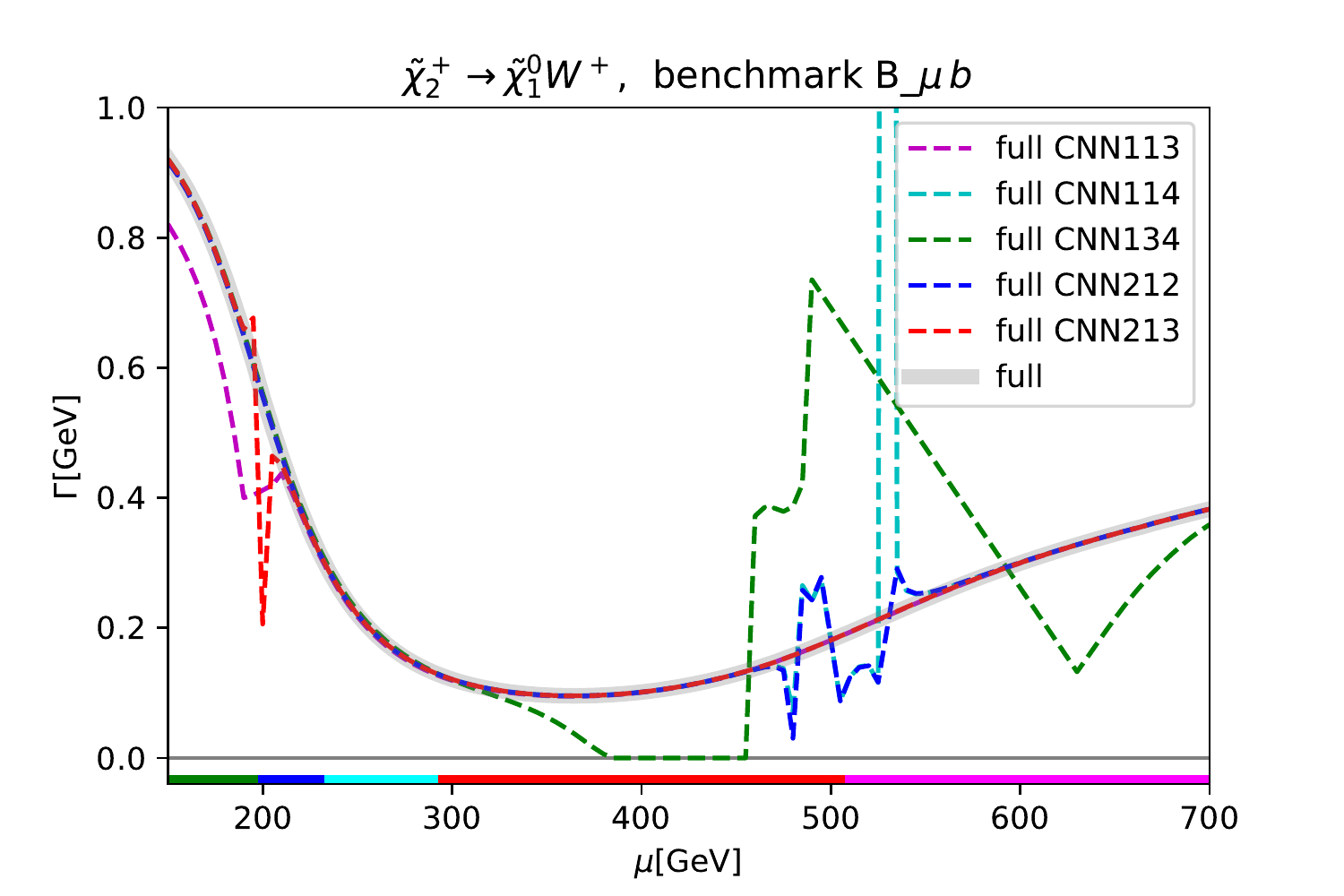}
  \includegraphics[width=0.45\textwidth]{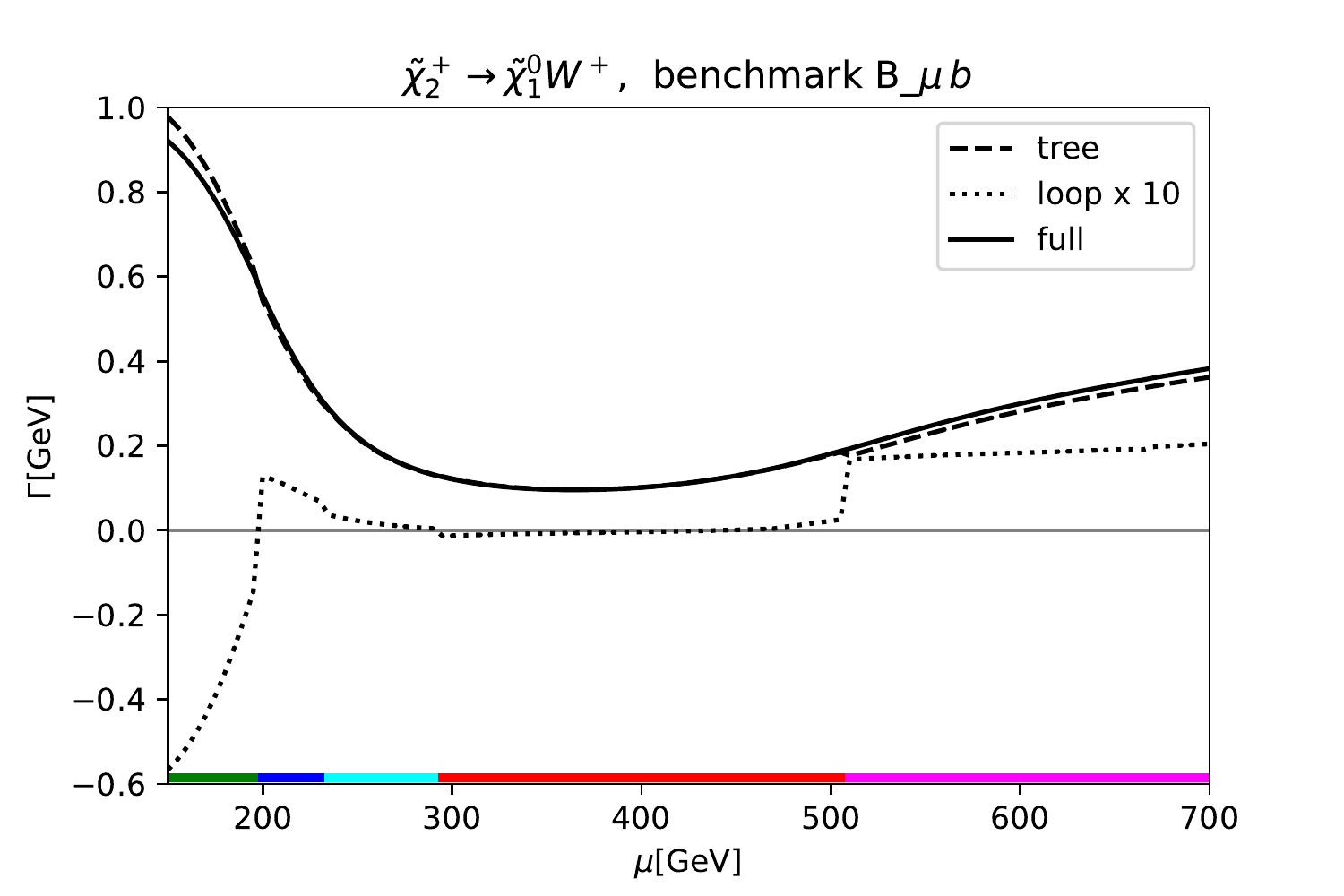}
\vspace{2em}
  \caption{
    Decay width for $\chap{2} \to \neu{1} W^+$ as a function of $\mu$
    in benchmark scenario B\_$\mu\,b$, see \refta{tab:bench1},
    with $M_1= -200\gev$, $M_2=500\gev$, $\tb= 10$. 
    The plots show the same quantities as in
    \protect\reffi{fig:2022_C2N1W_mu_a}. 
      Shown are the five ``best RS'' in this range of parameters (see text).
   The horizontal colored bar shows the best RS for the corresponding
   value of  $\mu$, 
   following  the same color coding as the curves:
   CNN$_{134}$ for $\mu\le 195\gev$ (green),
   CNN$_{212}$ for $200 \gev\le\mu\le 230\gev$ (blue),  
   CNN$_{114}$ for $235 \gev\le\mu\le 290\gev$ (cyan),  
   CNN$_{213}$ for $295 \gev\le\mu\le 505\gev$ (red), 
   CNN$_{113}$ for $510\gev\le\mu$ (pink).
  }
\label{fig:2022_C2N1W_mu_b}
\end{center}
\end{figure}

\bigskip
The next benchmark scenario B\_$\mu\,b$ differs from B\_$\mu\,a$ by the
sign of $M_1$, i.e.\ we set $M_1 = -200 \gev$, 
$M_2 = 500 \gev$, while $\mu$
varies from $150 \gev$ to $700 \gev$. In \reffi{fig:2022_C2N1W_mu_b} we
show the results for the same decay width as in
\reffi{fig:2022_C2N1W_mu_a}, $\Ga(\chap{2} \to \neu{1} W^+)$ as a
function of $\mu$
with the same arrangement of the plots. 
The upper plot shows the selection of the ``best RS'', where again 
the horizontal colored bar indicates this best RS for the corresponding
value of  $\mu$:
CNN$_{134}$ for $\mu\le 195\gev$ (green), 
CNN$_{212}$ for $200 \gev\le\mu\le 230\gev$ (blue), 
CNN$_{114}$ for $235 \gev\le\mu\le 290\gev$ (cyan), 
CNN$_{213}$ for $295 \gev\le\mu\le 505\gev$ (red), 
CNN$_{113}$ for $510\gev\le\mu$ (pink). 
These selections partially agree with the scheme selected in
B\_$\mu\,a$, but also some schemes are chosen that do not appear in
the previous scenario. As in B\_$\mu\,a$
the selected best scheme has determinants larger than
$\sim 0.5$, indicating that the 
counterterms
 can be determined reliably.

Overall, the result is qualitatively the same as for B\_$\mu\,a$: the
selected RS yields a stable tree-level result, stable loop corrections
and a smooth and stable full one-loop result is obtained. For the
negative sign of $M_1$ the tree-level coupling does not possess the
accidental zero crossing, yielding decay widths always larger than
$\sim 0.1 \gev$.

\begin{figure}[h!]
\vspace{2em}
\begin{center}
  \includegraphics[width=0.6\textwidth]{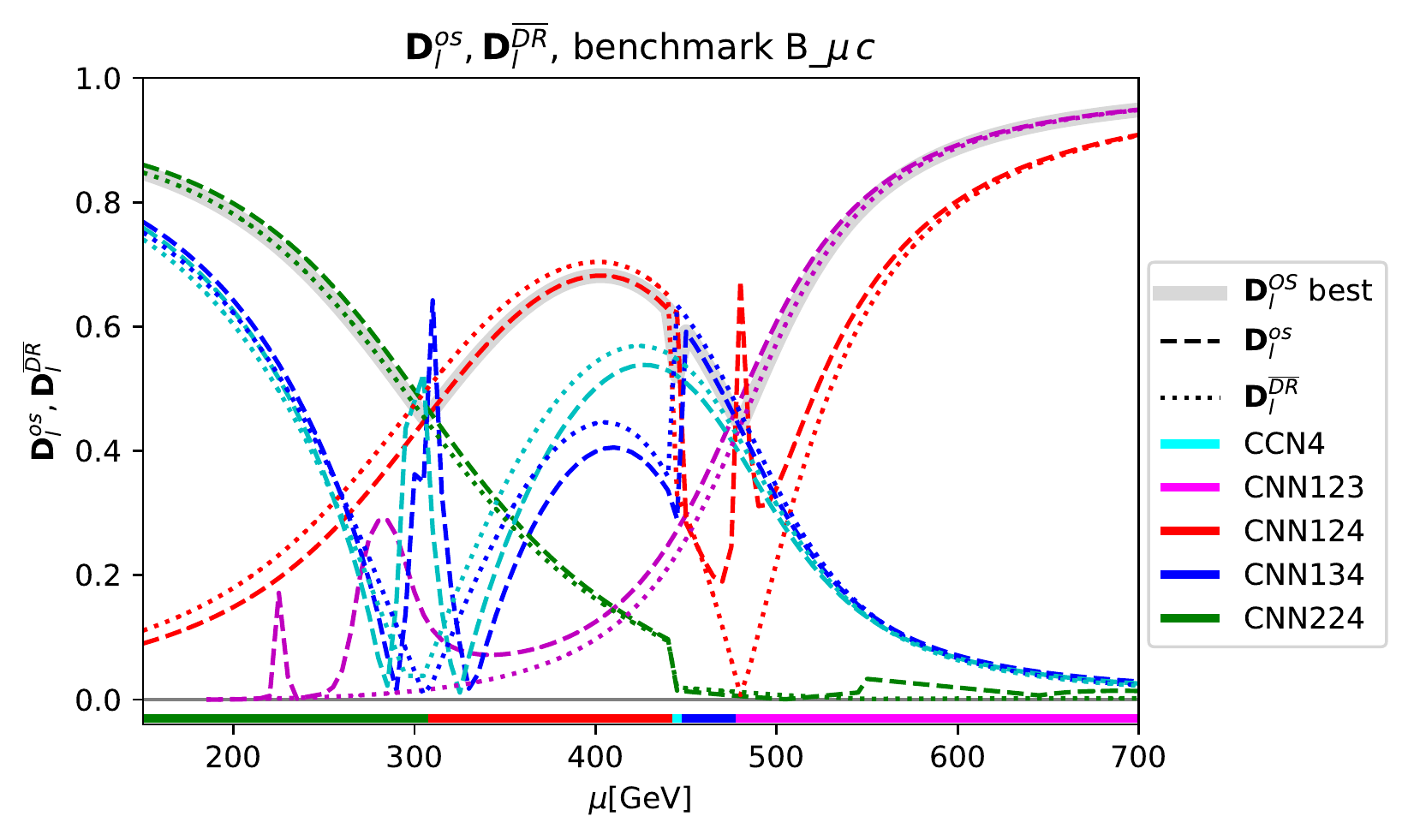}\\[1em]
  \includegraphics[width=0.45\textwidth]{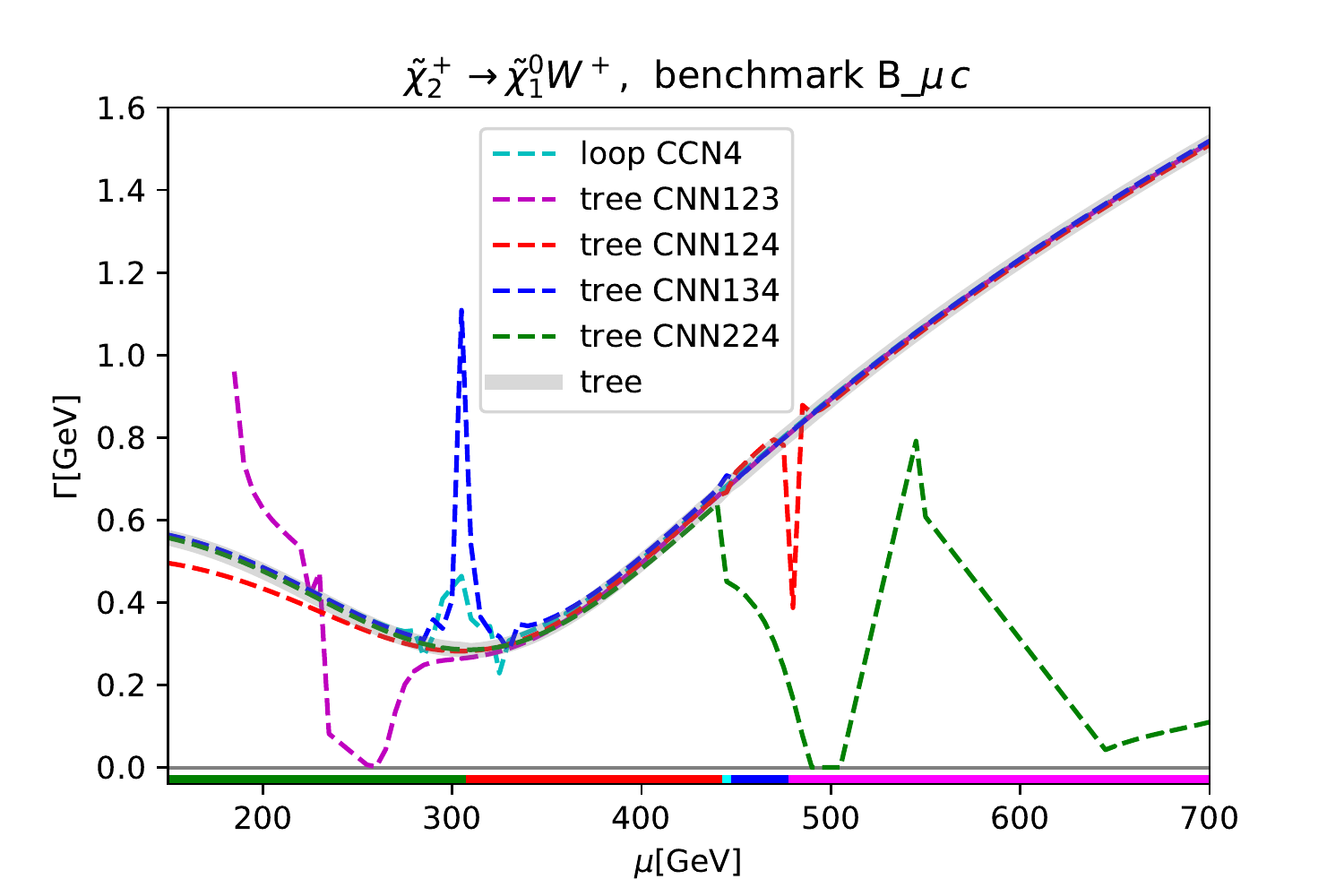}
  \includegraphics[width=0.45\textwidth]{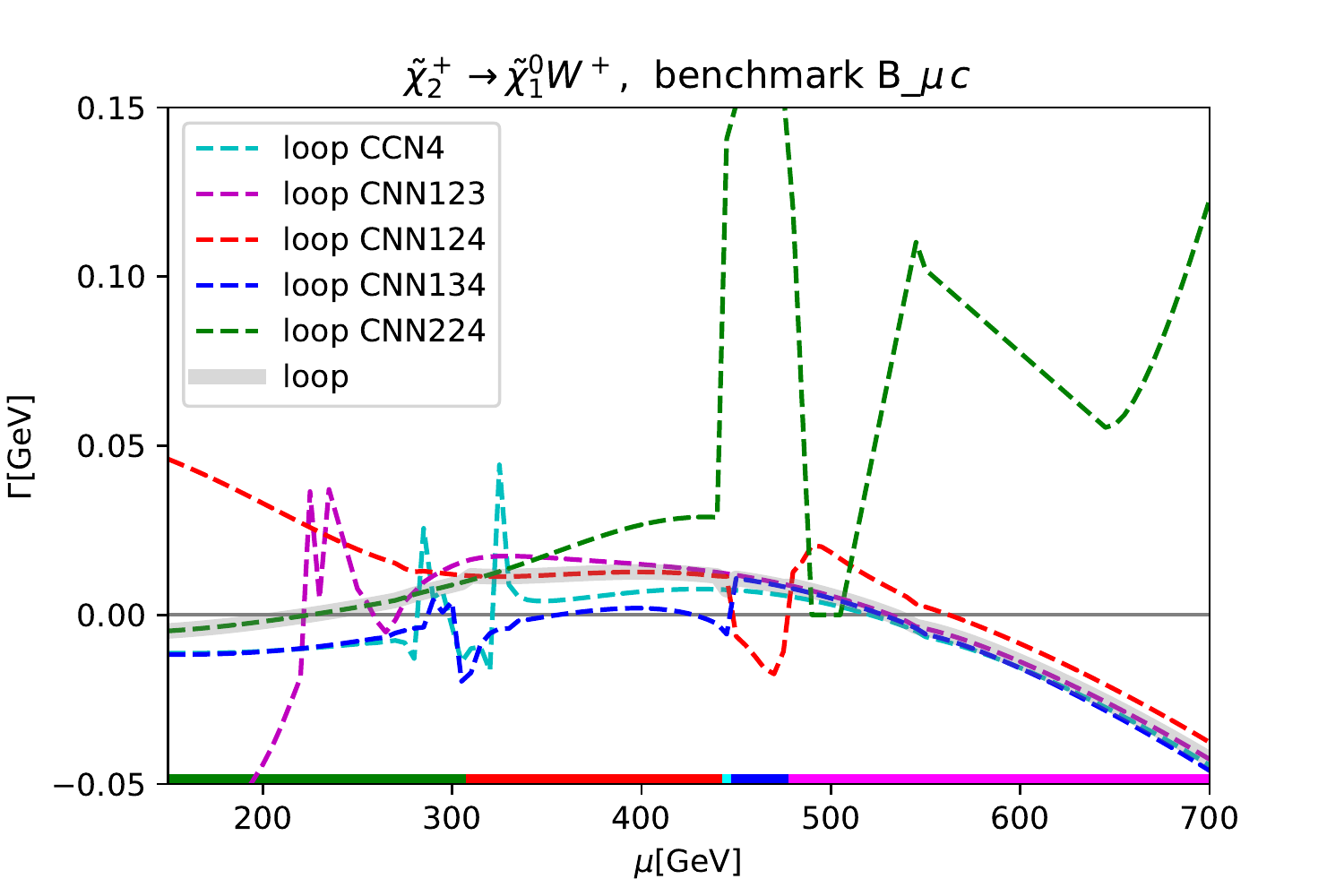}\\[1em]
  \includegraphics[width=0.45\textwidth]{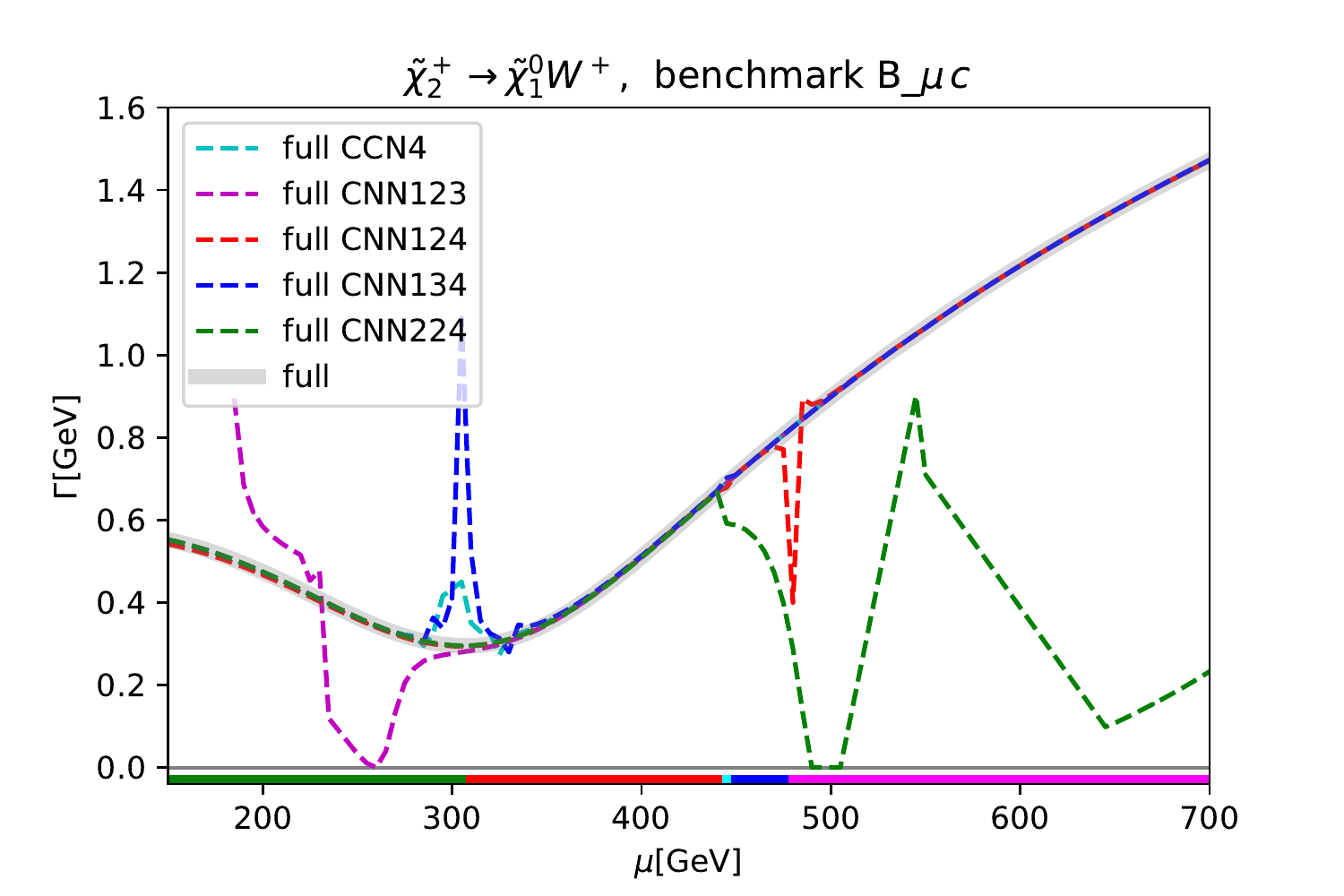}
  \includegraphics[width=0.45\textwidth]{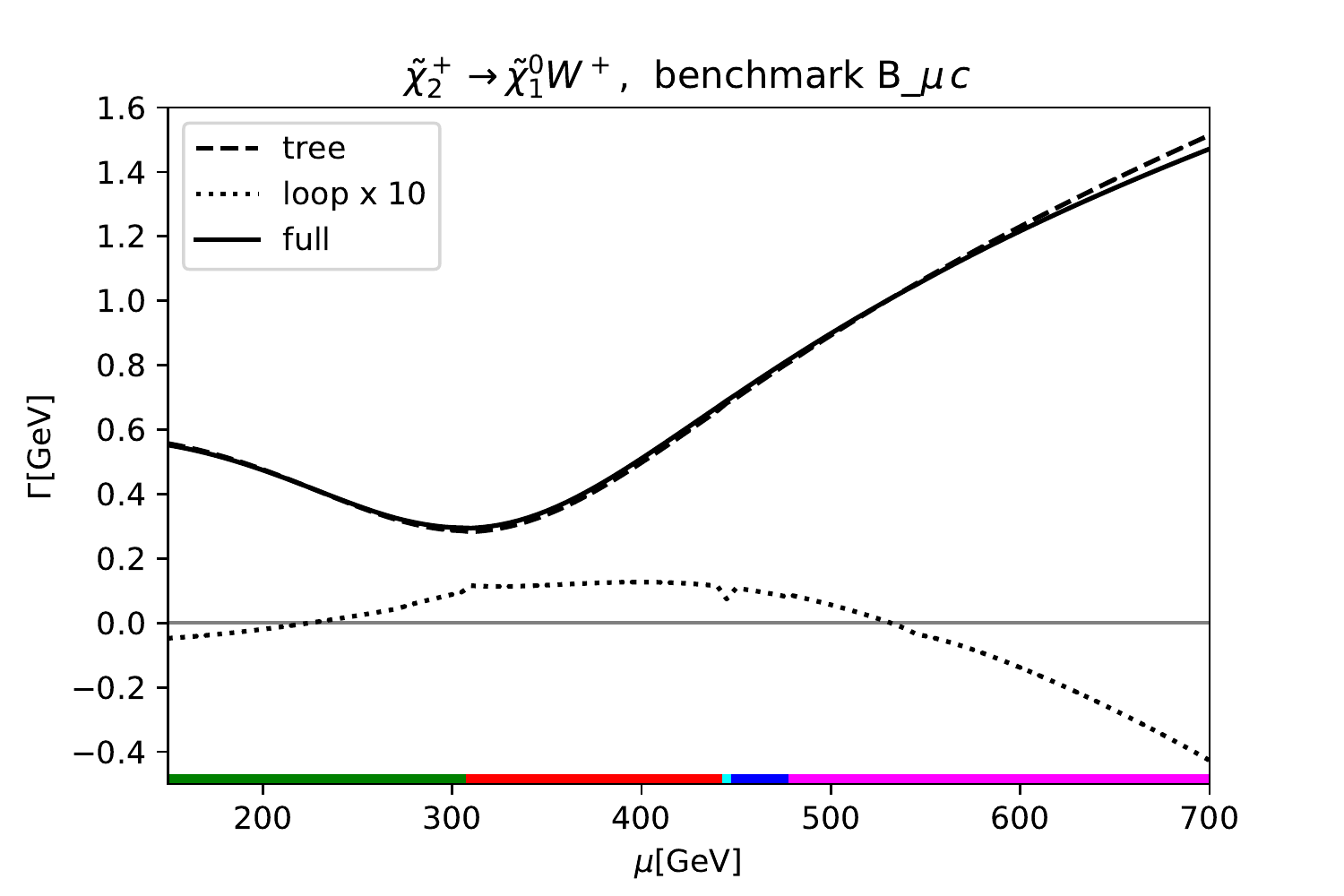}
\vspace{2em}
  \caption{Decay width for $\chap{2} \to \neu{1} W^+$ as a function of $\mu$
    in benchmark scenario B\_$\mu\,c$, see \refta{tab:bench1},
    with $M_1= 500\gev$, $M_2=200\gev$, $\tb= 10$. 
    Shown are the five
    ``best RS'' in this range of parameters (see text).
    The plots show the same quantities as in
    \protect\reffi{fig:2022_C2N1W_mu_a}. 
   The horizontal colored bar shows the best RS for the corresponding
   value of  $\mu$,  
   following  the same color coding as the curves: 
   CNN$_{224}$ for $\mu\le 305\gev$ (green), 
   CNN$_{124}$ for $310 \gev \le \mu\le 440\gev$ (red), 
   CCN$_{4}$ for $445 \gev \le \mu\le 445\gev$ (cyan), 
   CNN$_{123}$ for $480\gev\le \mu$ (pink).
   }
\label{fig:2022_C2N1W_mu_c}
\end{center}
\end{figure}

\smallskip
We finish the numerical analysis of the variation of $\mu$ with the
benchmark scenario B\_$\mu\,c$, see \refta{tab:bench1}, where the
hierarchy of $M_1$ and $M_2$ is inverted w.r.t.\ B\_$\mu\,a$:
$500 \gev = M_1 > M_2 = 200 \gev$ and $\tb = 10$, while $\mu$ is varied
as before from $150 \gev$ to $700 \gev$. The results are shown in 
\reffi{fig:2022_C2N1W_mu_c}, with the same ordering of plots as in
\reffi{fig:2022_C2N1W_mu_a}. The overall result is the same as before:
the automatically selected benchmark scenario exhibits a smooth
behavior, while each RS shown exhibits irregular behavior 
in a region of the parameter space where it is not selected.

In general ``similar'' RS as in B\_$\mu\,a$ are selected:
for small $\mu$ \cnn{2}{2}{4} is chosen here, whereas in B\_$\mu\,a$
\cnn{2}{2}{3} had been selected, reflecting the inverted hierarchy of
$M_1$ and $M_2$. Similarly, for large $\mu$ now \cnn{1}{2}{3} is
selected, whereas for B\_$\mu\,a$ it was \cnn{1}{1}{3}, again reflecting
the inverted hierarchy. 
In this benchmark scenario one can also observe that, albeit only for a
small part of the parameter space, a \ccn{} scheme, \ccn{4}, is
selected for $\mu \sim 450 \gev$, where all \cnn{}{}{} schemes turn
bad.
This is due to a ``level crossing'' of $\neu2$ and $\neu3$,
which occurs for slightly different values of $\mu$ when the masses
are computed with either the $\DRbar$ parameters or the $\OS$
parameters, i.e.\ reflected in $\mathbf{D}_l^{\DRbar}$ and
$\mathbf{D}_l^{\Os}$, respectively. Consequently,  the best value of
$\min\LV\mathbf{D}_l^{\Os}, \mathbf{D}_l^{\DRbar} \RV$ changes from
\cnn{1}{2}{4} to \cnn{1}{3}{4} 
for different values of $\mu$, leading to a small region in-between where
$\min \LV\mathbf{D}_l^{\Os}, \mathbf{D}_l^{\DRbar} \RV \sim 0.3$
and \ccn{4} is selected.


\subsection{Variation of \boldmath{$M_1$}}
\label{sec:m1-var}

\begin{figure}[h!]
\vspace{2em}
  \begin{center}
  \includegraphics[width=0.60\textwidth]{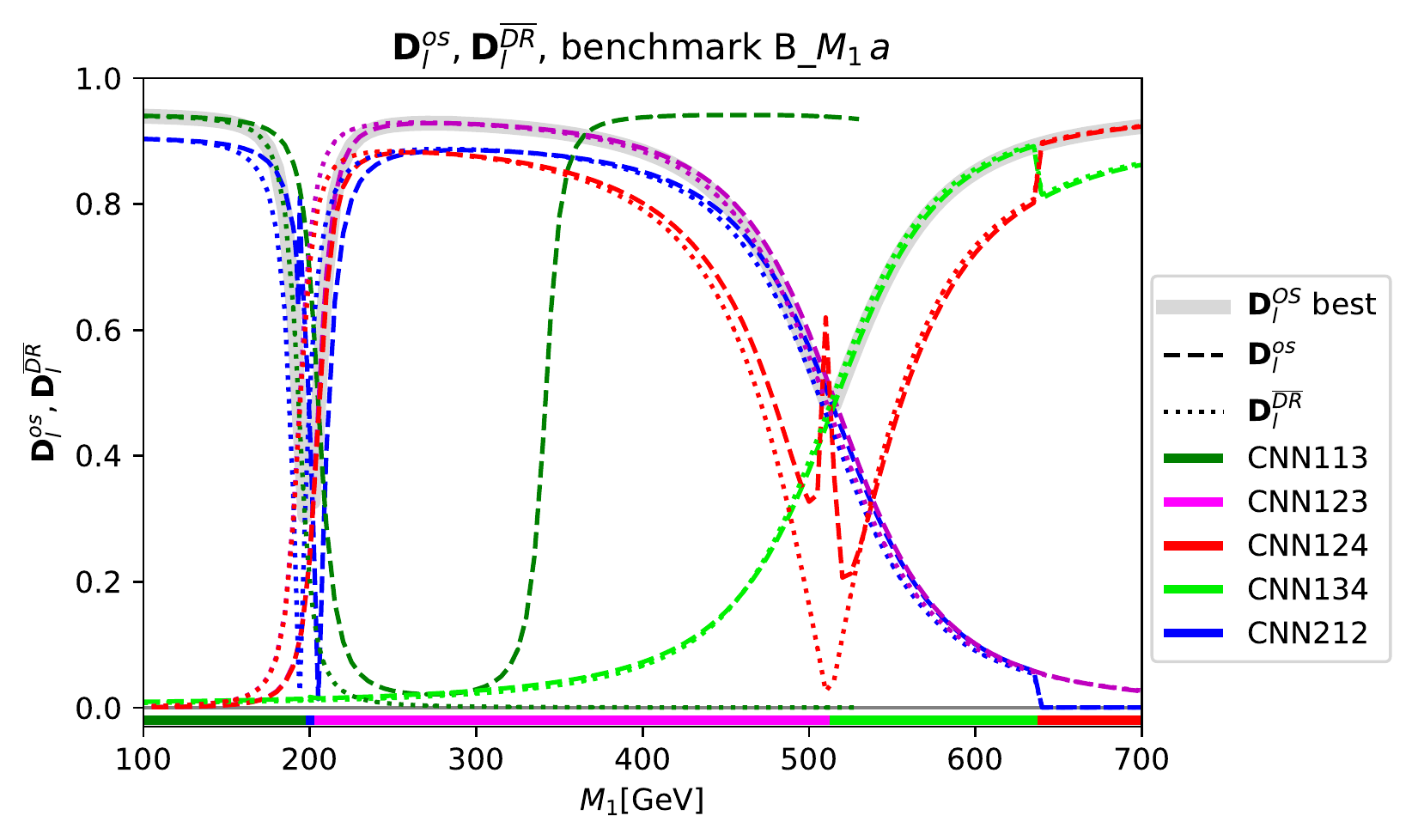}\\[1em]
  \includegraphics[width=0.45\textwidth]{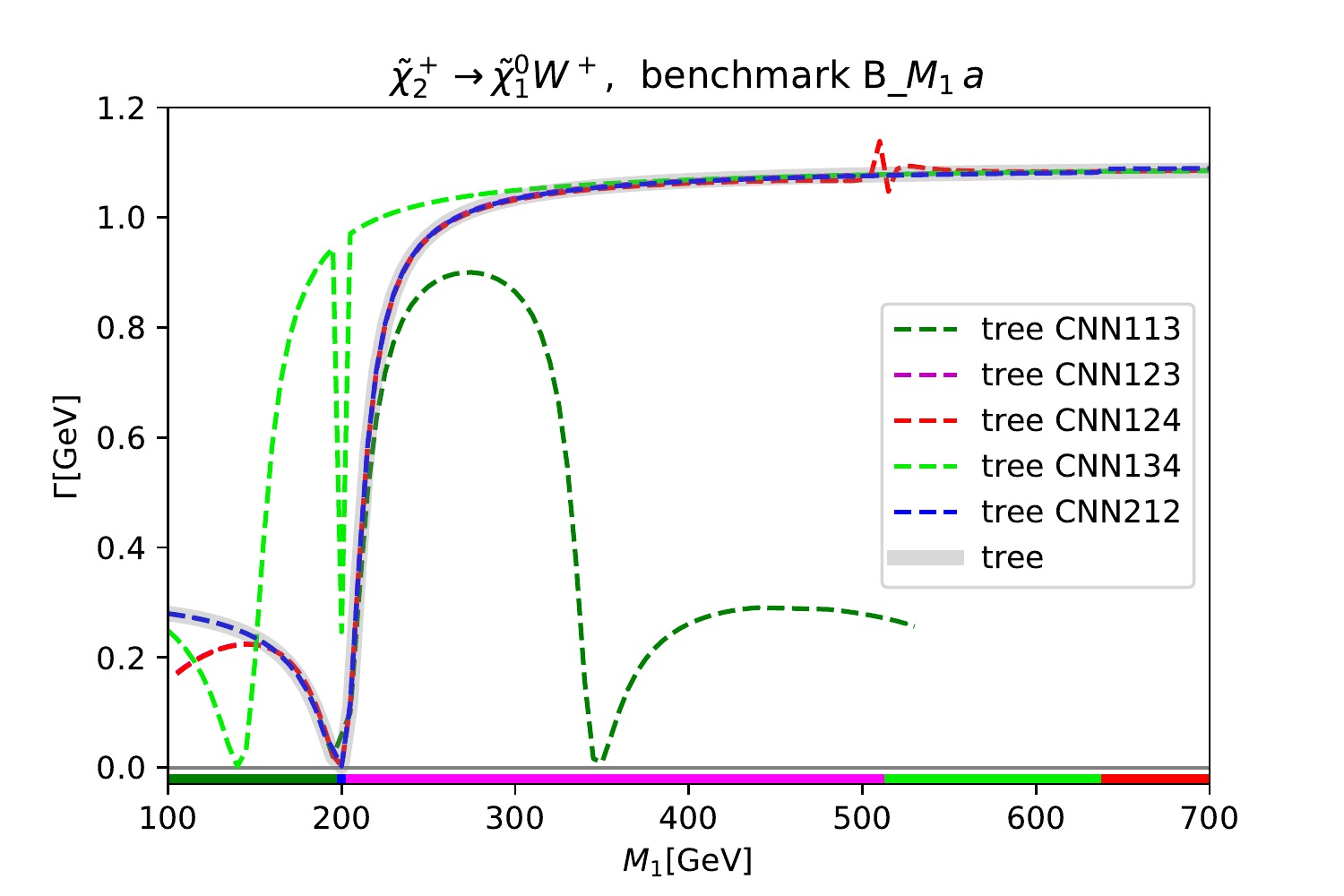}
  \includegraphics[width=0.45\textwidth]{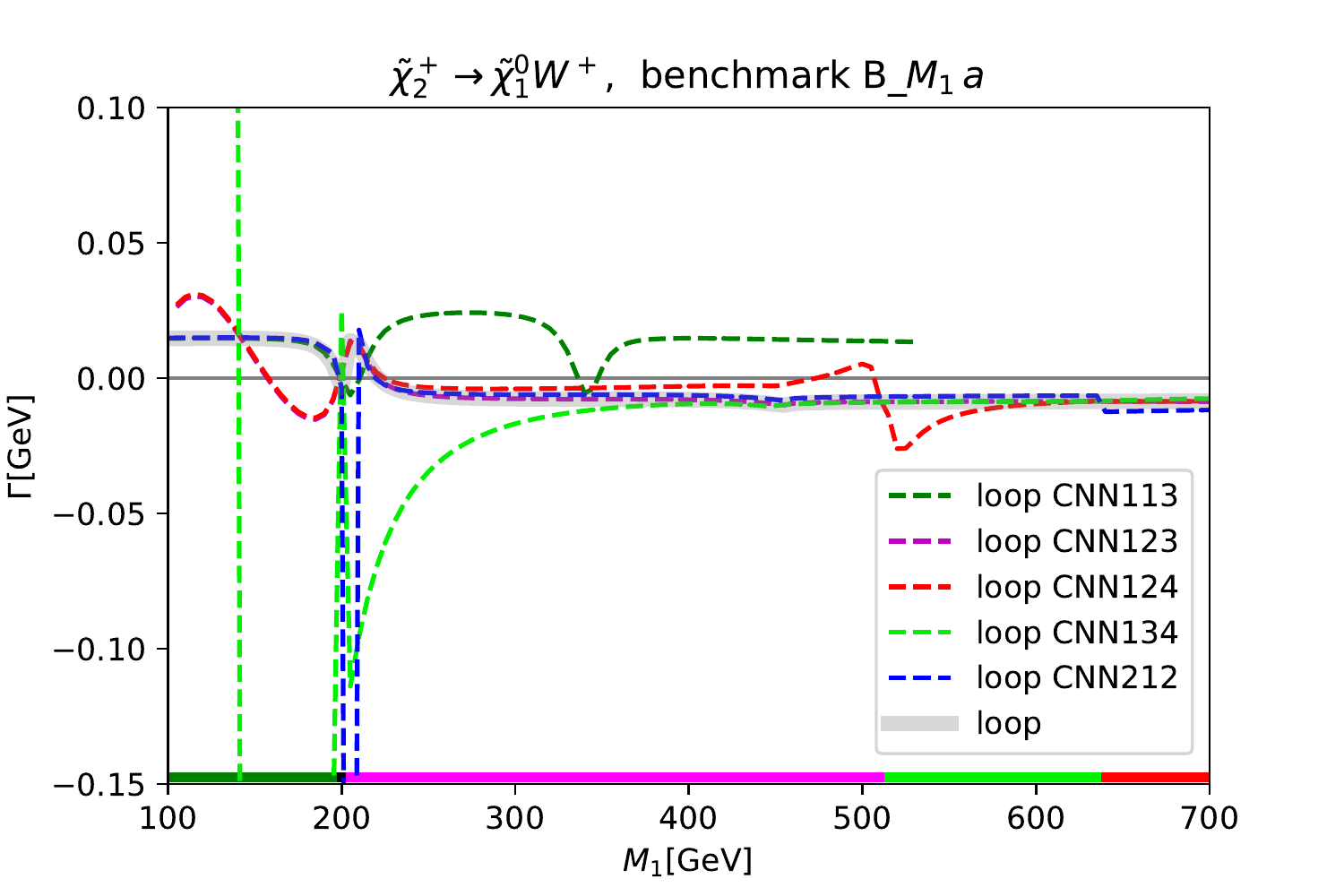}\\[1em]
  \includegraphics[width=0.45\textwidth]{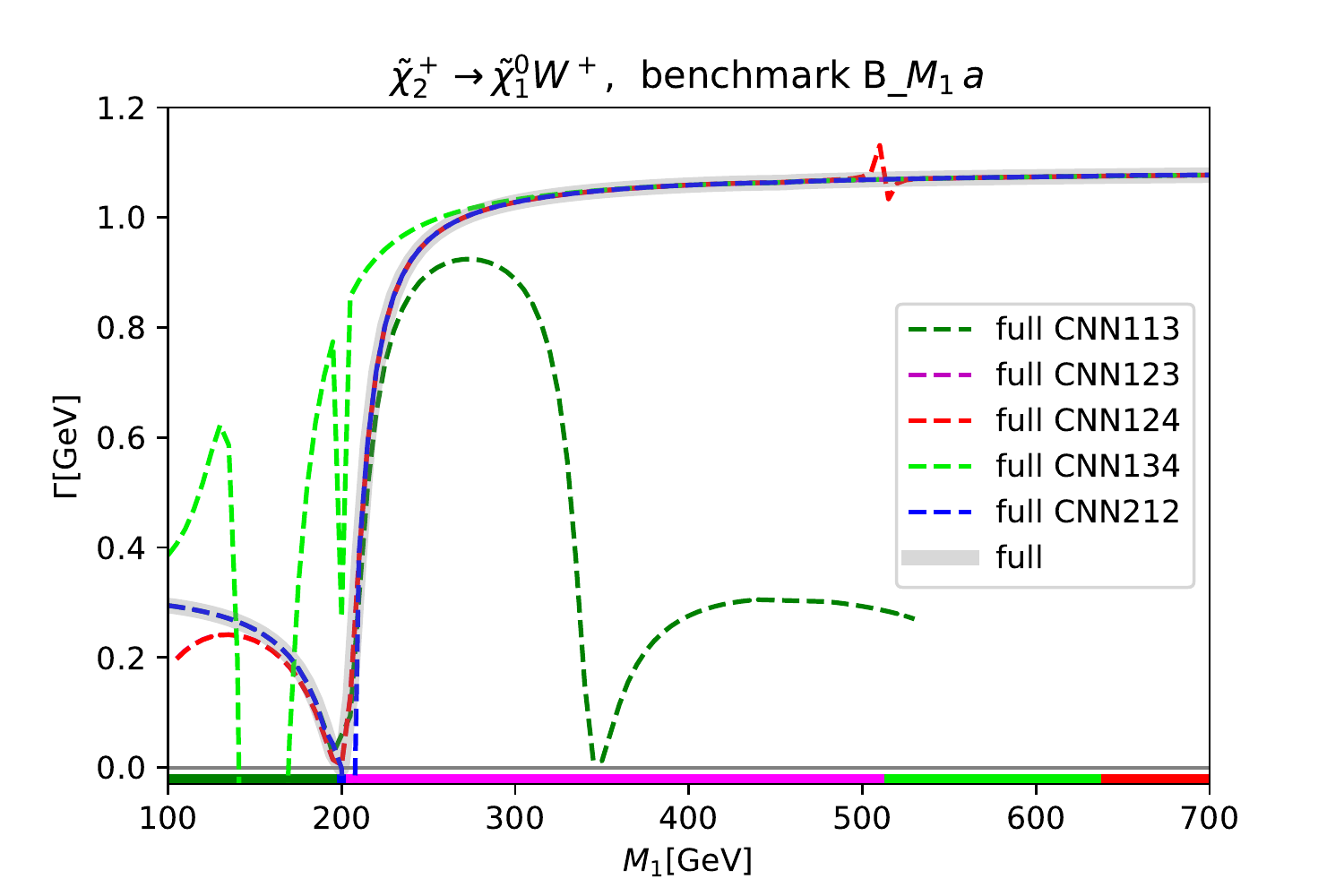}
  \includegraphics[width=0.45\textwidth]{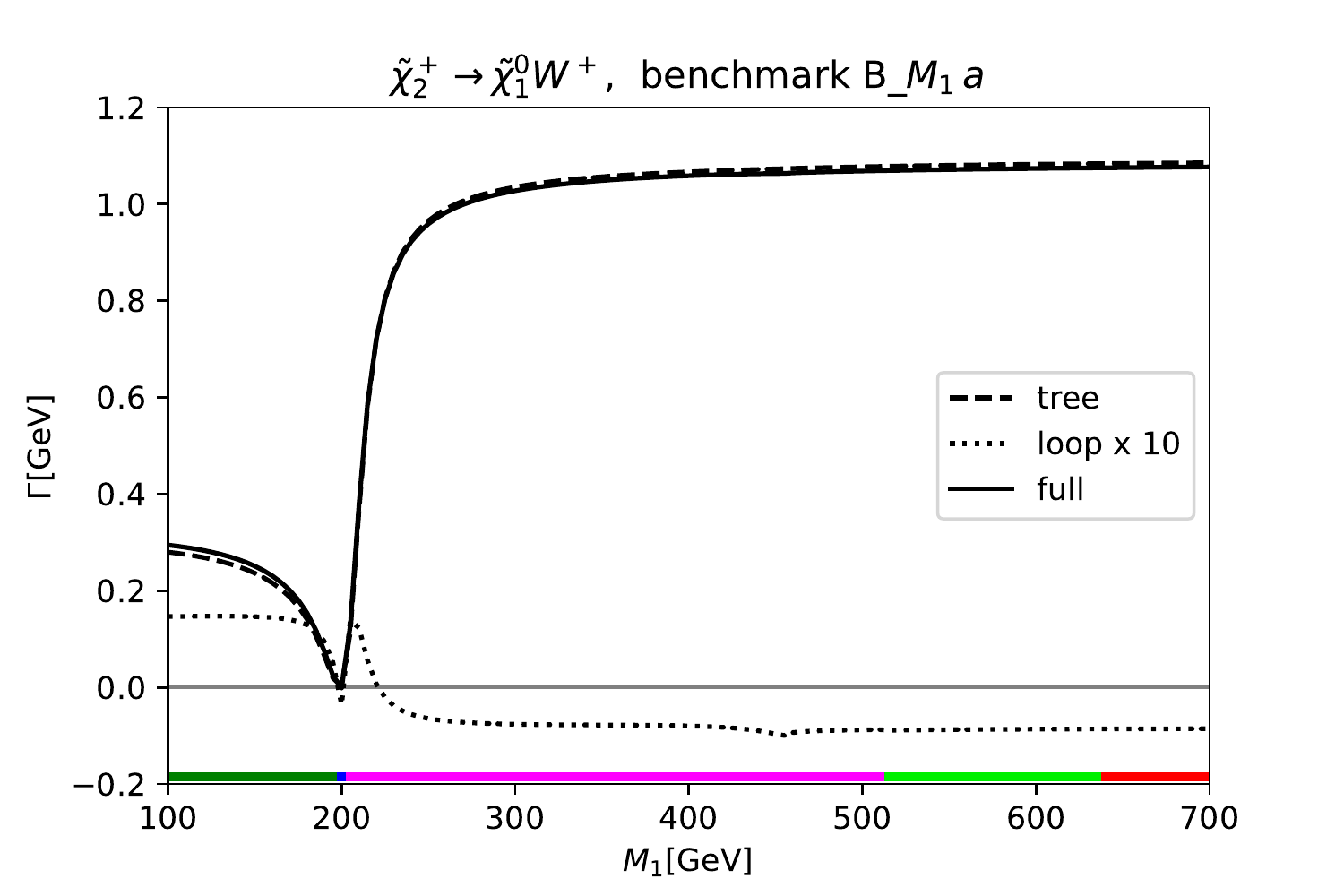}
\vspace{2em}
  \caption{
    Decay width for $\chap{2} \to \neu{1} W^+$ as a function of $M_1$ in
    benchmark scenario B\_$M_1\,a$,  
    see Table~\ref{tab:bench1},
    with $M_2= 200\gev$, $\mu=500\gev$, $\tb= 10$.  
    Shown are the five ``best RS'' in this range of parameters (see text).
    The plots show the same quantities as in
    \protect\reffi{fig:2022_C2N1W_mu_a}. 
   The horizontal colored bar shows the best RS for the corresponding
   value of  $M_1$,  
   following  the same color coding as the curves: 
   CNN$_{113}$ for $M_1\le 195\gev$, 
   CNN$_{212}$ for $200 \gev \le M_1\le 200\gev$, 
   CNN$_{123}$ for $205 \gev \le M_1\le 510\gev$, 
   CNN$_{134}$ for $515 \gev \le M_1\le 635\gev$, 
   CNN$_{124}$ for $640\gev\le M_1$.
  }
\label{fig:2022_C2N1W_M1a}
\end{center}
\end{figure}

\begin{figure}[h!]
\vspace{2em}
\begin{center}
  \includegraphics[width=0.60\textwidth]{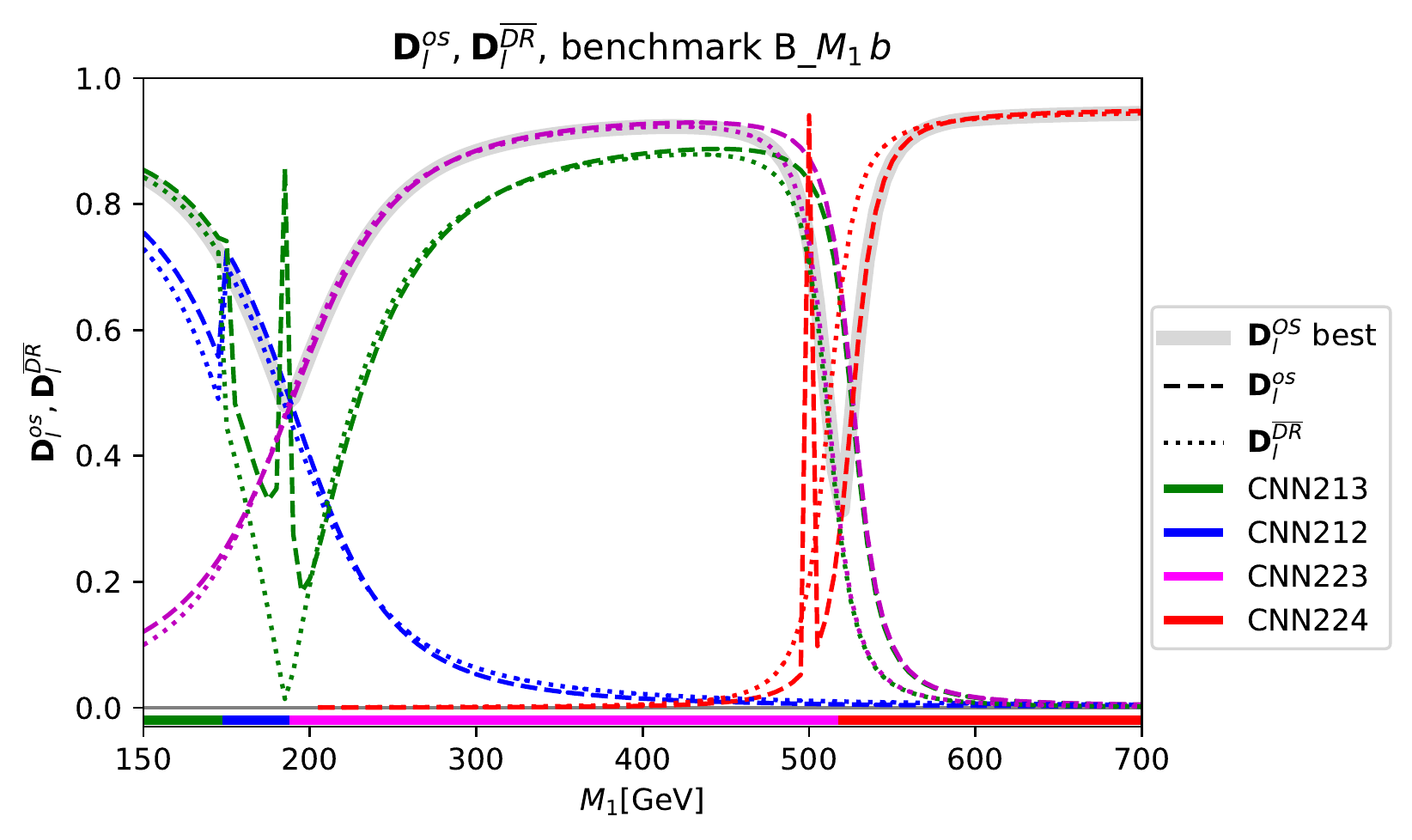}\\[1em]
  \includegraphics[width=0.45\textwidth]{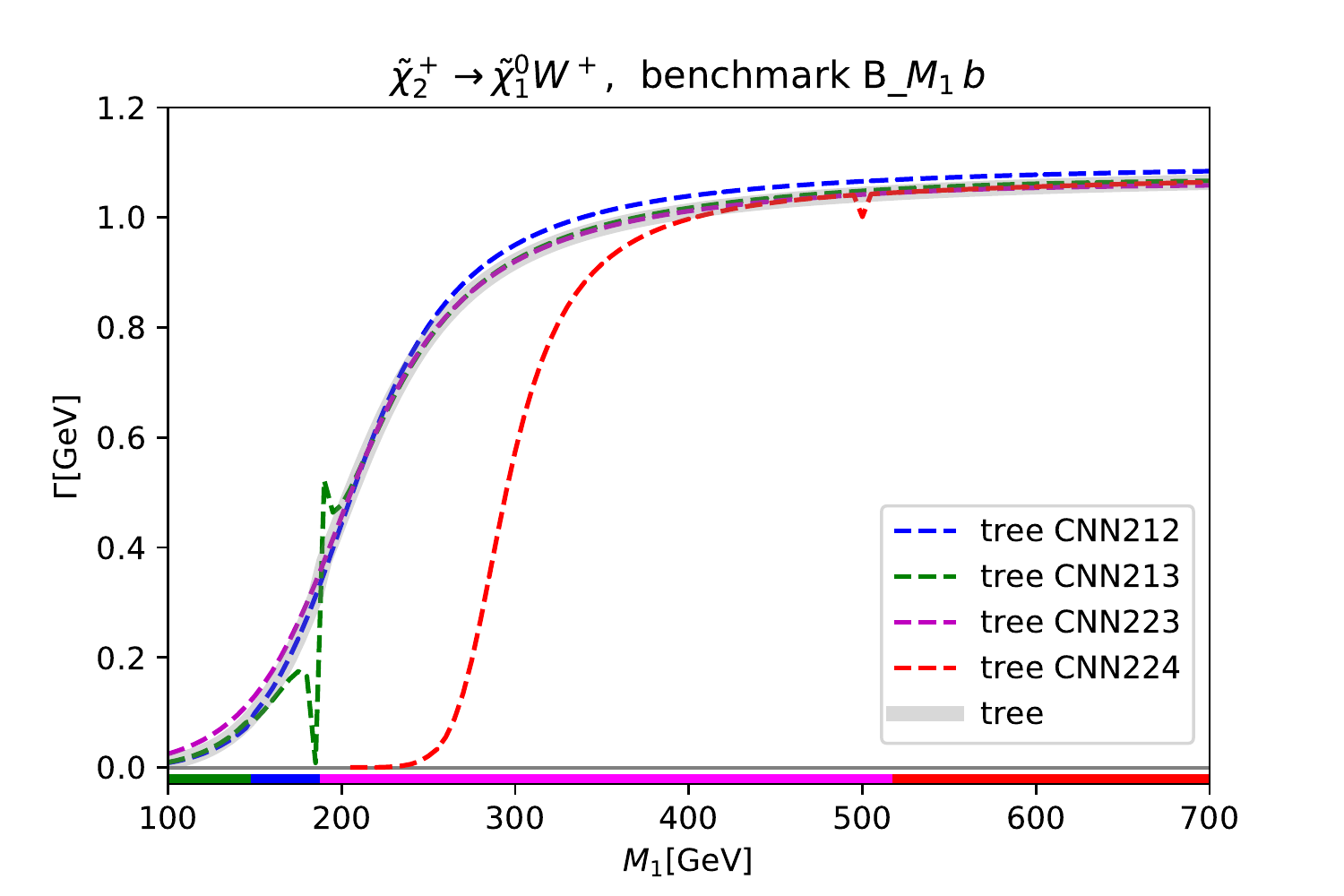}
  \includegraphics[width=0.45\textwidth]{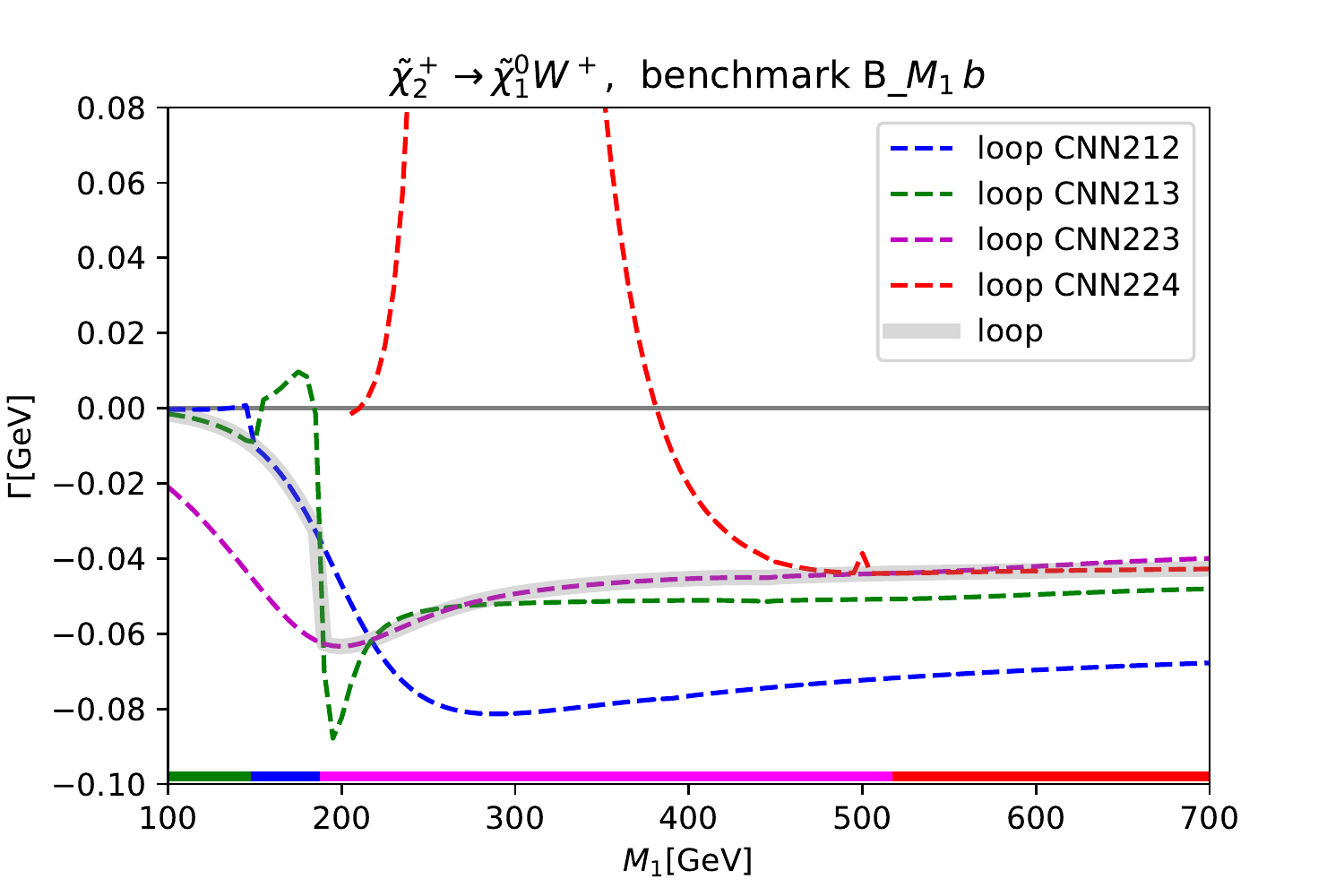}\\[1em]
  \includegraphics[width=0.45\textwidth]{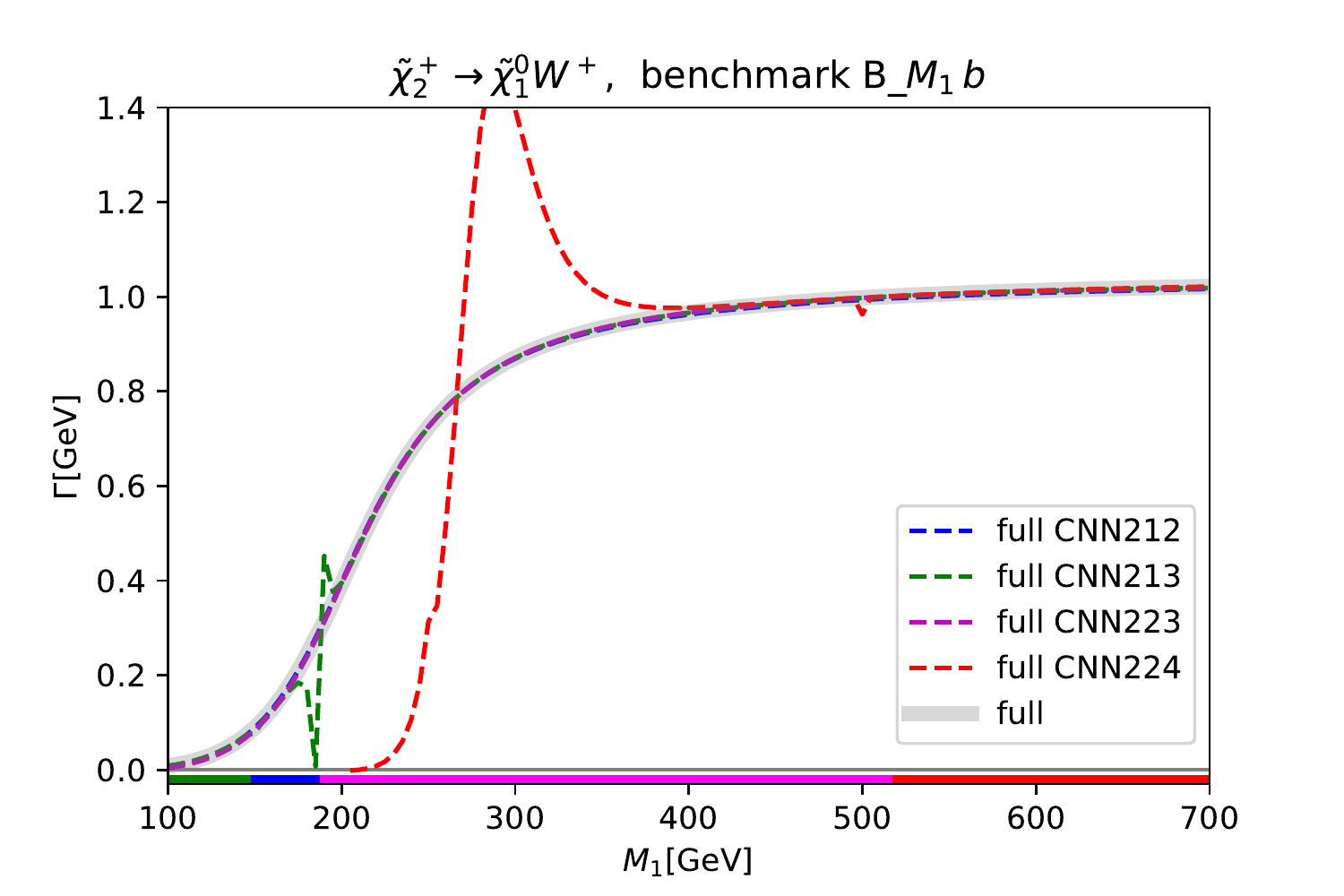}
  \includegraphics[width=0.45\textwidth]{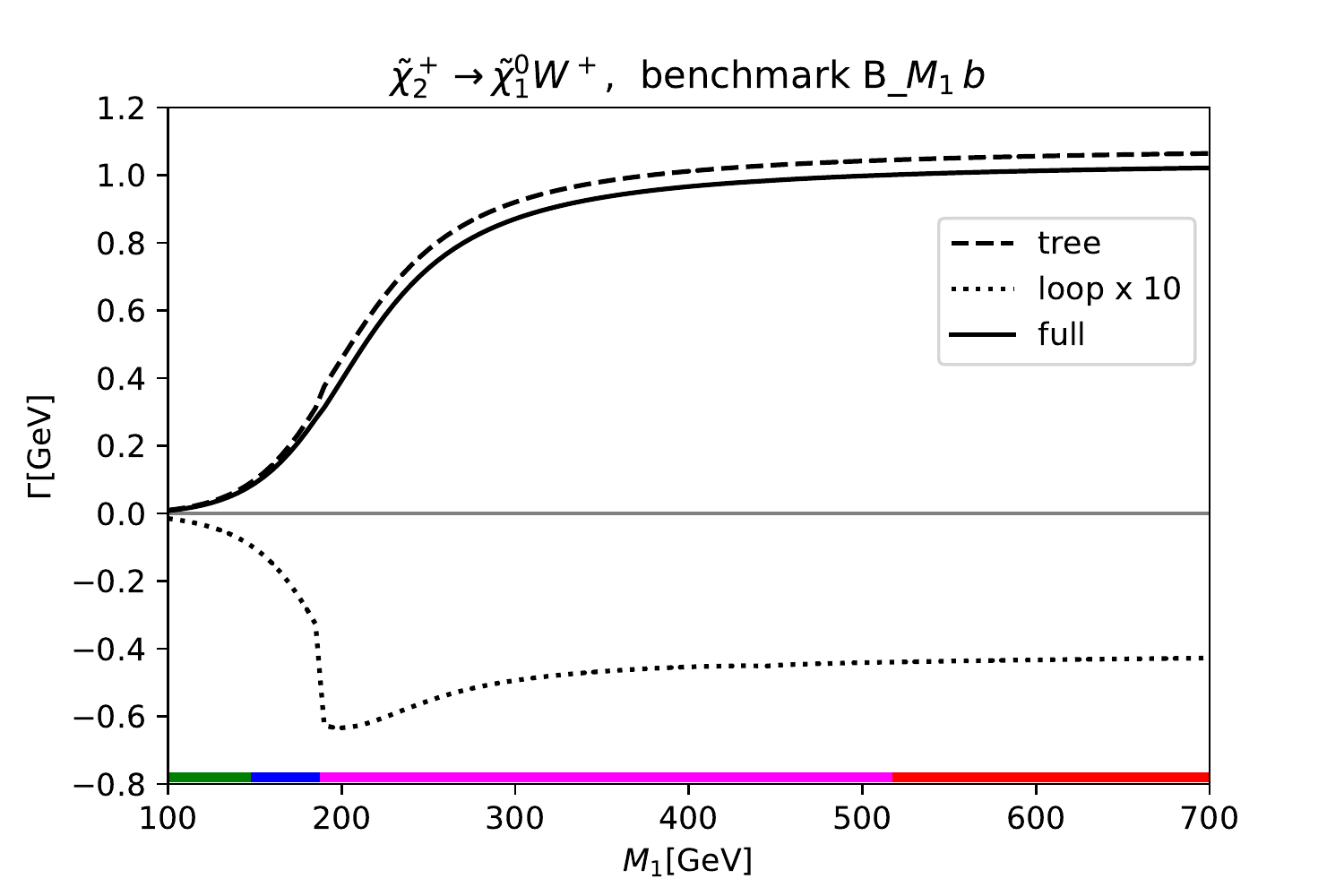}
\vspace{2em}
  \caption{
    Decay width for $\chap{2} \to \neu{1} W^+$ as a function of $M_1$ in
    benchmark scenario B\_$M_1\,b$,  
    see Table~\ref{tab:bench1},
    with  $M_2= 500\gev$, $\mu=200\gev$, $\tb= 10$. 
    Shown are the four ``best RS'' in this range of parameters (see text).
    The plots show the same quantities as in
    \protect\reffi{fig:2022_C2N1W_mu_a}. 
   The horizontal colored bar shows the best RS for the corresponding
   value of  $M_1$,  
   following  the same color coding as the curves: 
   CNN$_{213}$ for $M_1\le 145\gev$, 
   CNN$_{212}$ for $150 \gev \le M_1\le 185\gev$, 
   CNN$_{223}$ for $190 \gev \le M_1\le 515\gev$, 
   CNN$_{224}$ for $520\gev\le M_1$.
  }
\label{fig:2022_C2N1W_M1b}
\end{center}
\end{figure}

In this subsection we demonstrate that our proposed method for the
selection of an RS also works for the variation of the ``genuine
neutralino mass parameter'', $M_1$. The results are shown for the two
benchmark scenarios B\_$M_1\,a$ and B\_$M_1\,b$, see \refta{tab:bench1},
in \reffis{fig:2022_C2N1W_M1a} and \ref{fig:2022_C2N1W_M1b},
respectively. The plots in the two figures show the same quantities as in
\protect\reffi{fig:2022_C2N1W_mu_a}. 
The two scenarios differ in the hierarchy of $M_2$ and
$\mu$, where in the first (second) scenario we have chosen
$M_2 = 200 (500) \gev$ and $\mu = 500 (200) \gev$.

The overall results are found as for the variation of $\mu$, see
\refse{sec:mu-var}. The selected determinant, as shown in the upper rows of
\reffis{fig:2022_C2N1W_M1a} and \ref{fig:2022_C2N1W_M1b}, becomes
smallest for $M_1 \approx M_2$, but does not go below $\sim 0.30$.
For the rest of the parameter space the best
determinant is above $\sim 0.47$. For B\_$M_1\,a$ five
schemes, all of type CNN, are selected, while for B\_$M_1\,b$ only four
schemes are automatically chosen, also all of type CNN.
The different choices reflect, similar to the case discussed in
\refse{sec:mu-var}, the two different hierarchies of $M_2$ and $\mu$. 

Each of the
schemes shows good behavior where it is selected. The relative size of
the ``selected higher-order corrections'' does not exceed $\sim 15\%$
and mostly stays below $\sim 5\%$.
The only exceptions occur where
the Born contribution vanishes due to a zero crossing of the
respective coupling, as found in B\_$M_1\,a$ for $M_1 \sim 200 \gev$.
The final result, as shown in the
lower right plots of \reffis{fig:2022_C2N1W_M1a} and
\ref{fig:2022_C2N1W_M1b}, are smooth curves with a reliable higher-order
correction. Little kinks in the tree-level result, caused by a change of
the OS values of the input parameters (see, e.g.,
\reffi{fig:2022_C2N1W_M1b} at $M_1 \sim 190 \gev$) are smoothed out in
the full one-loop result. 

\subsection{Variation of \boldmath{$M_2$}}
\label{sec:m2-var}

\begin{figure}[h!]
\vspace{2em}
\begin{center}
  \includegraphics[width=0.60\textwidth]{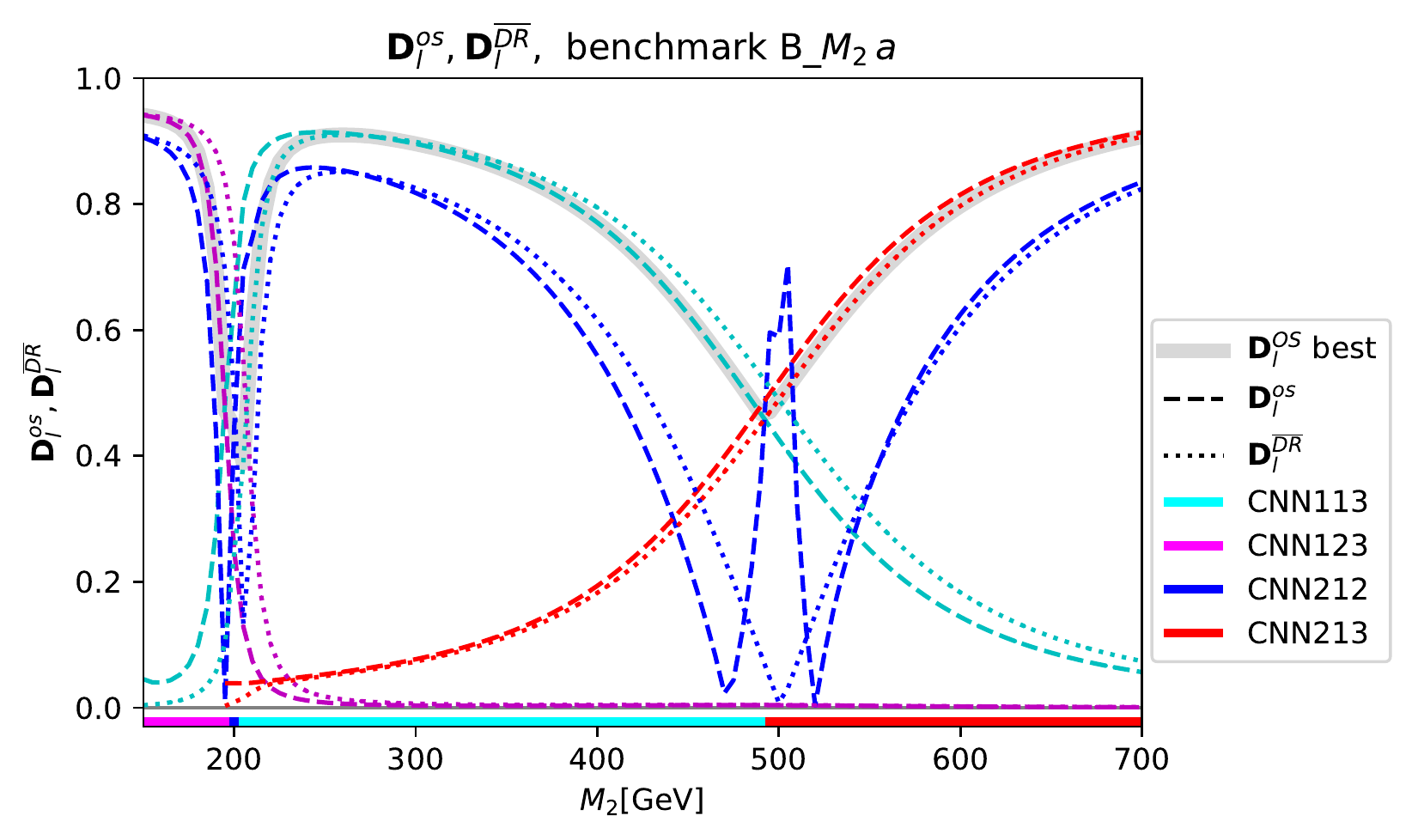}\\[1em]
  \includegraphics[width=0.45\textwidth]{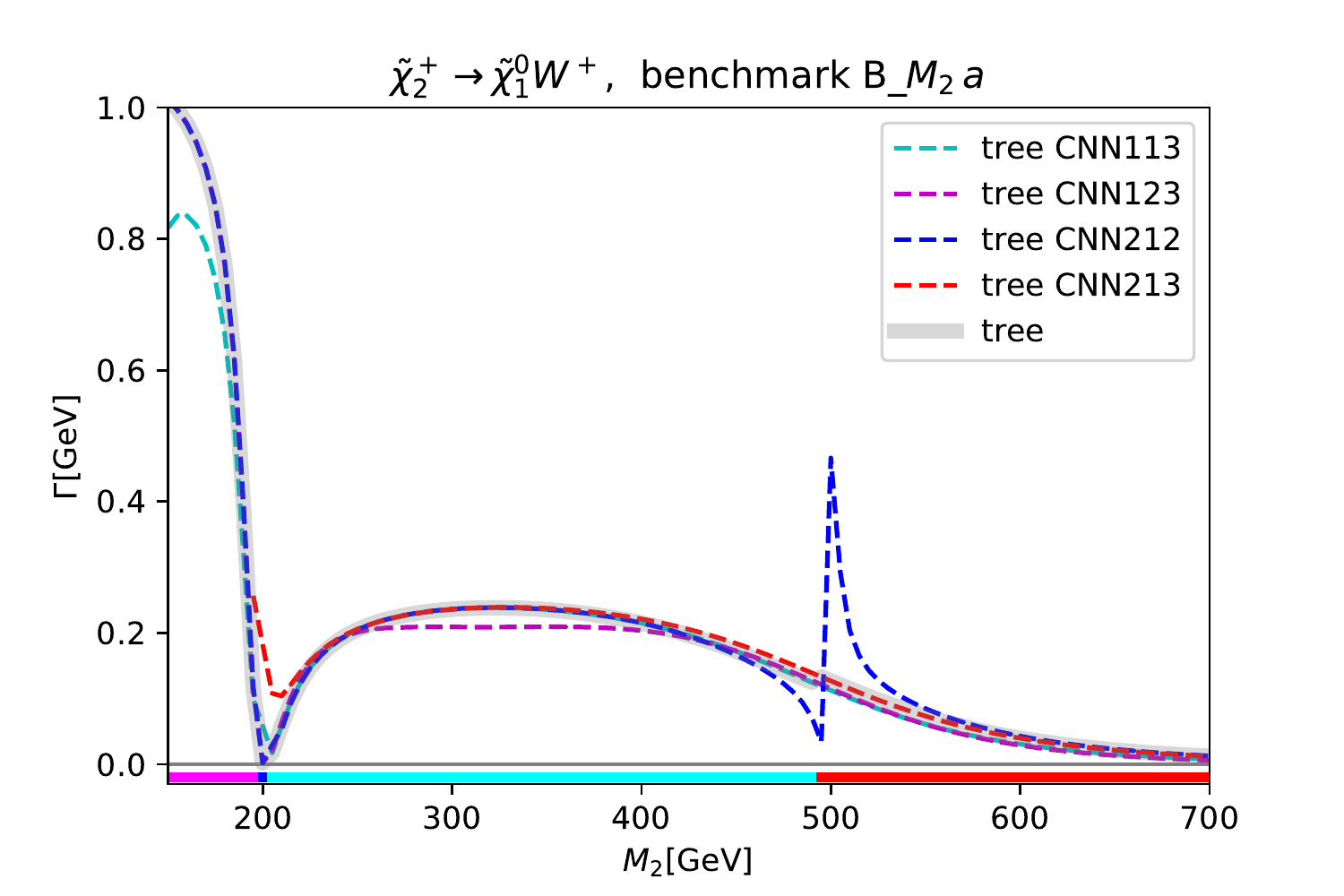}
  \includegraphics[width=0.45\textwidth]{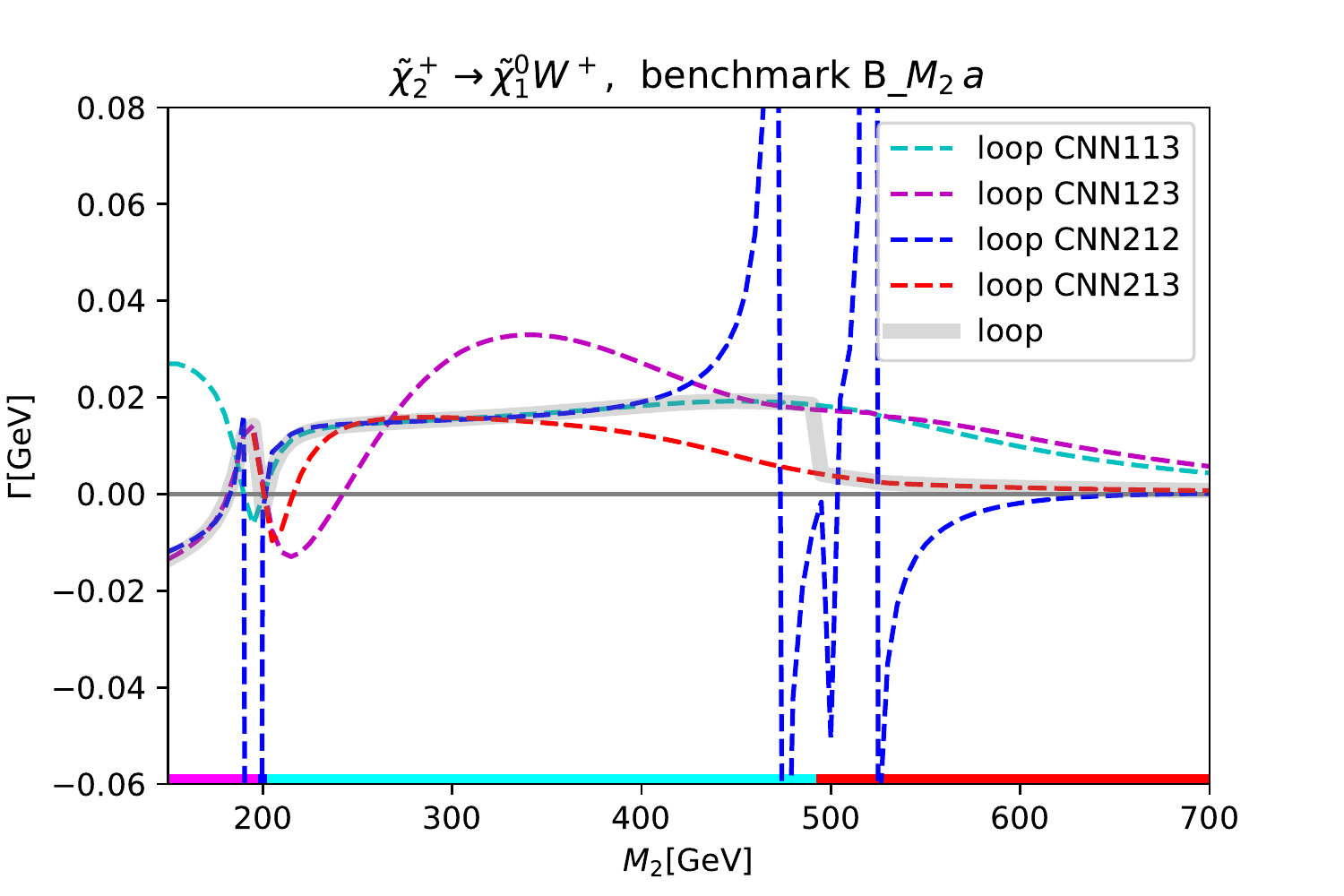}\\[1em]
  \includegraphics[width=0.45\textwidth]{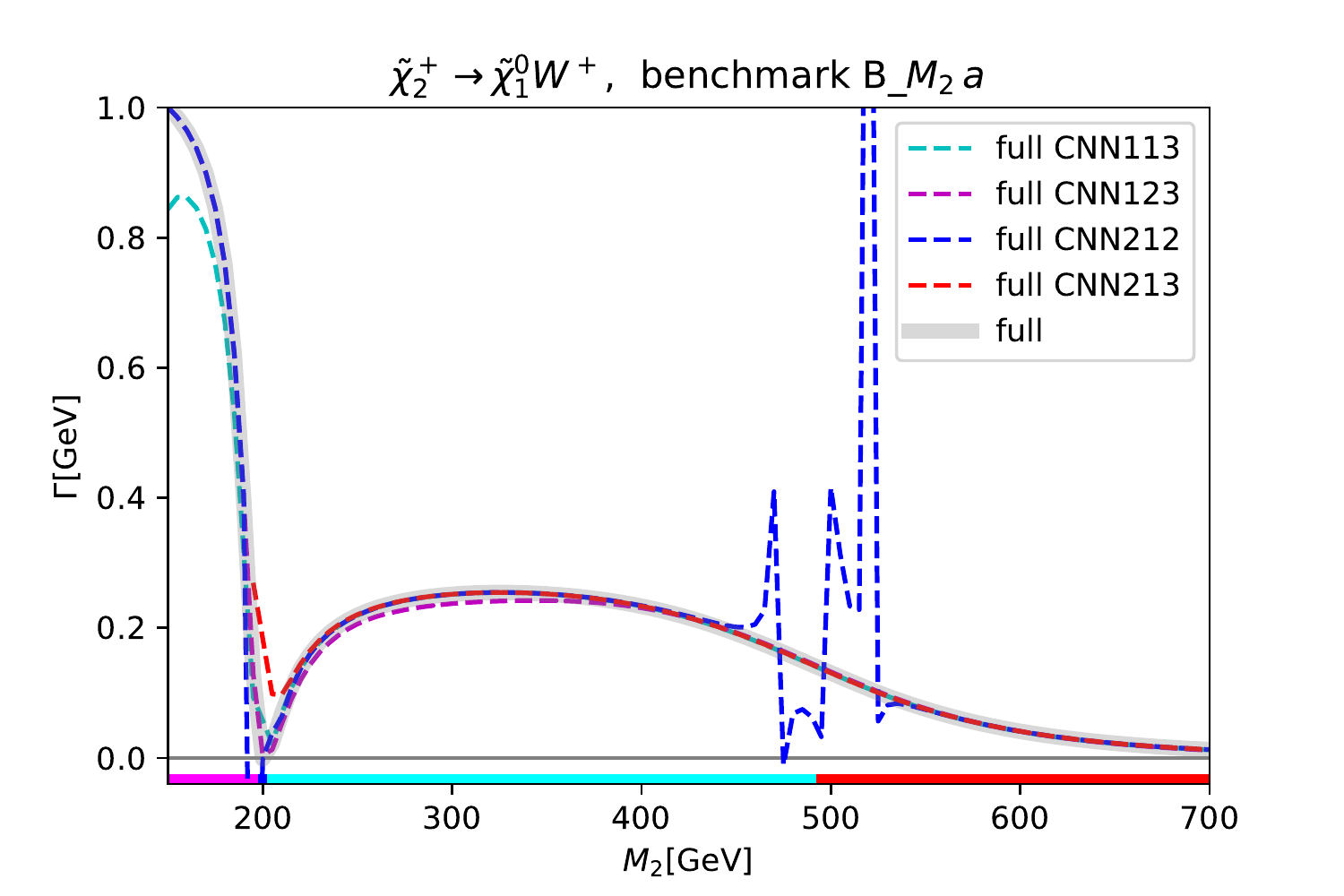}
  \includegraphics[width=0.45\textwidth]{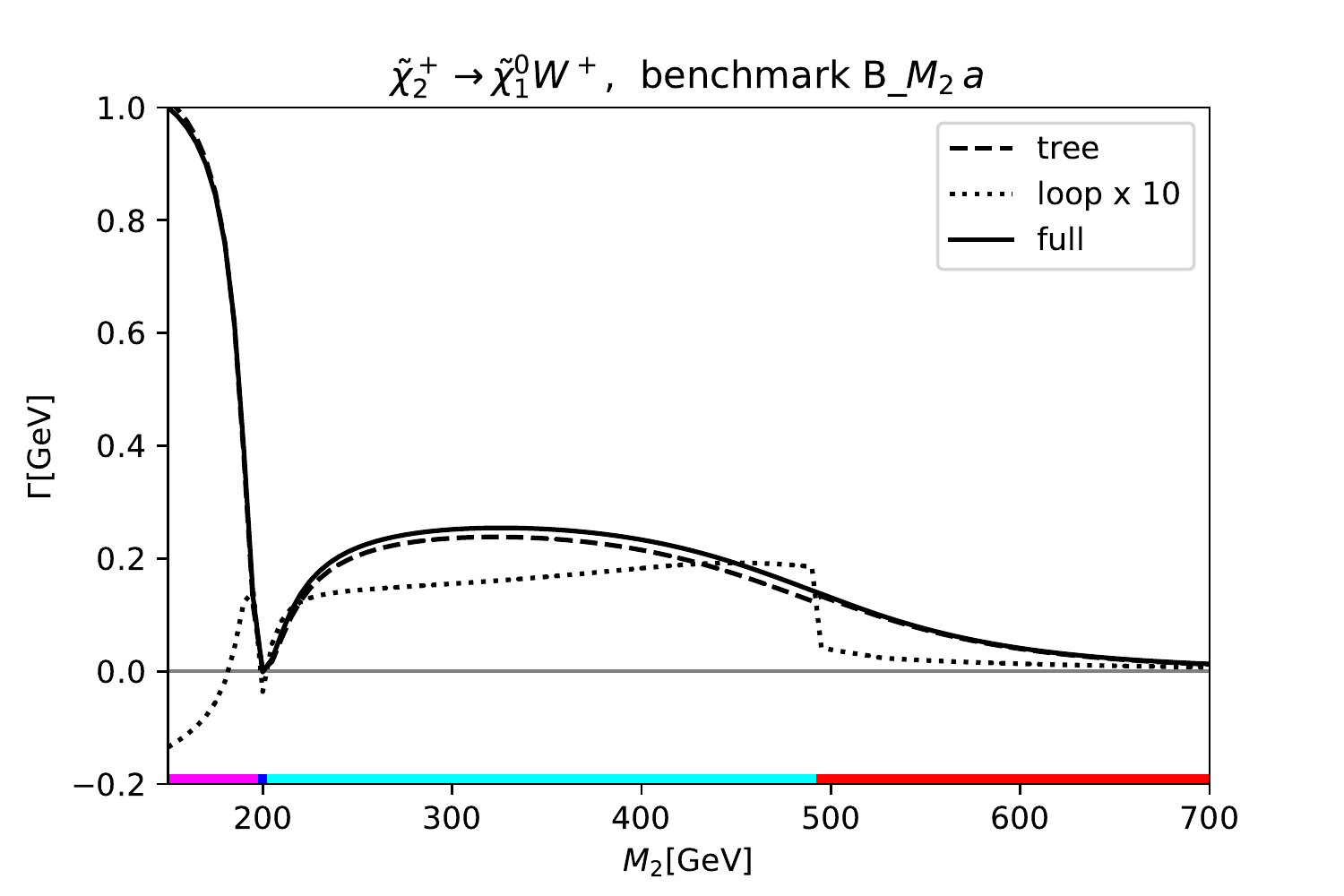}
\vspace{2em}
  \caption{Decay width for $\chap{2} \to \neu{1} W^+$
    as a function of $M_2$ in benchmark scenario B\_$M_2\,a$, 
    see \refta{tab:bench1},
    with $M_1= 200\gev$, $\mu=500\gev$, $\tb= 10$. 
    Shown are the four ``best RS'' in this range of parameters (see text).
    The plots show the same quantities as in
    \protect\reffi{fig:2022_C2N1W_mu_a}. 
   The horizontal colored bar shows the best RS for the corresponding
   value of  $M_2$,  
   following  the same color coding as the curves: 
   CNN$_{123}$ for $M_2\le 195\gev$, 
   CNN$_{212}$ for $200 \gev \le M_2\le 200\gev$, 
   CNN$_{113}$ for $205 \gev \le M_2\le 490\gev$, 
   CNN$_{213}$ for $495\gev\le M_2$.
}
\label{fig:2022_C2N1W_M2a}
\end{center}
\end{figure}

\begin{figure}[h!]
\vspace{2em}
\begin{center}
  \includegraphics[width=0.60\textwidth]{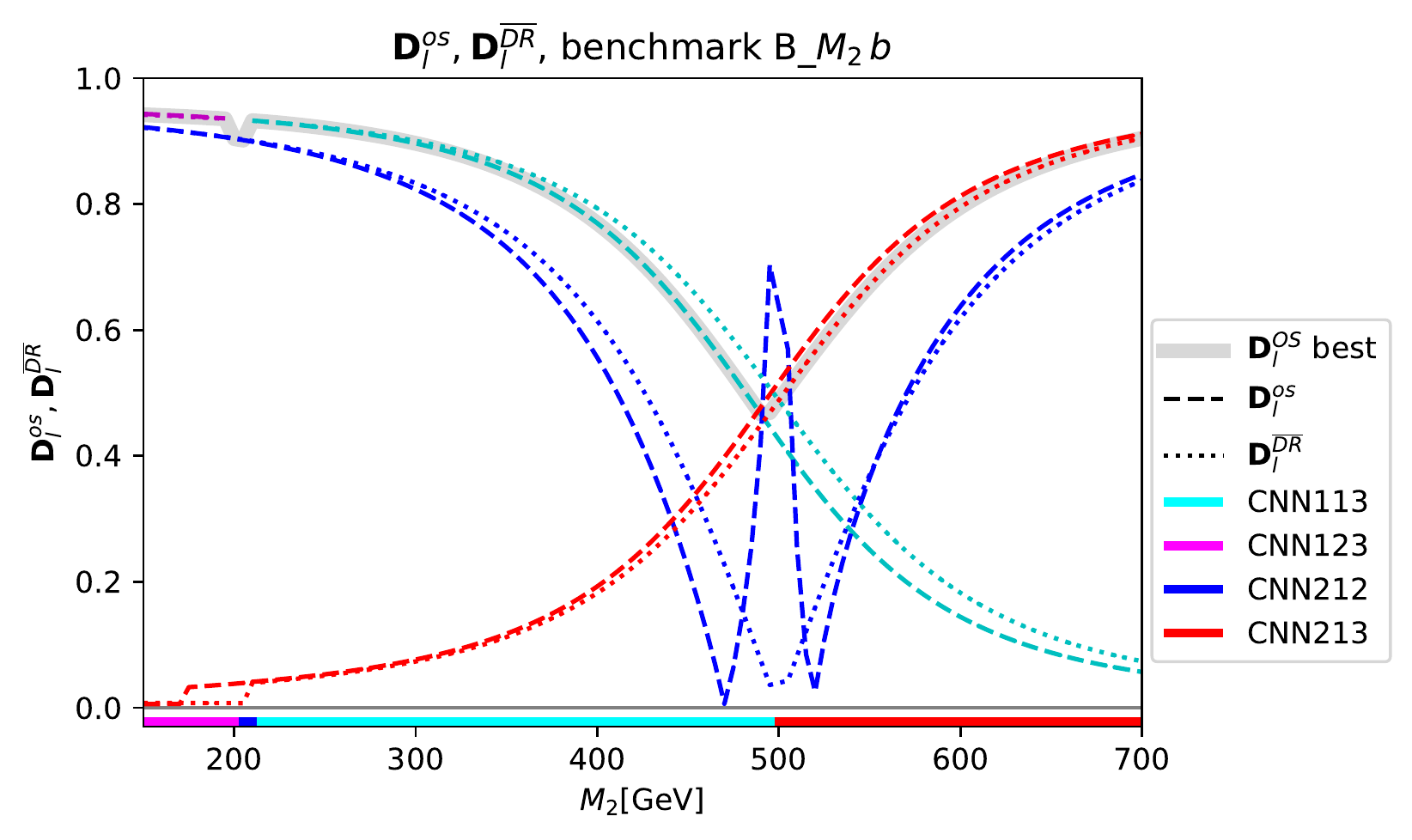}\\[1em]
  \includegraphics[width=0.45\textwidth]{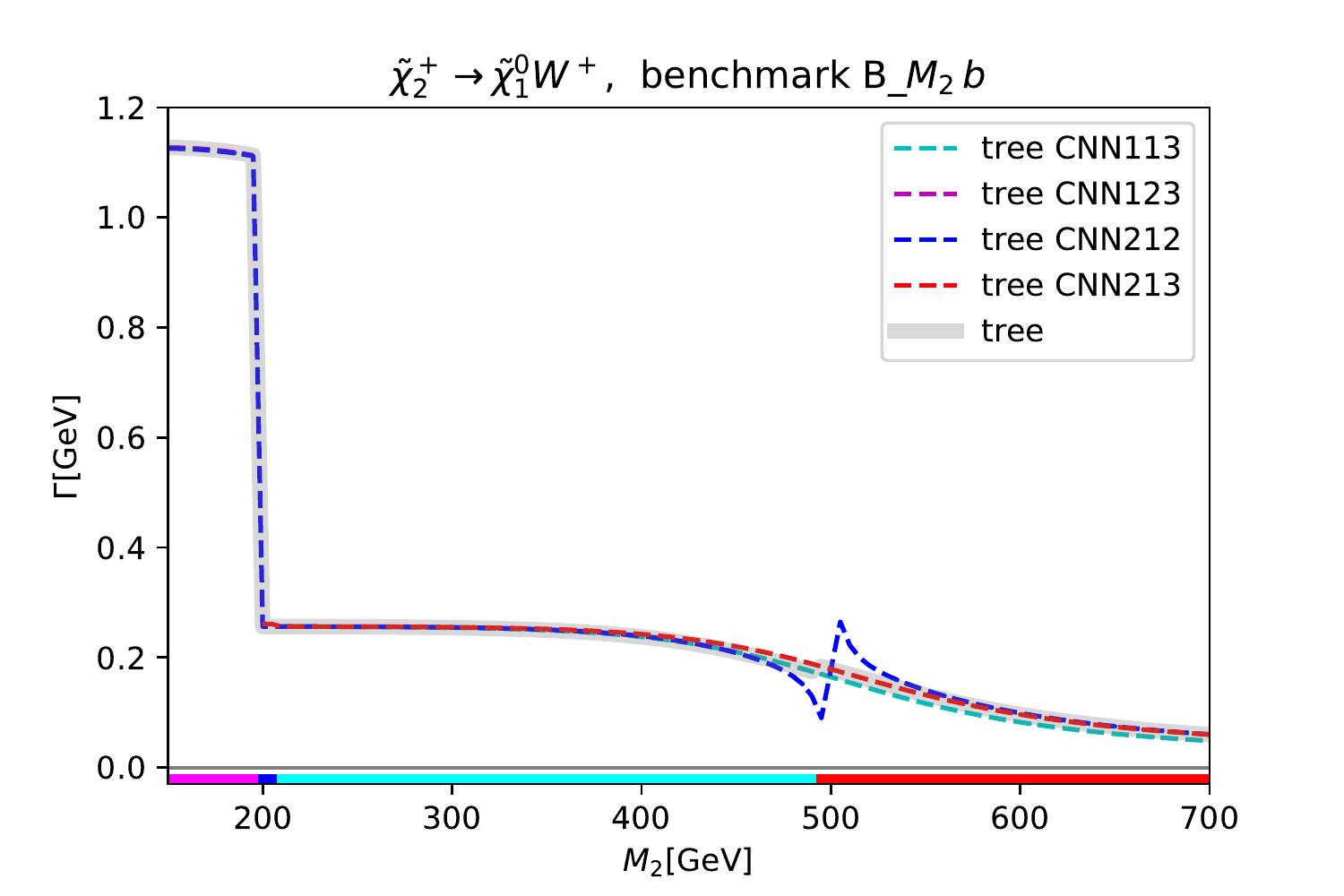}
  \includegraphics[width=0.45\textwidth]{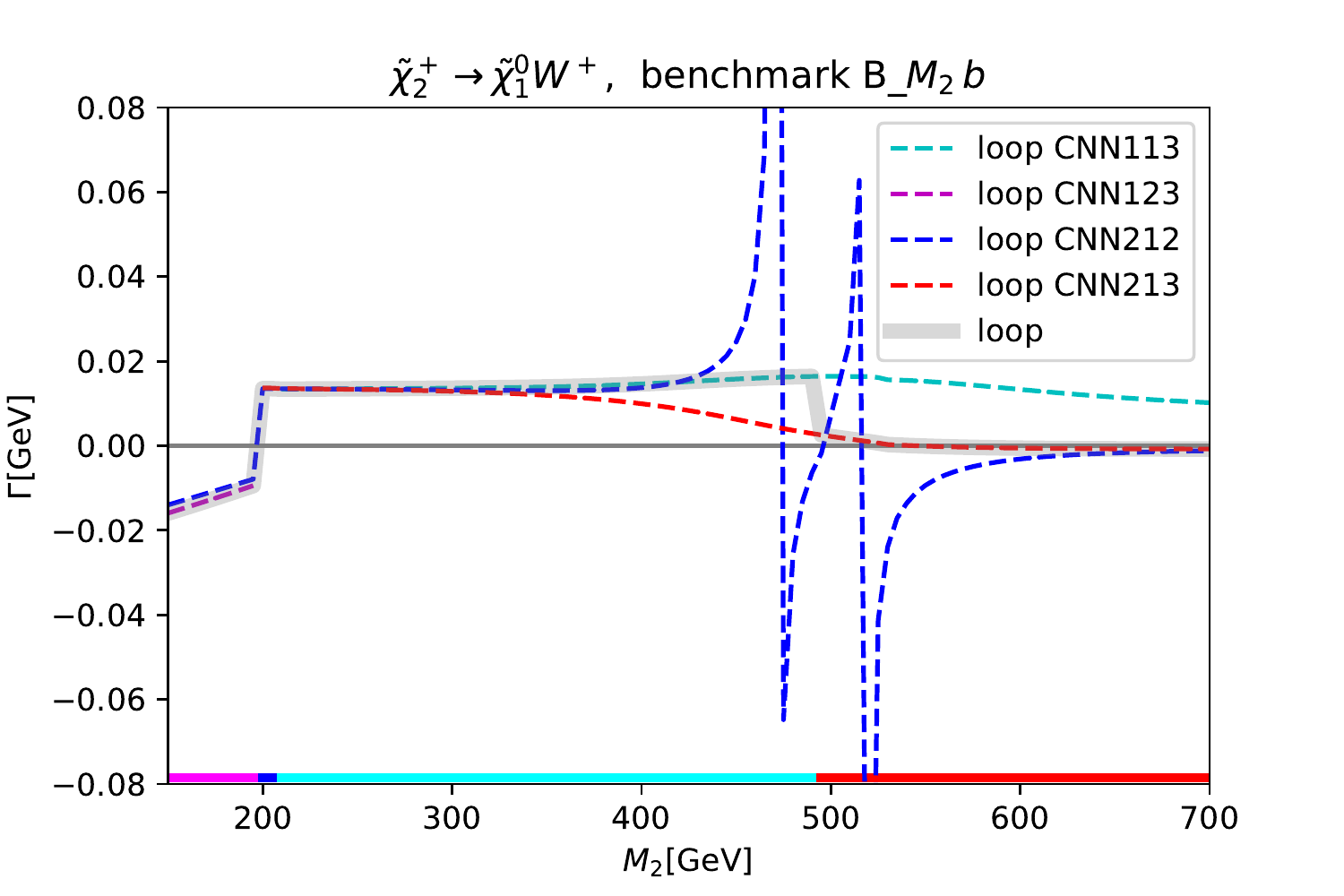}\\[1em]
  \includegraphics[width=0.45\textwidth]{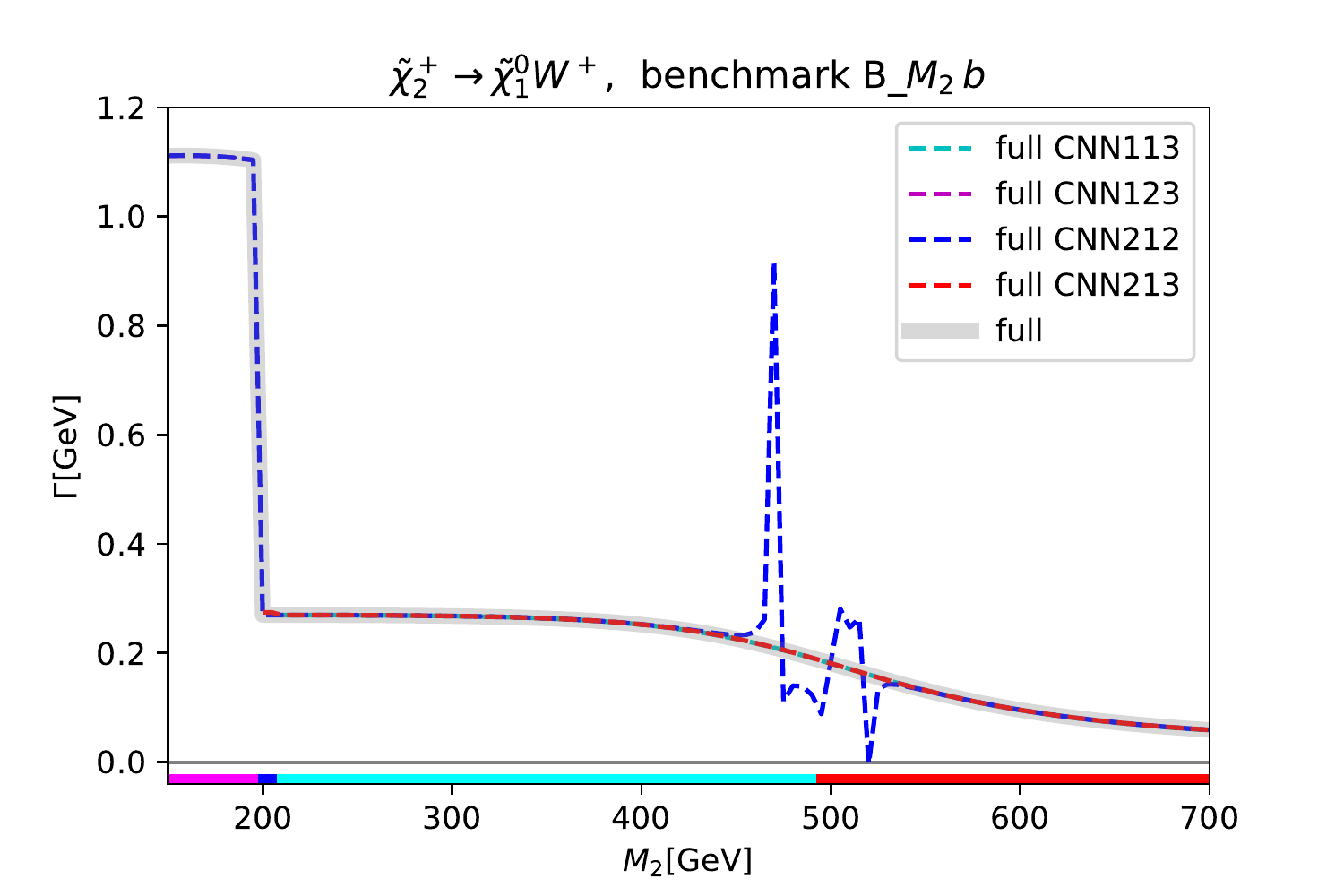}
  \includegraphics[width=0.45\textwidth]{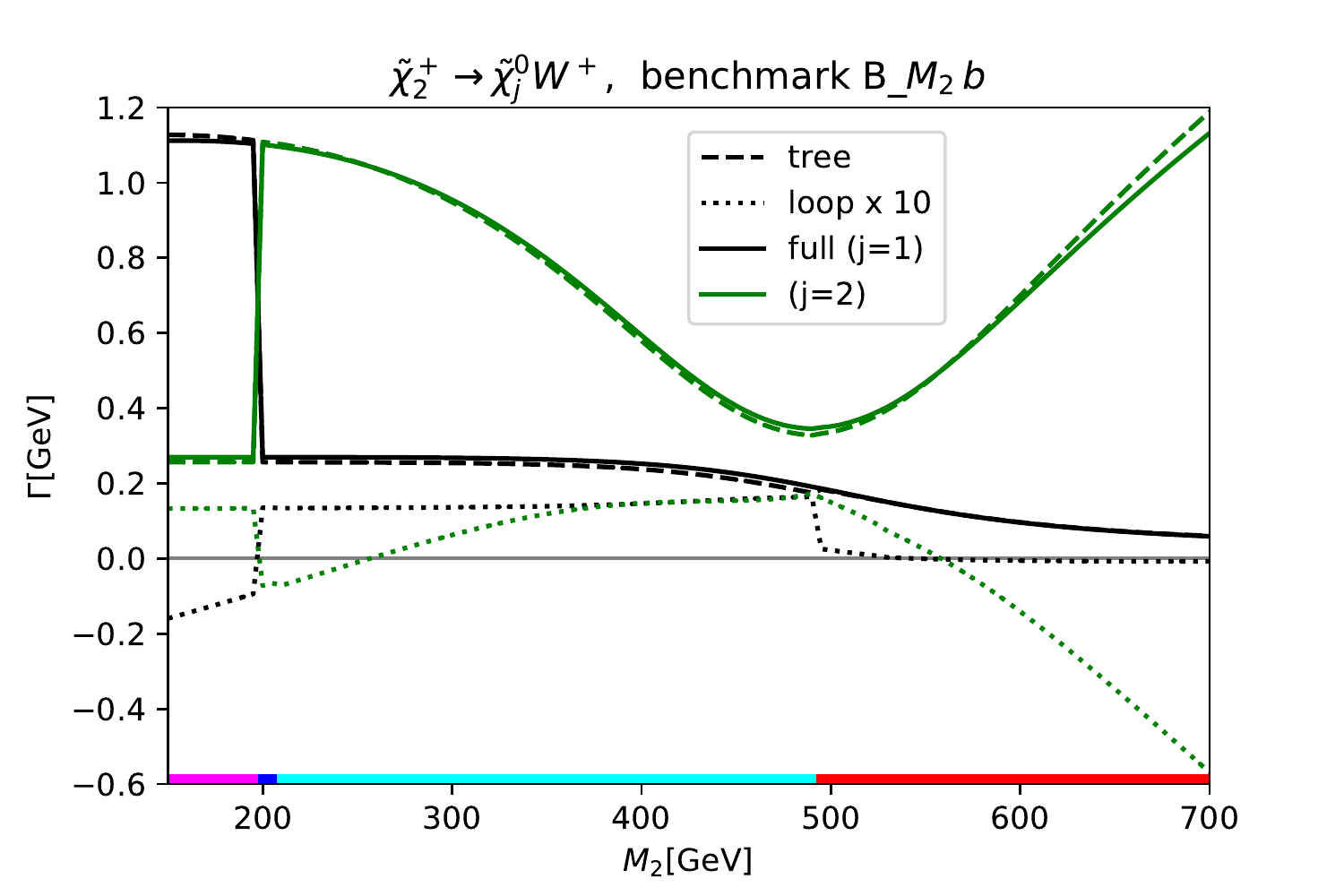}
  \caption{Decay width for $\chap{2} \to \neu{1} W^+$
    as a function of $M_2$ in benchmark scenario B\_$M_2\,b$, 
    see \refta{tab:bench1},
    with $M_1= -200\gev$, $\mu=500\gev$, $\tb= 10$.
    Shown are the four ``best RS'' in this range of parameters (see text).
    The plots show the same quantities as in
    \protect\reffi{fig:2022_C2N1W_mu_a}. 
    In the lower right plot we show the results for
      $\Ga(\chap\ \to \neu{j} W^+$ for $j=1,2$ (see text).
    The horizontal colored bar shows the best RS for the corresponding
   value of  $M_2$,  
   following  the same color coding as the curves: 
   CNN$_{123}$ for $M_2\le 195\gev$, 
   CNN$_{212}$ for $200 \gev \le M_2\le 205\gev$, 
   CNN$_{113}$ for $210 \gev \le M_2\le 490\gev$, 
   CNN$_{213}$ for $495\gev\le M_2$.
  }
\label{fig:2022_C2N1W_M2b}
\end{center}
\end{figure}

\begin{figure}[h!]
\vspace{2em}
\begin{center}
  \includegraphics[width=0.6\textwidth]{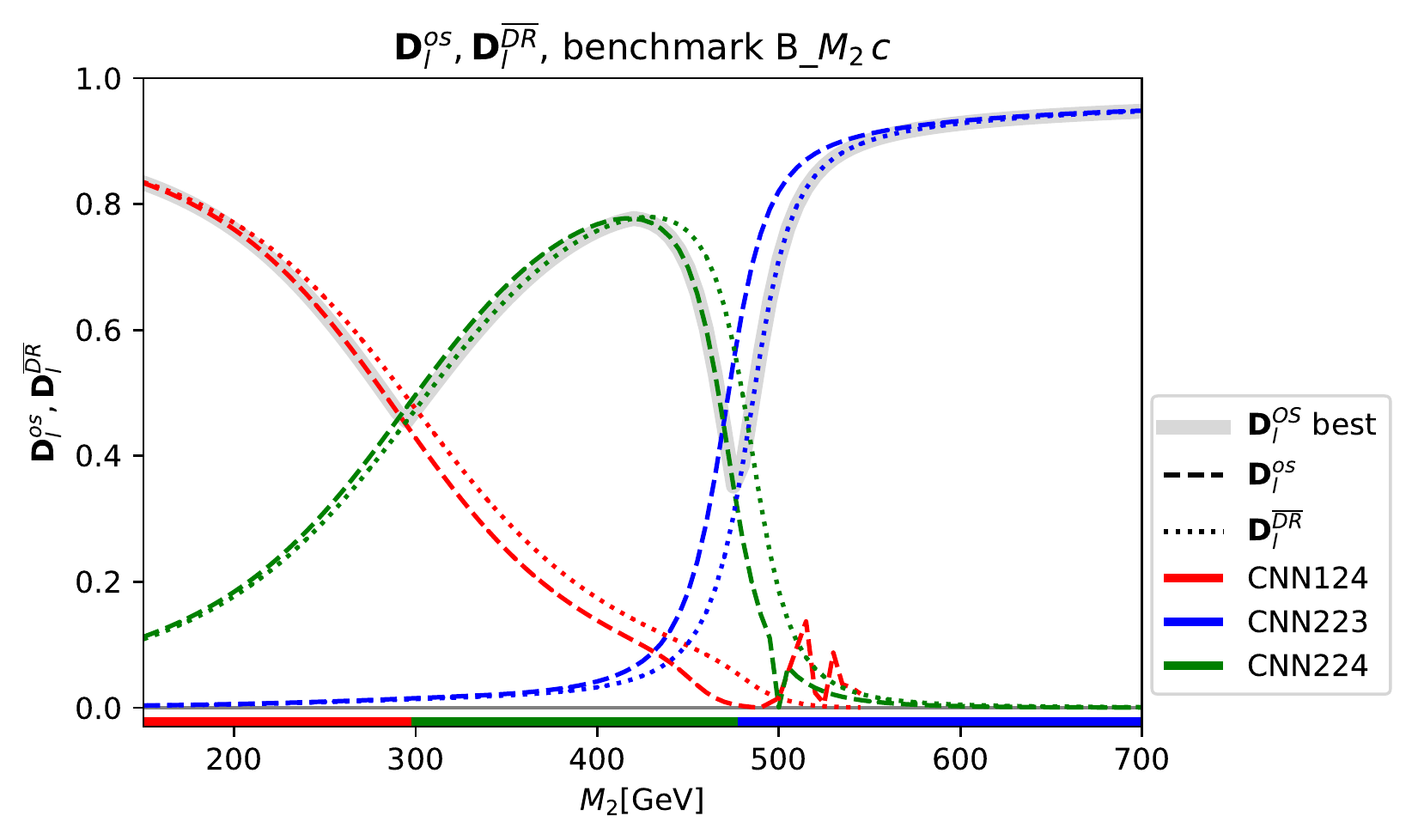}\\[1em]
  \includegraphics[width=0.45\textwidth]{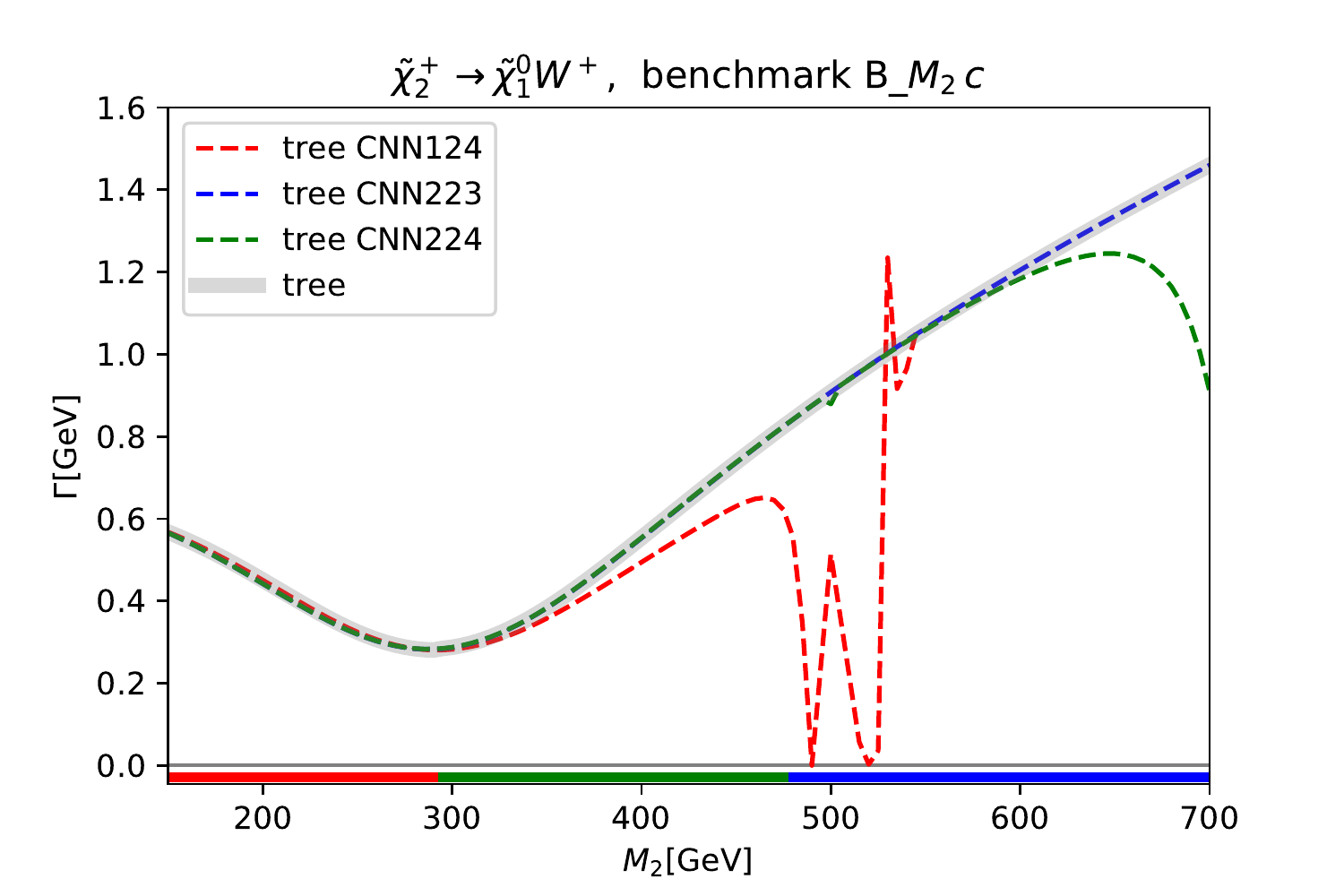}
  \includegraphics[width=0.45\textwidth]{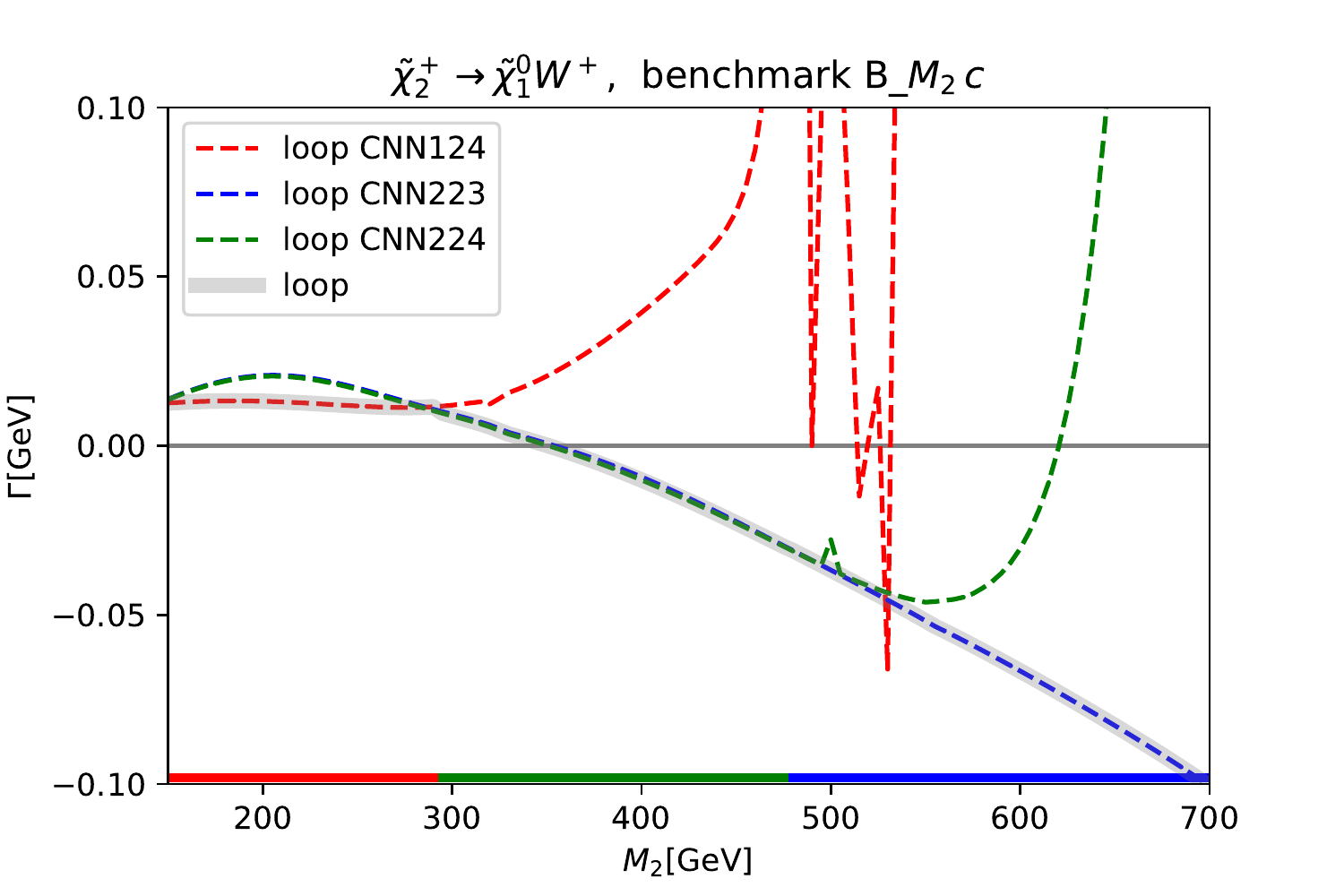}\\[1em]
  \includegraphics[width=0.45\textwidth]{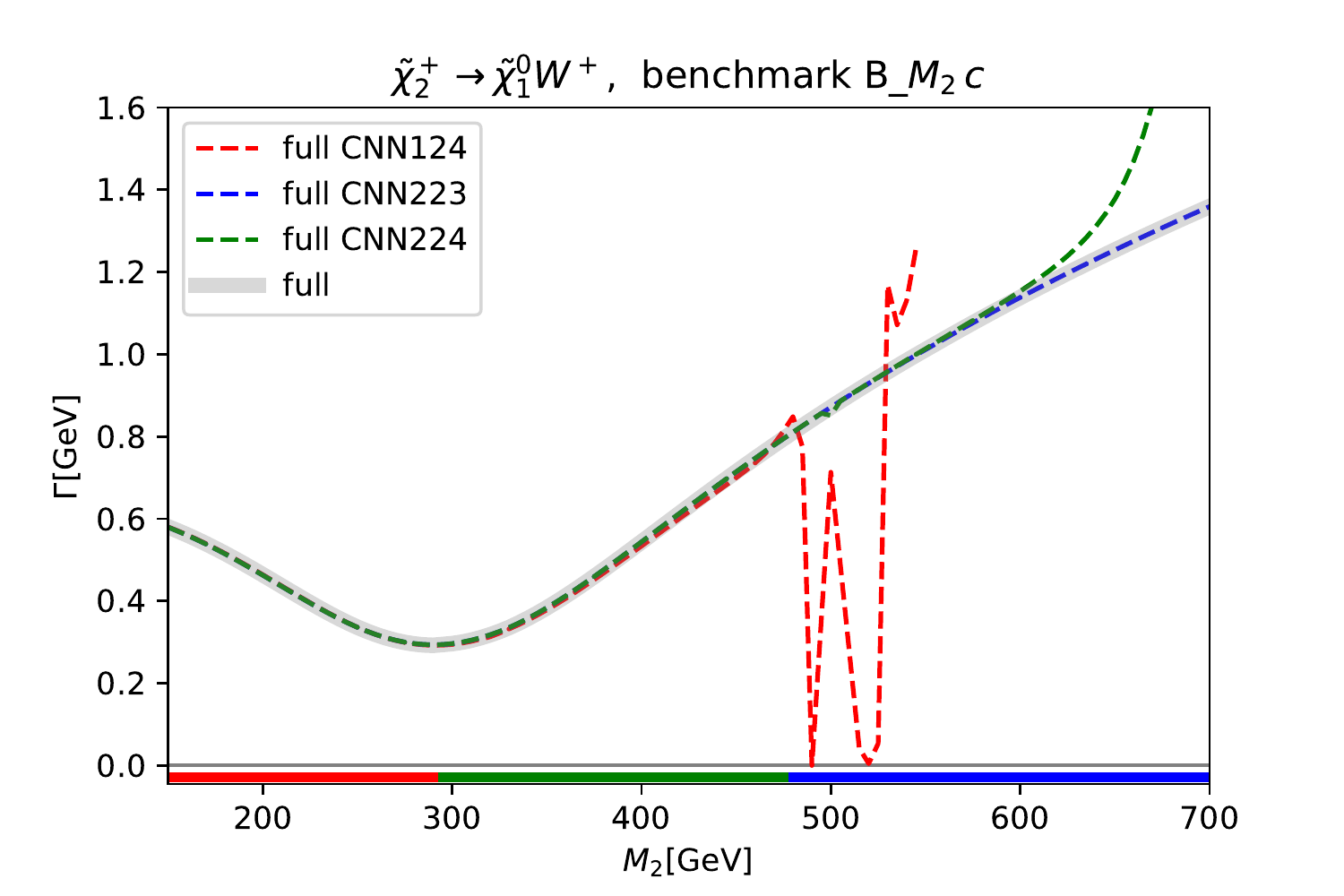}
  \includegraphics[width=0.45\textwidth]{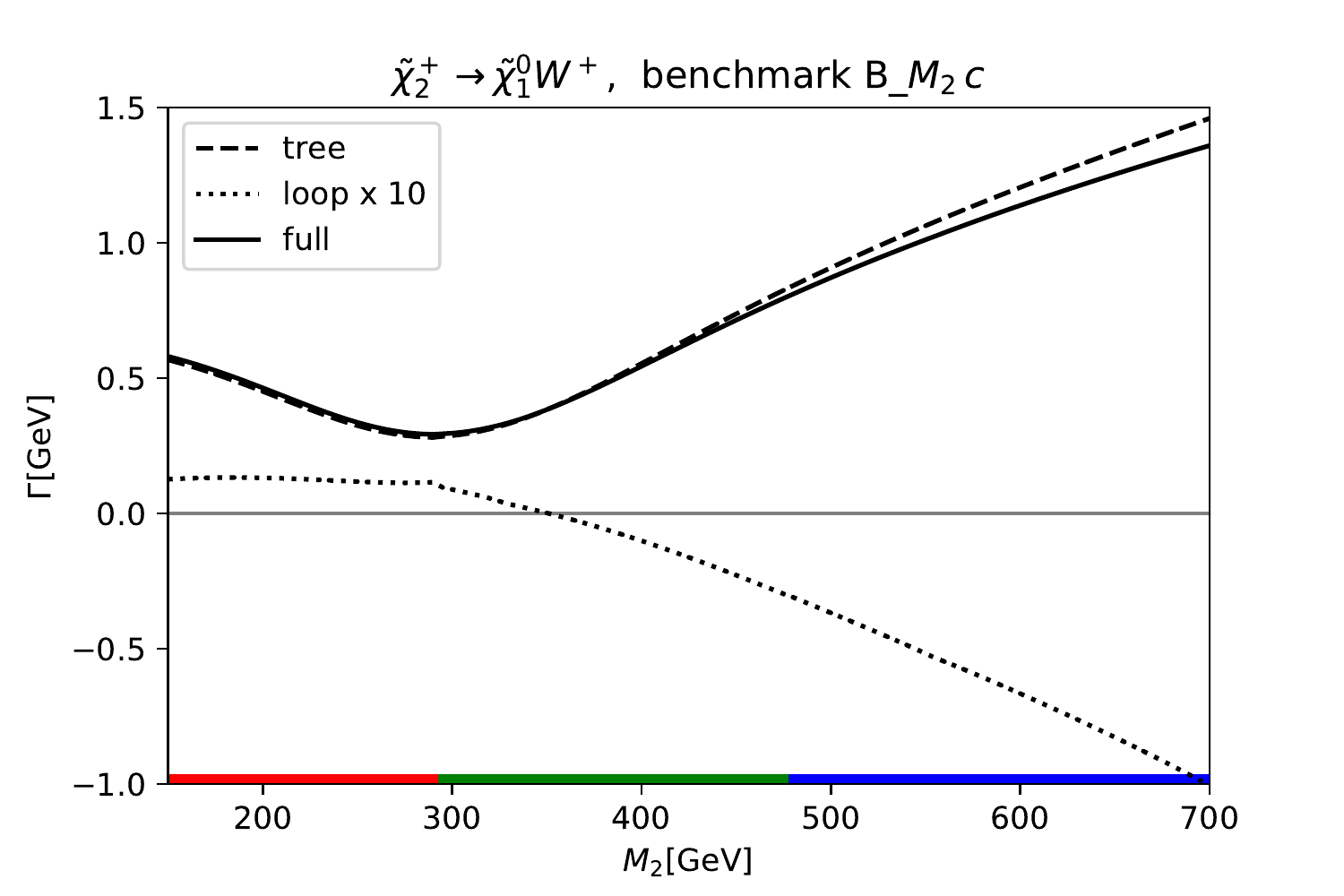}
\vspace{2em}
  \caption{
    Decay width for $\chap{2} \to \neu{1} W^+$ as a function of $M_2$ in
    benchmark scenario B\_$M_2\,c$, see \refta{tab:bench1},
    with $M_1= 500\gev$, $\mu=300\gev$, $\tb= 10$.
    Shown are the three ``best RS'' in this range of parameters (see text).
    The plots show the same quantities as in
    \protect\reffi{fig:2022_C2N1W_mu_a}. 
   The horizontal colored bar shows the best RS for the corresponding
   value of  $M_2$,  
   following  the same color coding as the curves: 
   CNN$_{124}$ for $M_2\le 290\gev$, 
   CNN$_{224}$ for $295 \gev \le M_2\le 475\gev$, 
   CNN$_{223}$ for $480\gev\le M_2$.
  }
\label{fig:2022_C2N1W_M2c}
\end{center}
\end{figure}

We finish the investigation of all possible different mass hierarchies
in the chargino/neutralino sector with an analysis of the benchmark
scenarios with varied~$M_2$. The results are shown for the three
benchmark scenarios B\_$M_2\,a$, B\_$M_2\,b$ and B\_$M_2\,c$,
see \refta{tab:bench1},
in \reffis{fig:2022_C2N1W_M2a}\, \ref{fig:2022_C2N1W_M2b} and
\ref{fig:2022_C2N1W_M2c}, respectively. The plots in the three figures
show the same quantities as in 
\protect\reffi{fig:2022_C2N1W_mu_a}. 
The three scenarios differ in the hierarchy and sign of $M_1$ and
$\mu$, where in the first (second, third) scenario we have chosen
$M_1 = 200 (-200, 500) \gev$ and $\mu = 500 (500, 200) \gev$.

In principle, the
overall results are found as for the variation of $\mu$ and $M_1$, see
\refse{sec:mu-var} and \ref{sec:m1-var}.
However, in the benchmark scenario with negative $M_1$ a
``discontinuous feature'' in $\Ga(\chap2 \to \neu1 W^+)$ can be
observed at $M_2 \sim 200 \gev$. 
At this mass value a ``level crossing'' of $\neu1$ and $\neu2$ takes
place. For $M_2 \lsim 200 \gev$ the $\neu1$ ($\neu2$) has wino (bino)
character, whereas for $M_2 \gsim 200 \gev$ this changes to bino
(wino) character. The $\cha2-\neu1-W$ coupling changes from
higgsino-higgsino-gauge to wino-wino-gauge coupling, where the former
is substantially larger than the latter, resulting in the strong drop
of the decay width at $M_2 \sim 200 \gev$.
A similar pattern is observed at $\sim 200 \gev$ in the B\_$M_2\,a$ 
benchmark scenario, where also a strong, but continuous increase of the
decay rate can be observed. 
The difference between B\_$M_2\,b$ and B\_$M_2\,a$ 
is the $\cp$ character of the $\neu1$ and $\neu2$. 
The
negative value of $M_1$ in the scenario B\_$M_2\,b$ results in
opposite values for
the intrinsic $\cp$-parity of the two neutralinos,
preventing a mixture of the two
particles (we assume $\cp$-conservation throughout the paper). This
results in a sharp transition at $M_1 = M_2$. On the other hand, in
scenario B\_$M_2\,a$ 
the two states have the same intrinsic $\cp$ parities
and can mix, resulting in a smooth transition. This is also reflected in
the mass patterns of the neutralinos, as shown in \reffi{fig:chimasses},
see the left plot in the second (fourth) row for B\_$M_2\,b$ (B\_$M_2\,a$).
Correspondingly, the ``best RS'' changes from \cnn{1}{2}{3} for
$M_2 \lsim 200 \gev$ to \cnn{1}{1}{3}, such that always one wino-like
state (the $\cha1$), a higgsino-like state (the $\neu3$) and a bino-like
state (either $\neu2$ for $M_2 \lsim 200 \gev$, or $\neu1$ for
$M_2 \gsim 200 \gev$) is renormalized OS.
In order to demonstrate this level crossing, in the lower right plot
of \reffi{fig:2022_C2N1W_M2b} we show the results for
$\Ga(\chap2 \to \neu{j} W)$ (tree, loop~($\times 10$) and full) for
$j = 1,2$. It can clearly be observed that by switching from $j=1$ to
$j=2$ at $M_2 \sim 200 \gev$ effectively smooth results are obtained
for the decay width as a function of $M_2$.

As stated above, in general the overall results for the variation
of $M_2$ are found effectively identical as for
the variation of $\mu$ and $M_1$.
The selected determinant, as shown in the upper rows of
\reffis{fig:2022_C2N1W_M2a}, \ref{fig:2022_C2N1W_M2b} and
\ref{fig:2022_C2N1W_M2c}, becomes 
smallest for $M_1 \approx M_2$, but does not go below $\sim 0.30$.
For the rest of the parameter space the best
determinant is above $\sim 0.46$. For B\_$M_2\,a$ four
schemes, all of type CNN, are selected, for B\_$M_2\,b$ (the scenario
with negative $M_1$), four schemes are selected
while for B\_$M_2\,c$ only three
schemes are automatically chosen, all of them are of type CNN.
The different choices reflect, similar to the case discussed in
\refse{sec:mu-var}, the two different hierarchies of $M_1$ and $\mu$. 

Each of the
schemes shows good behavior where it is selected. The relative size of
the ``selected higher-order corrections'' does not exceed $\sim 15\%$
and mostly stays below $\sim 10\%$.
As before, the only exceptions occur where
the Born contribution vanishes due to a zero crossing of the
respective coupling, as for instance $M_2\sim M_1$ in benchmark B\_$M_2\,a $.
The final result, as shown in the
lower right plots of \reffis{fig:2022_C2N1W_M2a}, \ref{fig:2022_C2N1W_M2b} and
\ref{fig:2022_C2N1W_M2c}, are smooth curves with a reliable higher-order
correction (taking into account the level crossing in B\_$M_2\,b$).

\medskip
Additional figures with other decay widths involving $Z$ and $h$ in the
final state ($\Ga(\neu4 \to \neu1 h)$, $\Ga(\neu4 \to \neu1 Z)$) can be
found in appendix~\ref{sec:plots}.

\medskip
For the benchmark scenario B\_$M_2\,a$ we also discuss another physics
case: having one (or more) external charged particles makes it desirable
to have them renormalized OS. This ensures the cancellation of IR
divergences to all orders. If, on the other hand, such a RS
cannot be chosen, chargino mass shifts are employed, see the
discussion in \refse{sec:notation}.
~From a perturbation theory point of view both solutions are
acceptable. However, a full OS renormalization of external charged
particles still appears preferred.

\begin{figure}[h!]
  \begin{center}
  \includegraphics[width=0.60\textwidth]{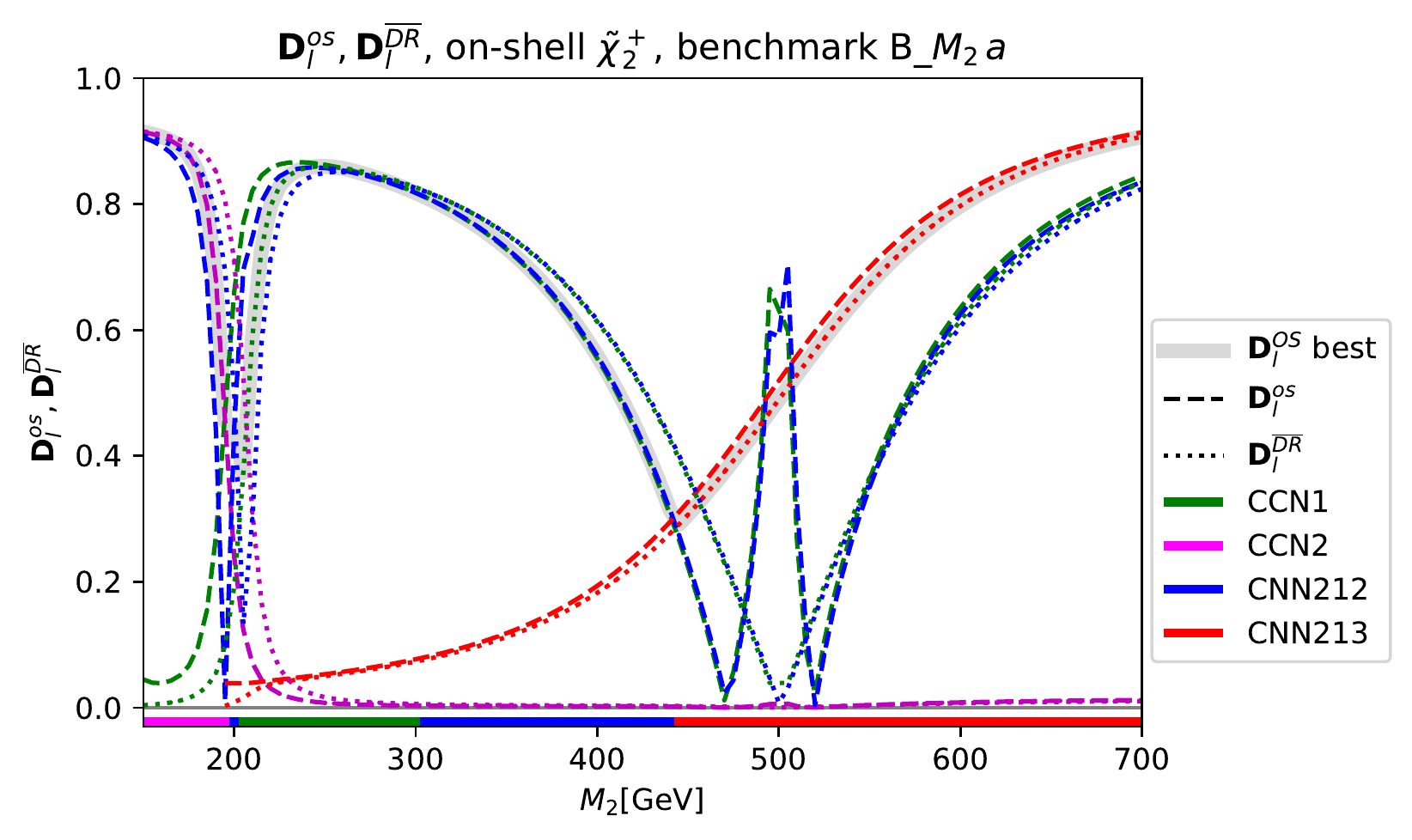}\\[1em]
  \includegraphics[width=0.45\textwidth]{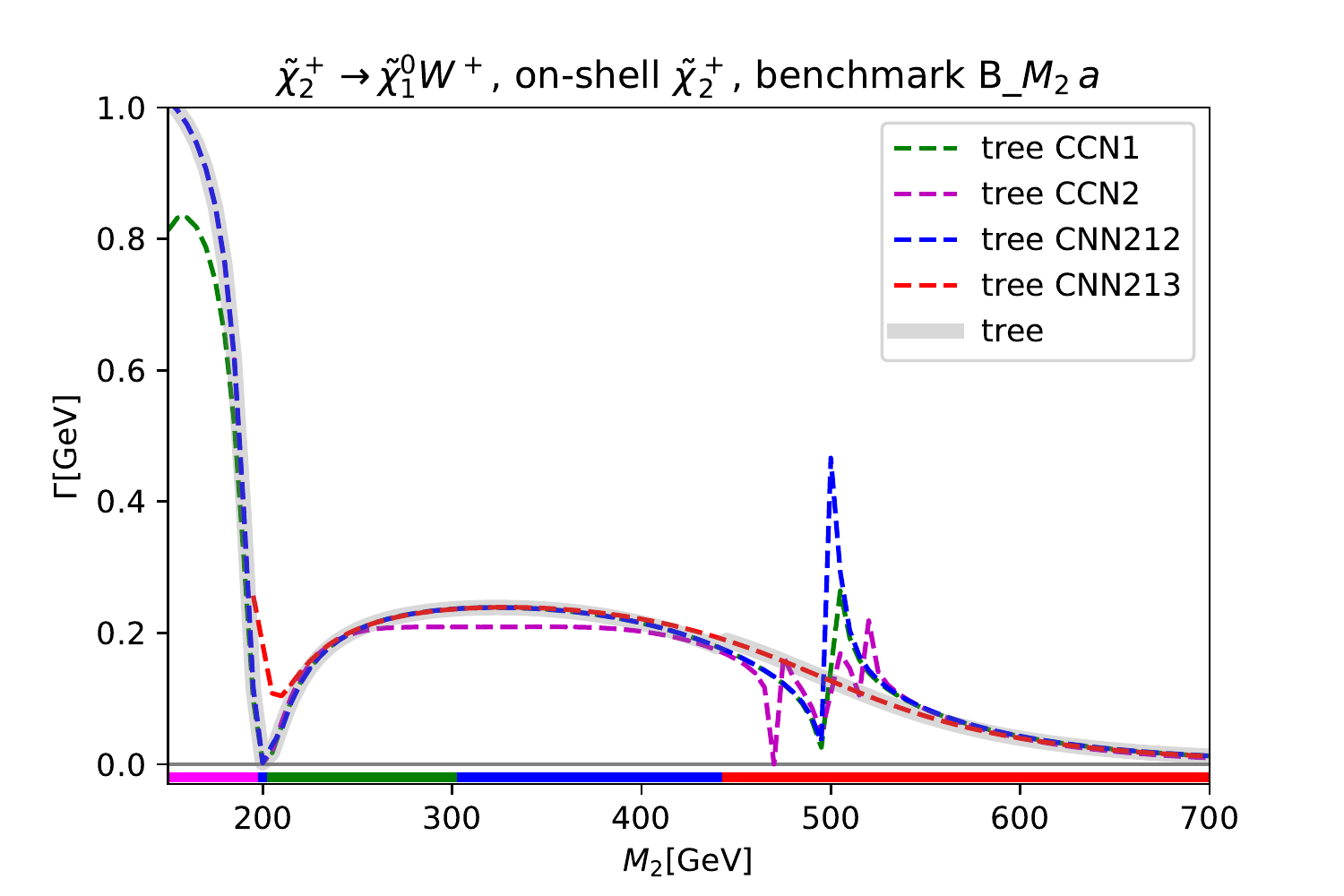}
  \includegraphics[width=0.45\textwidth]{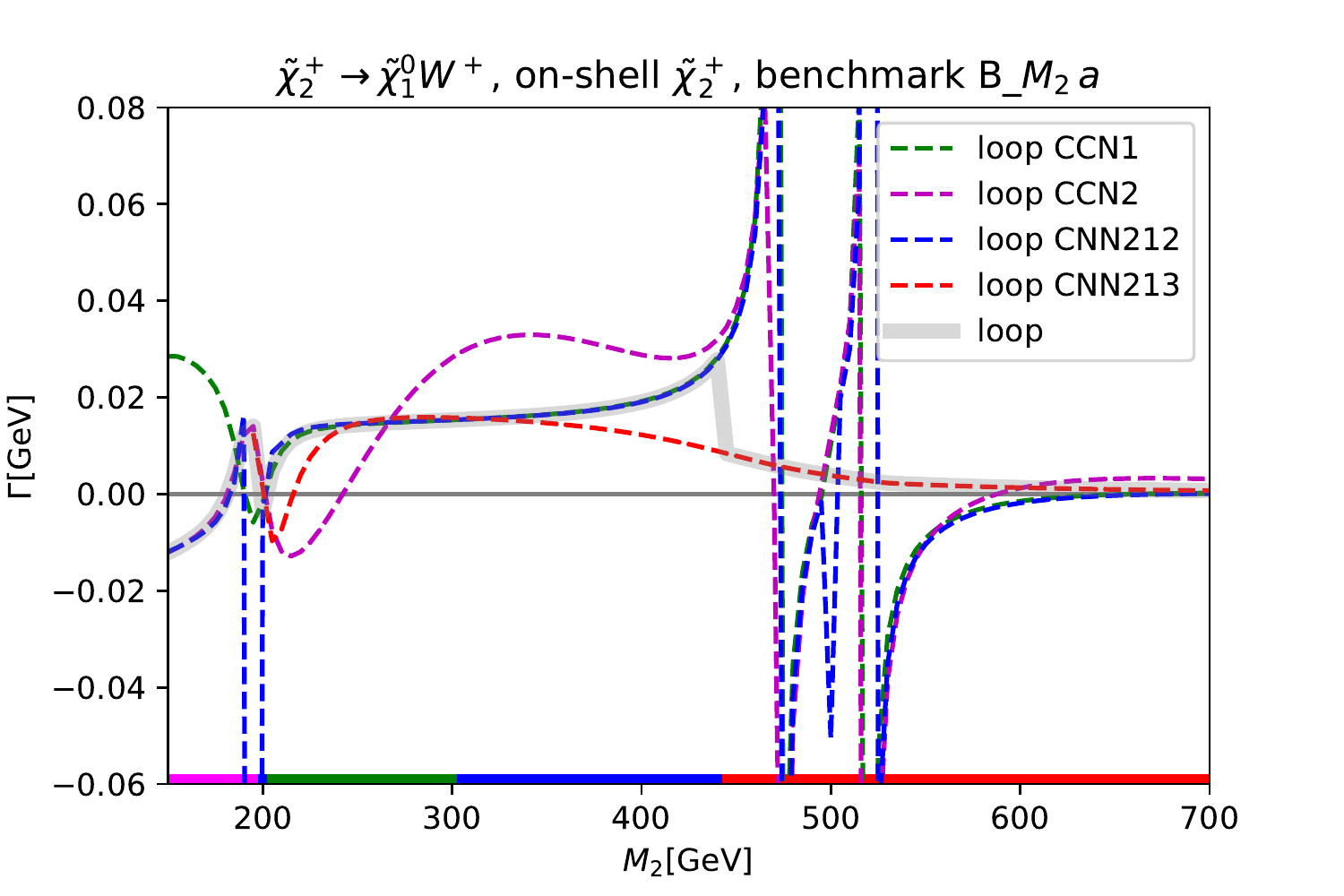}\\[1em]
  \includegraphics[width=0.45\textwidth]{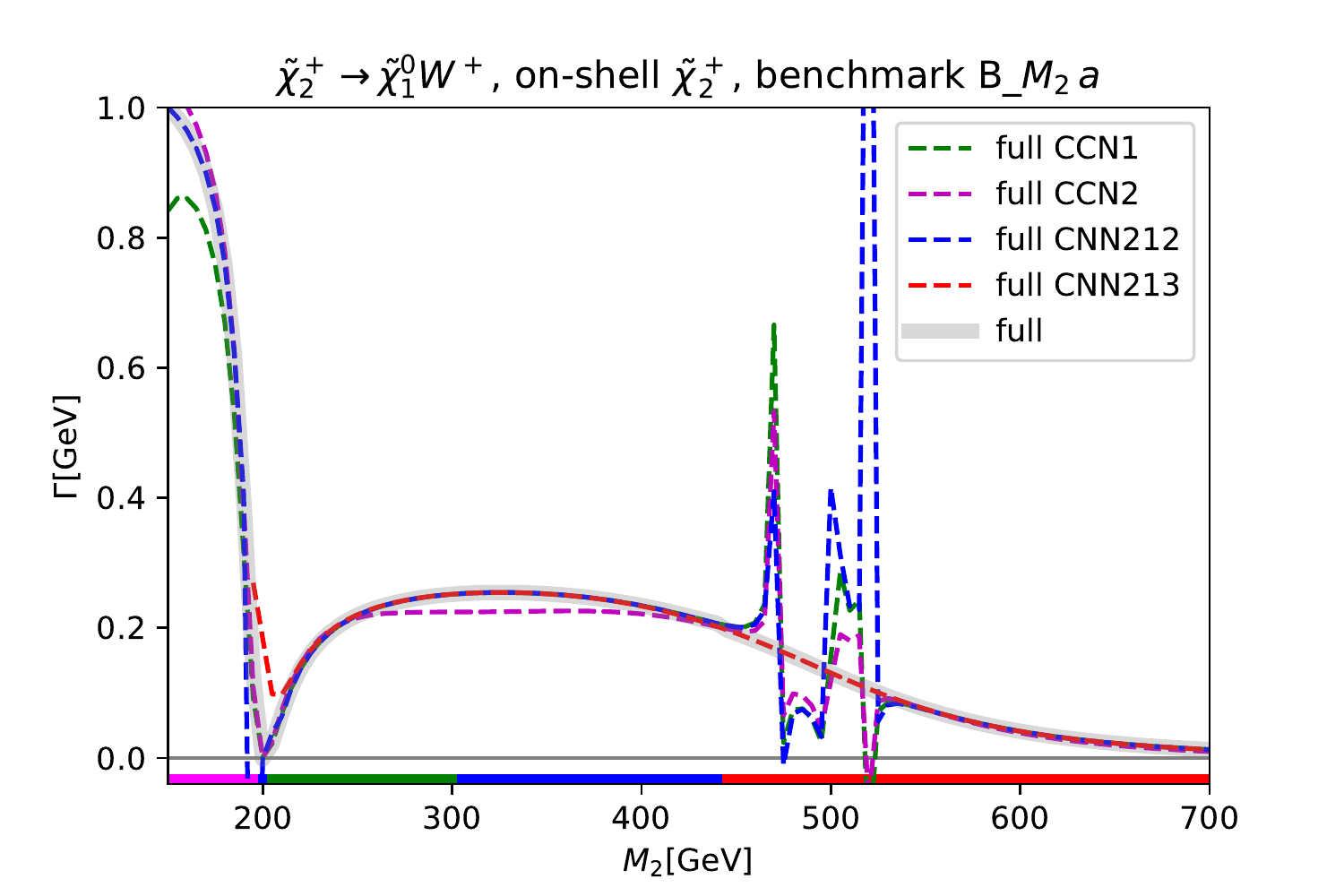}
  \includegraphics[width=0.45\textwidth]{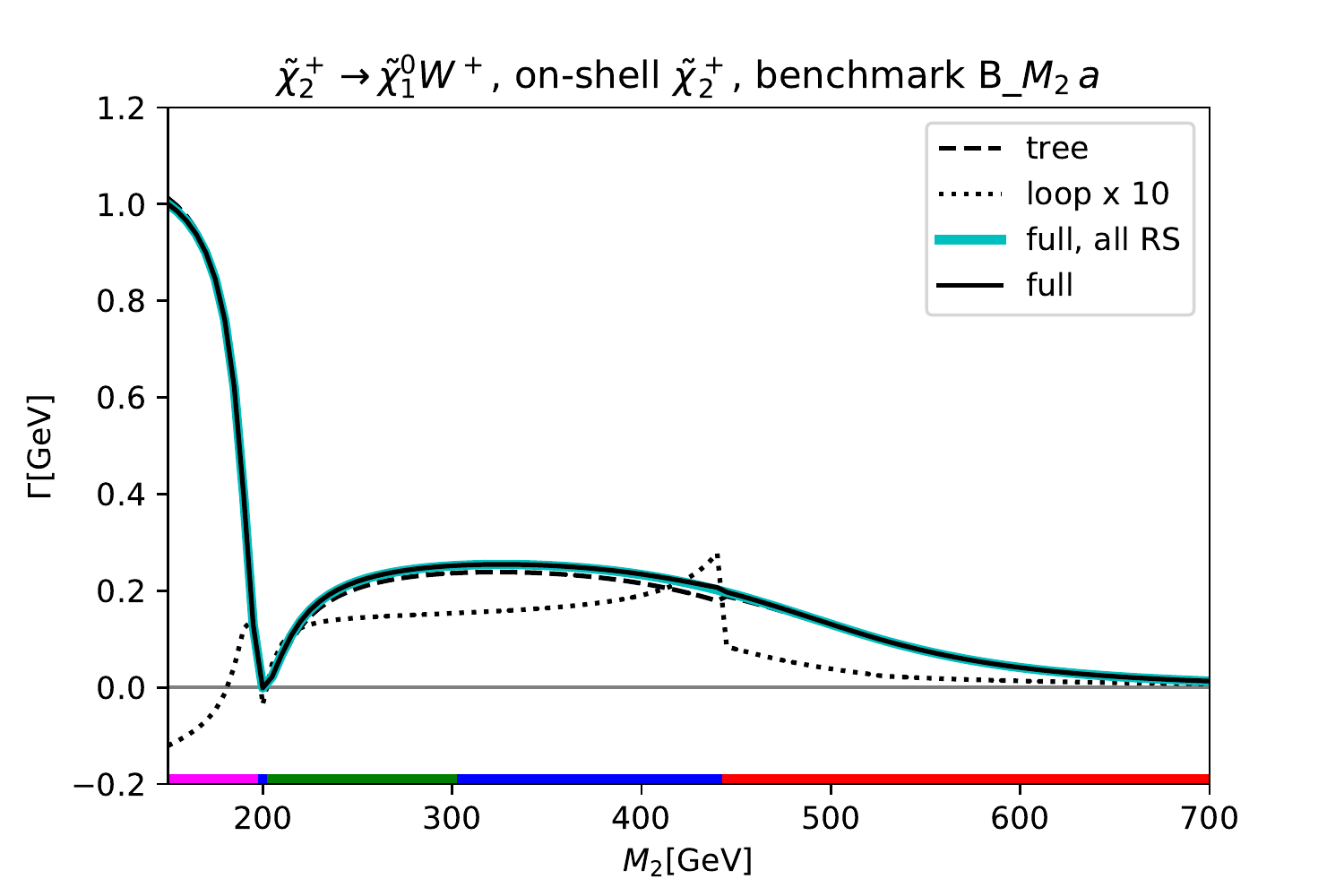}
\vspace{2em}
  \caption{
    Decay width for $\chap{2} \to \neu{1} W^+$
    as a function of $M_2$ in benchmark scenario B\_$M_2\,a$, 
    see \refta{tab:bench1},
    with $M_1= 200\gev$, $\mu=500\gev$, $\tb= 10$. 
    Shown are the four ``best RS'' in this range of parameters, where
    the $\cha2$ is forced to be renormalized OS.
    The plots show the same quantities as in
    \protect\reffi{fig:2022_C2N1W_mu_a}. 
    Also shown in the lower right plot
    is the ``full'' result with all RS allowed
    (cyan line, reproduced from the lower right plot of
    \protect\reffi{fig:2022_C2N1W_M2a}).
   The horizontal colored bar shows the best RS for the corresponding
   value of  $M_2$,  
   following  the same color coding as the curves: 
   CCN$_{2}$ for $M_2\le 195\gev$, 
   CNN$_{212}$ for $200 \gev \le M_2\le 200\gev$, 
   CCN$_{1}$ for $205 \gev \le M_2\le 300\gev$, 
   CNN$_{212}$ for $305 \gev \le M_2\le 440\gev$, 
   CNN$_{213}$ for $445\gev\le M_2$.
  }
\label{fig:2022_C2N1W_M2_forcedC2}
\end{center}
\end{figure}

In \reffi{fig:2022_C2N1W_M2_forcedC2} we show the results, analogous to 
\reffi{fig:2022_C2N1W_M2a} where all
benchmark scenarios were allowed to be 
chosen, for the decay $\chap2 \to \neu1 W^+$, but now enforcing the
$\cha2$ to be renormalized OS. In \reffi{fig:2022_C2N1W_M2a} two out of
the four chosen RS had the $\cha2$ not renormalized OS (\cnn{1}{1}{3}
and \cnn{1}{2}{3}), which are now forbidden. Forcing the $\cha2$ to be
renormalized OS now results in two CCN to be selected (\ccn{1} and
\ccn{2}), along with the already previously chosen \cnn{2}{1}{2} and
\cnn{2}{1}{3}. However, also the $M_2$ values where the latter ones are
selected change strongly w.r.t.\ \reffi{fig:2022_C2N1W_M2a}. However,
also in this case the selected determinant is larger than
$\sim 0.30$.
The overall behavior of the selected loop corrections and the final
obtained decay width as shown in \reffi{fig:2022_C2N1W_M2_forcedC2} is
nearly identical to the result shown in \reffi{fig:2022_C2N1W_M2a},
included for a better comparison as cyan line in the lower right plot of
\reffi{fig:2022_C2N1W_M2_forcedC2}, showing the ``best'' full result.
For $M_2 > 490 \gev$ the chosen RS are the same, i.e.\ leading to
identical results, while for 
$M_2 \le 490 \gev$ the difference is too small to be
appreciated, except for $M_2 \sim 440 \gev$, where 
a slight difference can be observed. This
demonstrates that the overall idea to choose a ``good RS'' also works in
case of a restricted allowed set of RS (in this case forcing the $\cha2$
to be renormalized OS to avoid problems with the cancellation of IR
divergences).

\begin{figure}[h!]
\vspace{1em}
  \begin{center}
  \includegraphics[width=0.60\textwidth]{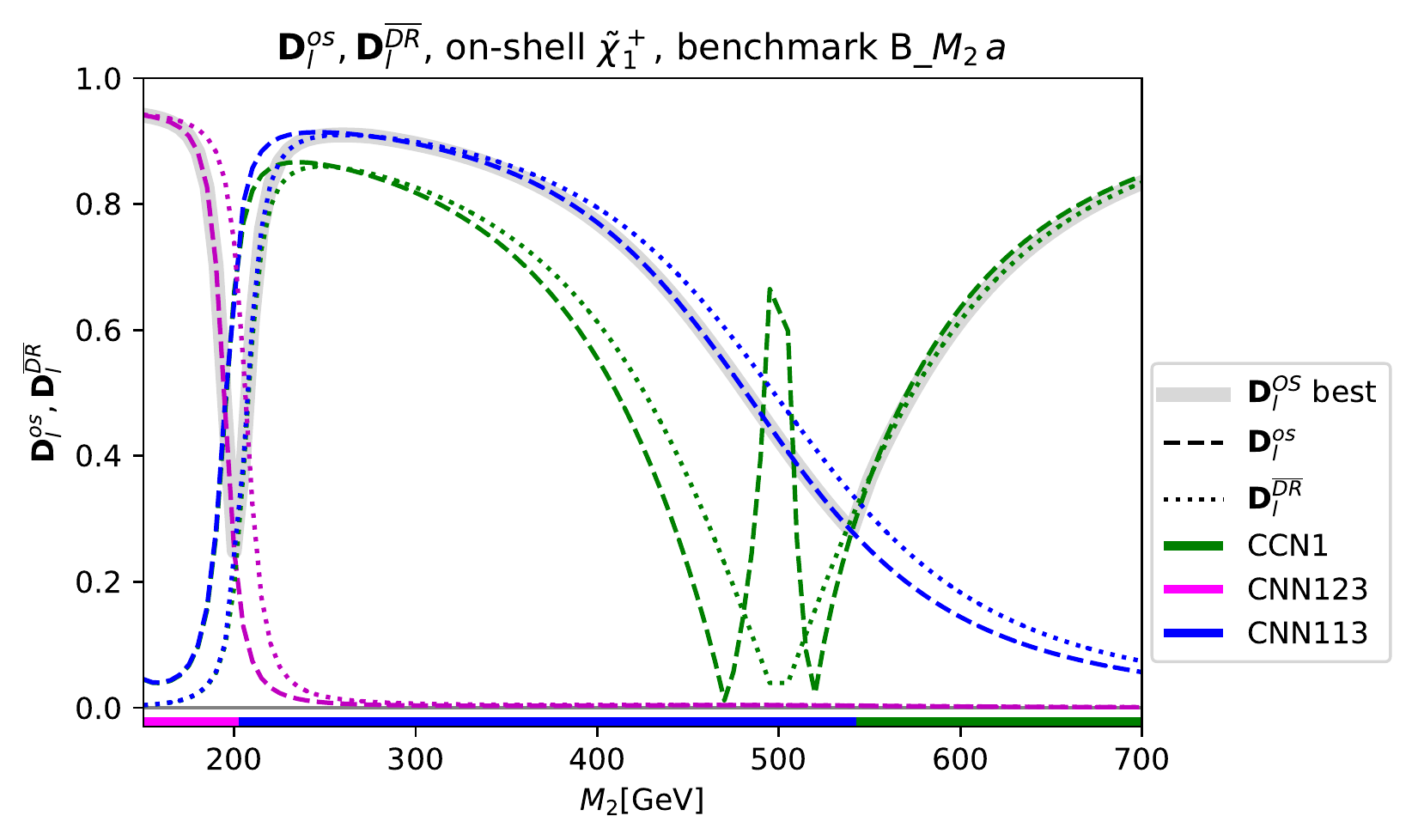}\\[1em]
  \includegraphics[width=0.45\textwidth]{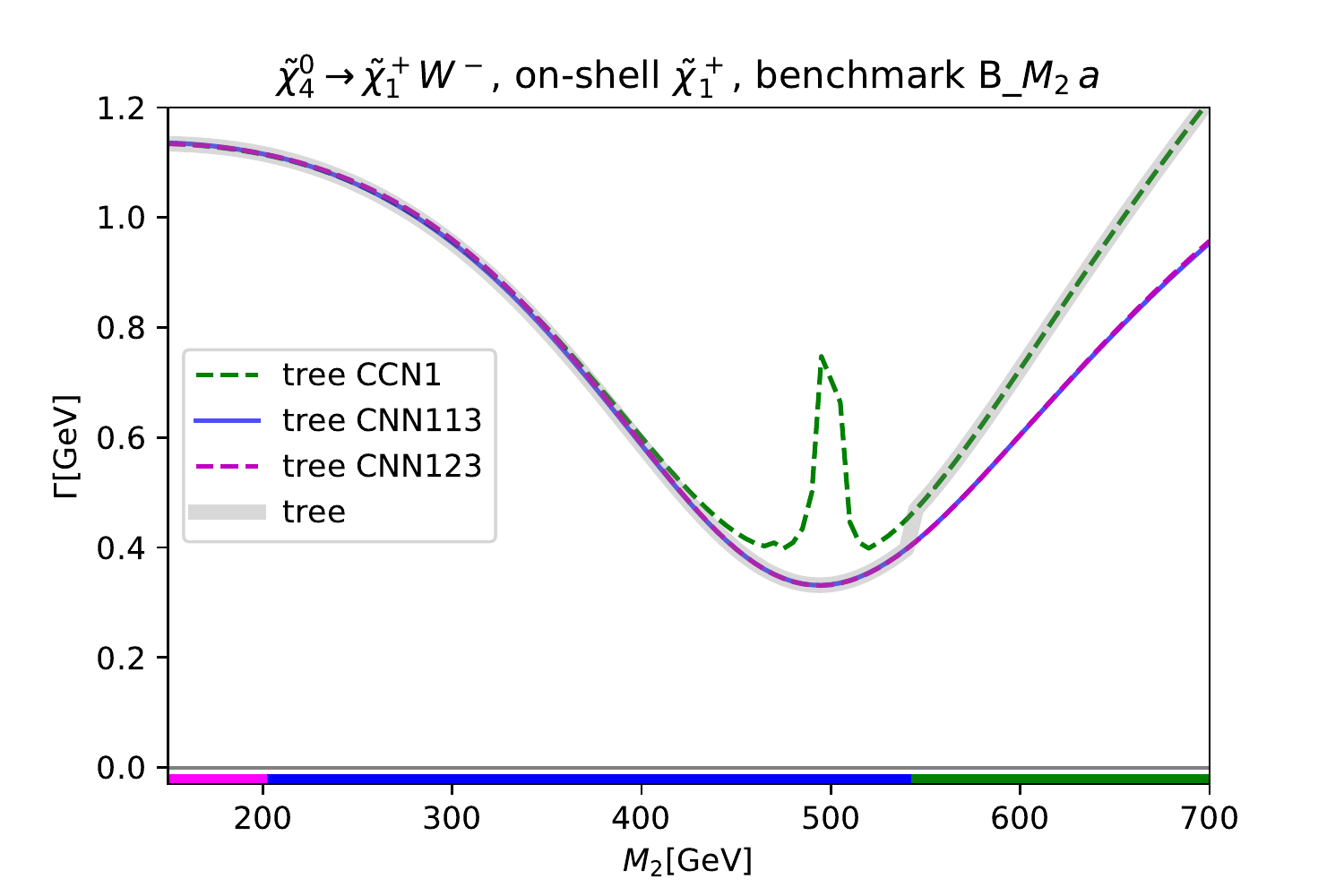}
  \includegraphics[width=0.45\textwidth]{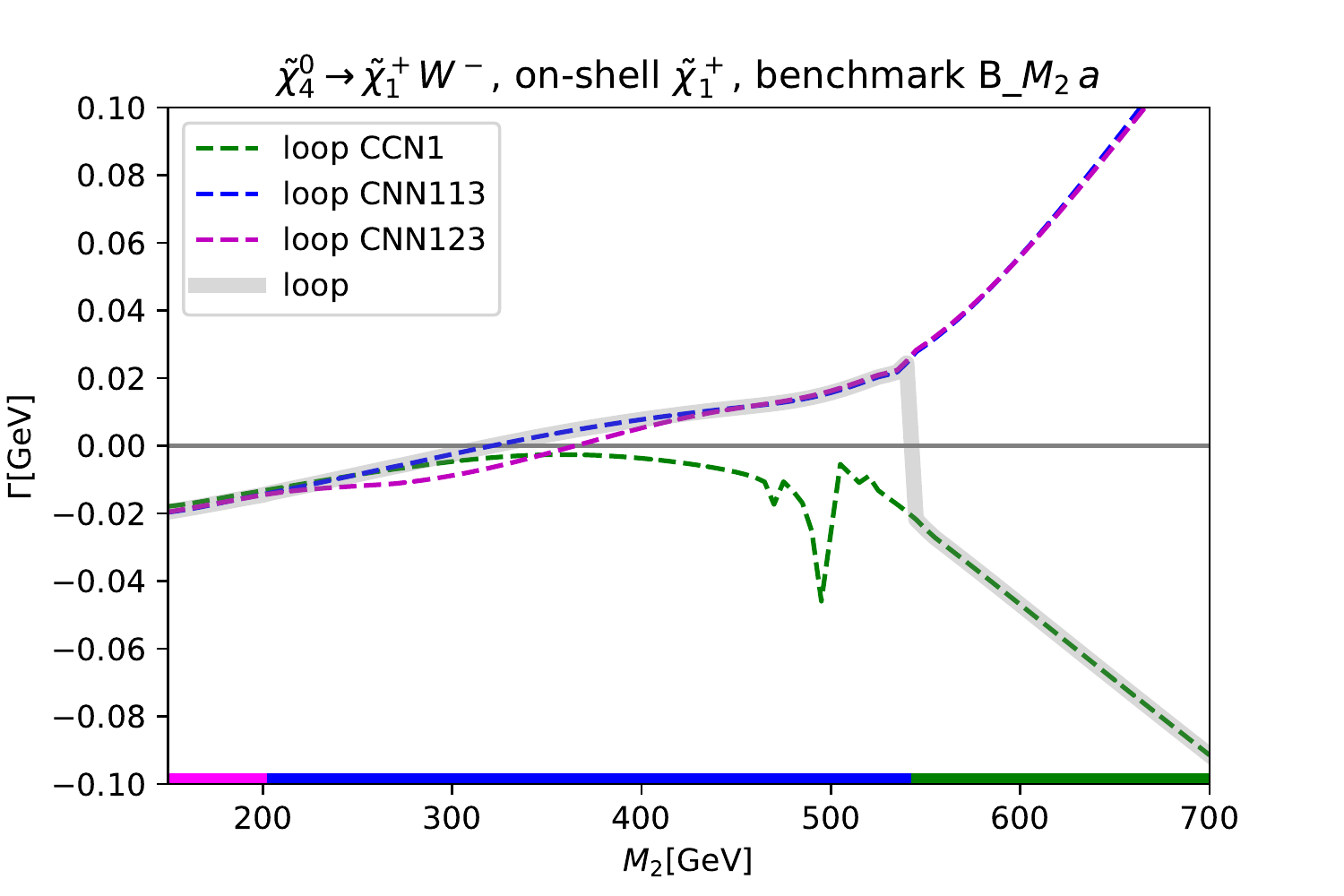}\\[1em]
  \includegraphics[width=0.45\textwidth]{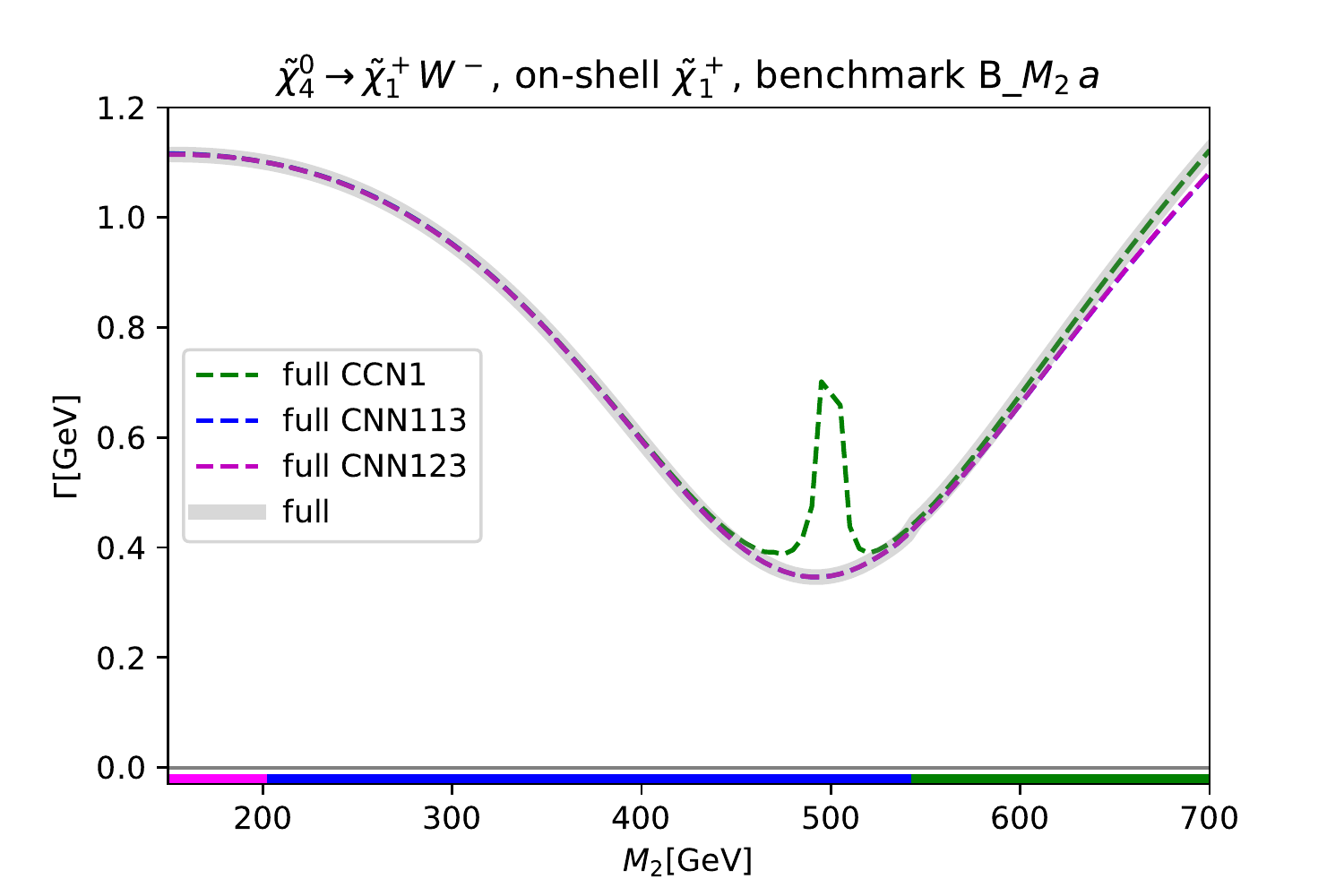}
  \includegraphics[width=0.45\textwidth]{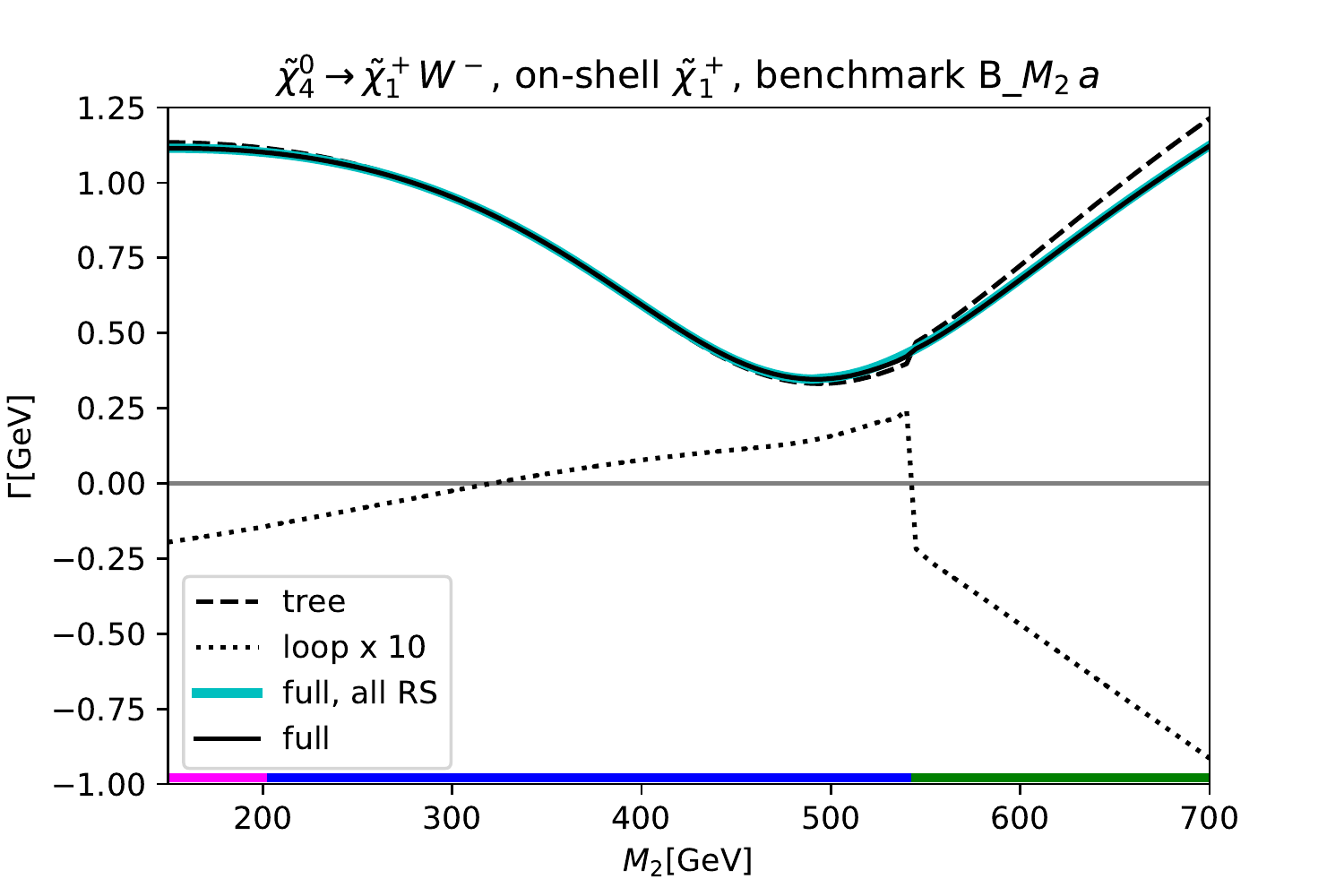}
\vspace{2em}
  \caption{
    Decay width for $\neu{4} \to \chap{1} W^-$
    as a function of $M_2$ in benchmark scenario B\_$M_2\,a$, 
    see \refta{tab:bench1},
    with $M_1= 200\gev$, $\mu=500\gev$, $\tb= 10$. 
    Shown are the four ``best RS'' in this range of parameters, where
    the $\cha1$ is forced to be renormalized OS.
    The plots show the same quantities as in
    \protect\reffi{fig:2022_C2N1W_mu_a}.
    Also shown in the lower right plot 
    is the ``full'' result with all RS allowed (cyan line).
   The horizontal colored bar shows the best RS for the corresponding
   value of  $M_2$,  
   following  the same color coding as the curves: 
   CNN$_{123}$ for $M_2\le 200\gev$, 
   CNN$_{113}$ for $205 \gev \le M_2\le 540\gev$, 
   CCN$_{1}$ for $545\gev\le M_2$.
 }
\label{fig:2022_N4C1W_M2_forcedC1}
\end{center}
\end{figure}

Similarly, in \reffi{fig:2022_N4C1W_M2_forcedC1} we show the
results for the decay $\neu4 \to \chap1 W^-$ in the benchmark scenario
B\_$M_2\,a$, and with the $\cha1$ forced to be renormalized OS. Now one
CCN scheme (\ccn{1}) and two CNN schemes (\cnn{1}{2}{3} and
\cnn{1}{1}{3}) are selected. As in the case of the $\cha2$ forced to be
renormalized OS, also here we find the smallest selected determinant
larger than $\sim 0.25$, and in the end an effectively smooth
loop-corrected decay width, as can be observed the lower right plot
of \reffi{fig:2022_N4C1W_M2_forcedC1}. 
Only at $M_2 \sim 550 \gev$ can a very
small kink due to the change from one \cnn{1}{1}{3} to \ccn{1} 
be
observed. As in the 
previous figure, for the sake of clarity, we
also include the ``best'' full results as obtained without any
restrictions on the set of allowed RS, shown as cyan line in the
lower right plot. Only a tiny difference can be observed for $M_2$
between $500$ and $540\gev$, as expected from different renormalization
schemes, resulting in differences at the two-loop level.

\subsection{Variation of two parameters}
\label{sec:2par-var}

\begin{figure}[h!]
\vspace{4em}
  \begin{center}
  \includegraphics[width=0.48\textwidth]{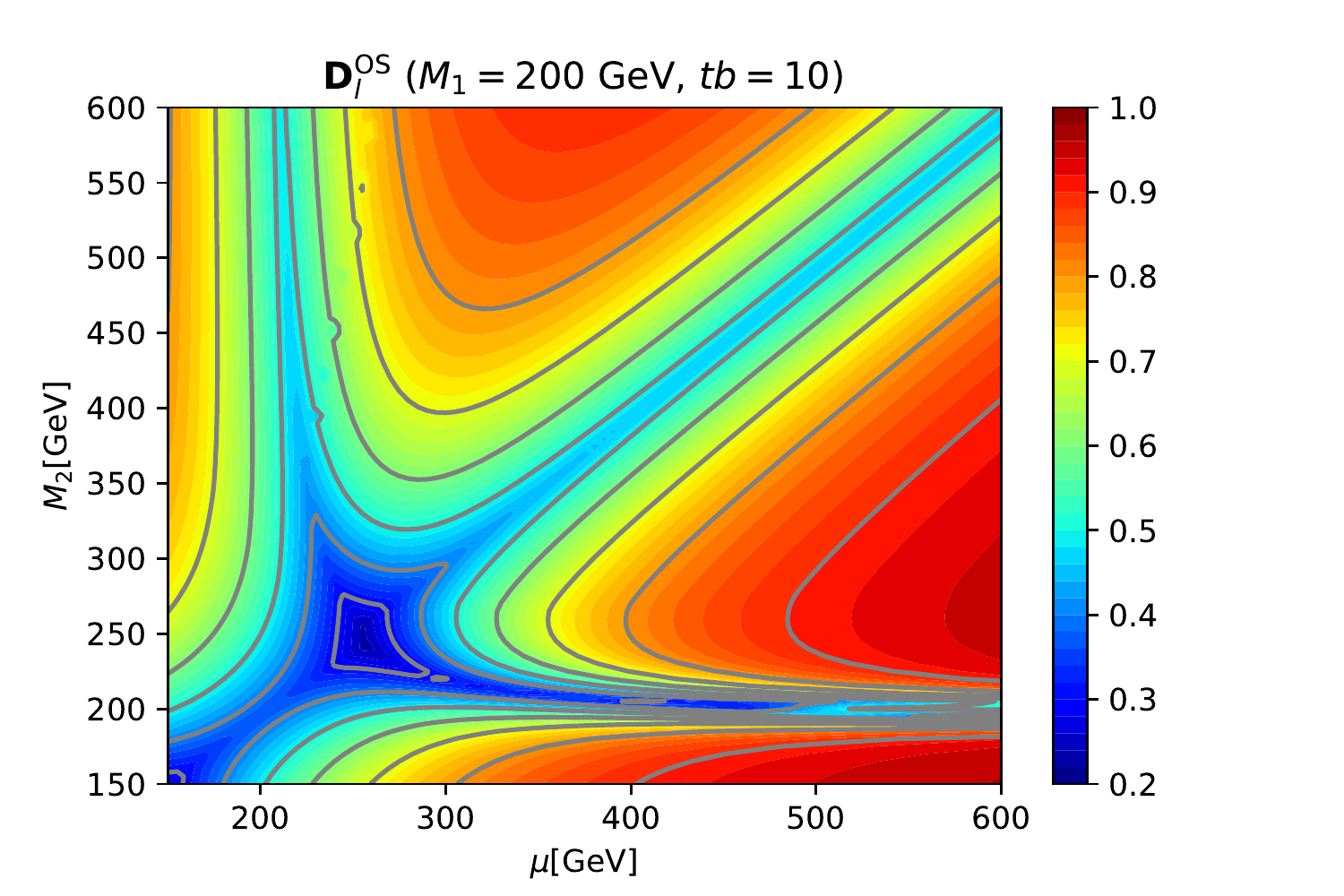}
  \includegraphics[width=0.48\textwidth]{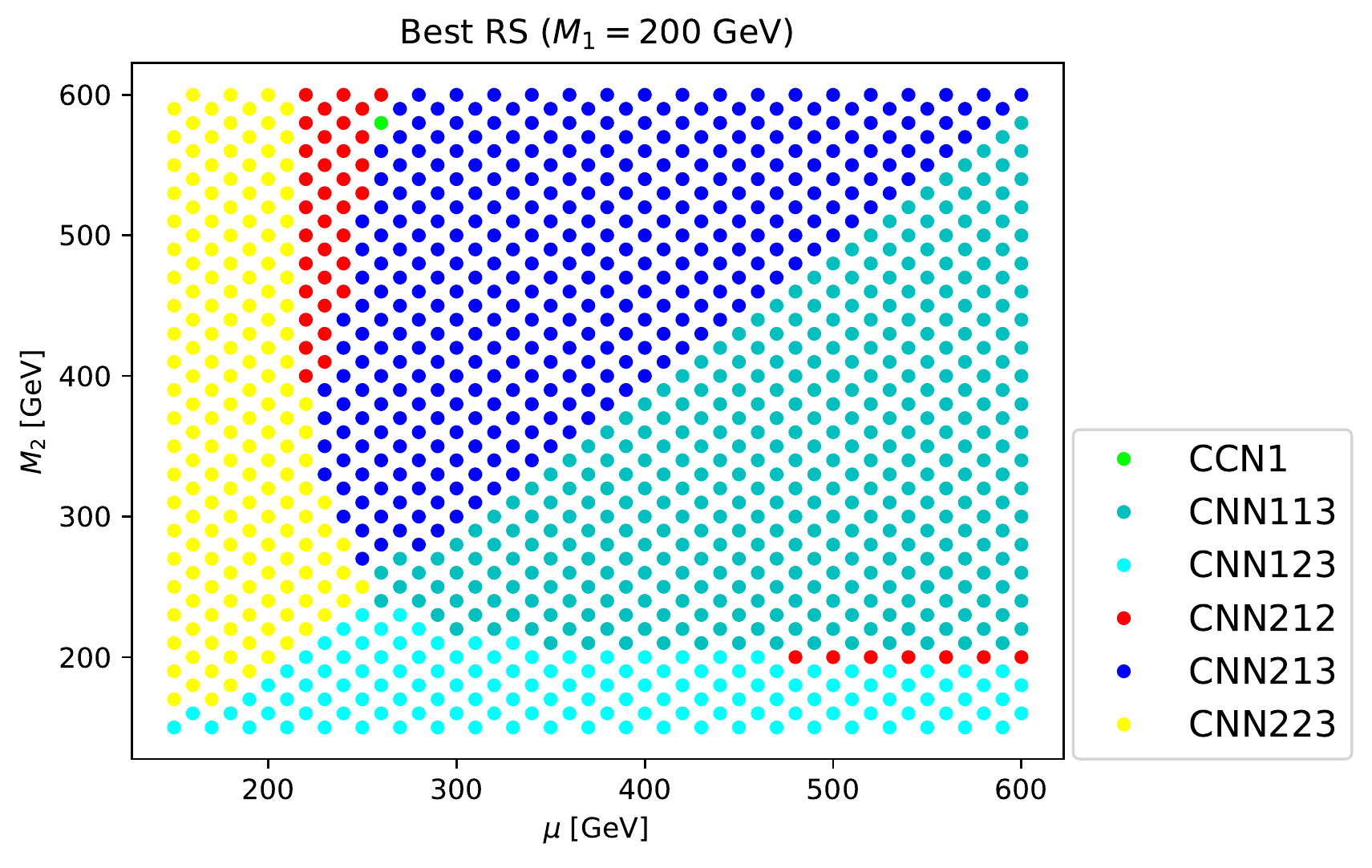}\\[2em]
  \includegraphics[width=0.48\textwidth]{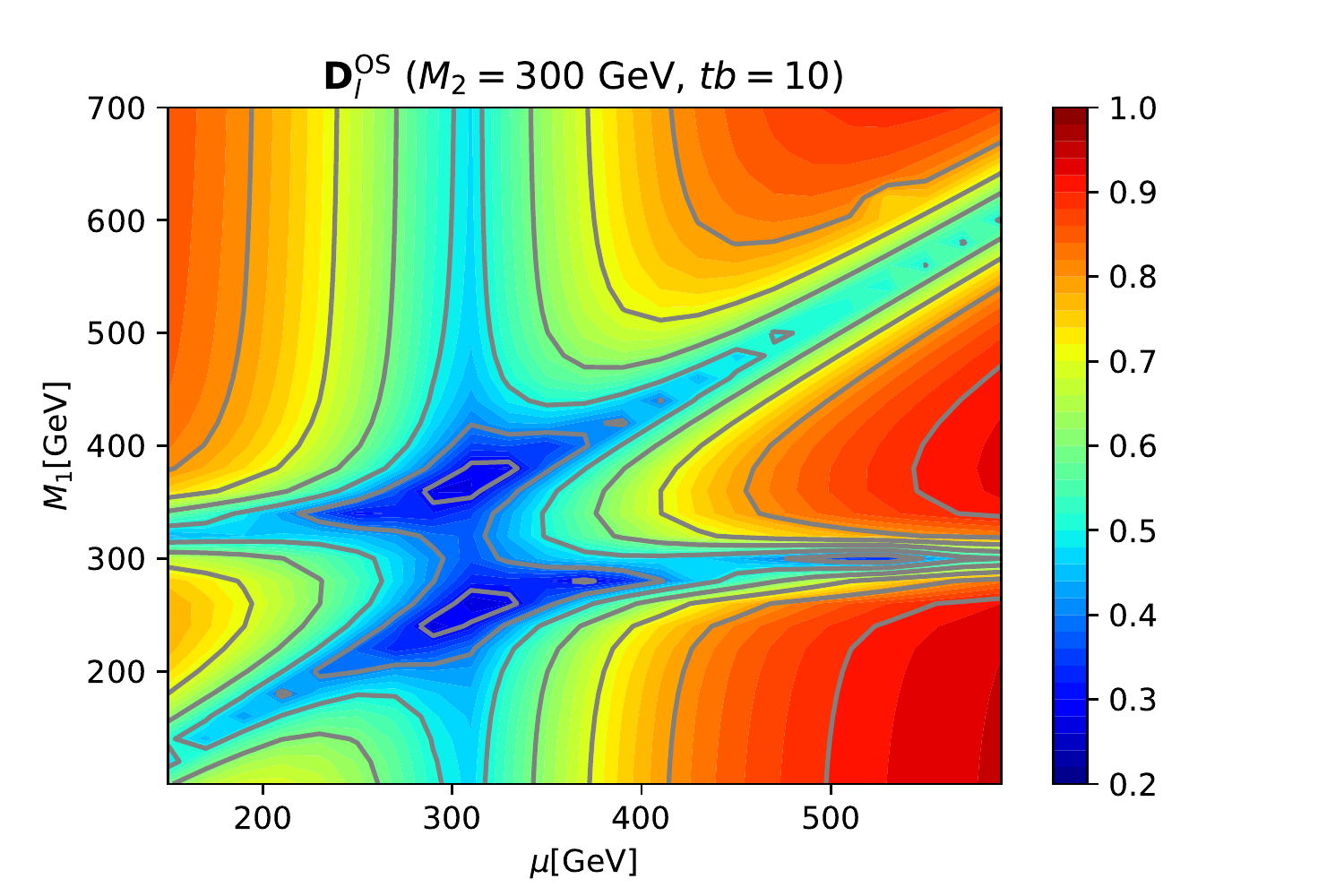}
  \includegraphics[width=0.48\textwidth]{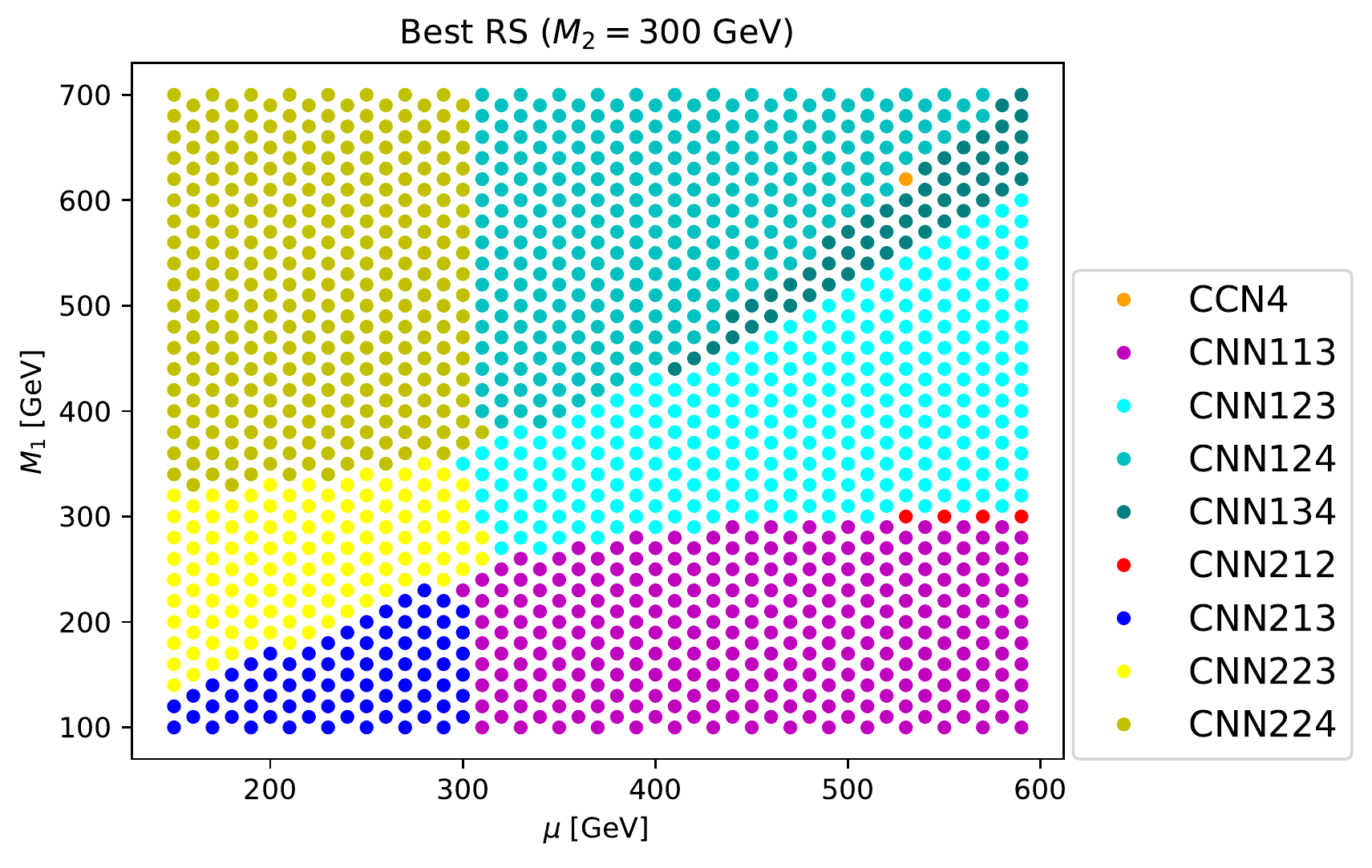}\\[2em]
  \includegraphics[width=0.48\textwidth]{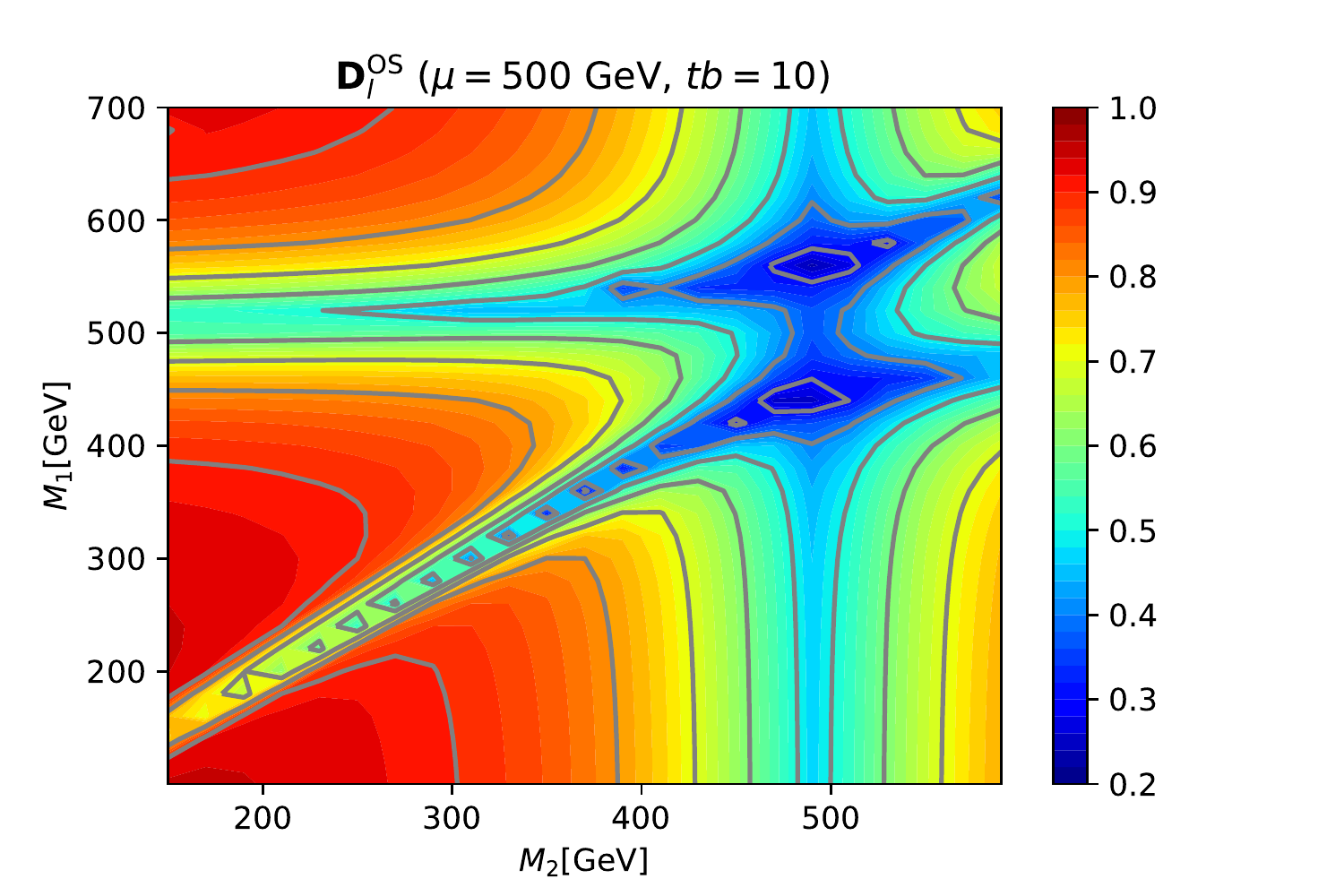}
  \includegraphics[width=0.48\textwidth]{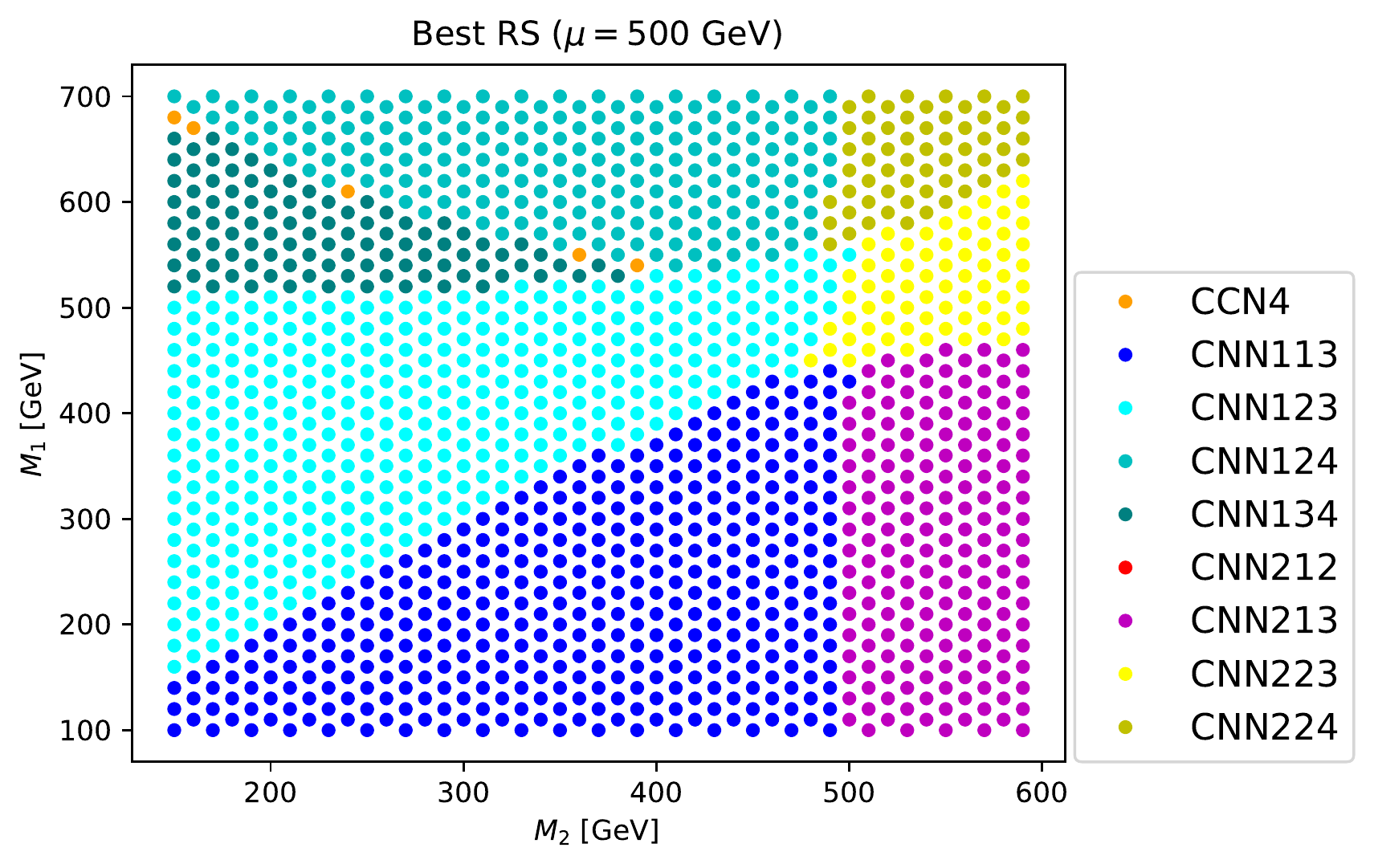}
\vspace{2em}
  \caption{
    The normalized determinant
    $\matr{D}^{\OS}_l $, 
    see \refeq{eq:matrAOS},
    for the best RS (left) and the corresponding RS (right) in: 
    the $\mu - M_2$ plane for $M_1=200 \gev$ (top),
    the $\mu - M_1$ plane for $M_2=300 \gev$ (middle), k
    and in the $M_2 - M_1$ plane for $\mu=500 \gev$ (bottom).
    In all scenarios $\tb=10$.
}
\label{fig:2022_C2N1W_scanM1}
\end{center}
\end{figure}

We finish our analysis of the automated RS choice with the results for
2-dimensional planes. In the previous subsections we demonstrated that
our proposed algorithm is capable of selecting a good RS, resulting in a
smooth result over the analyzed one-dimensional parameter interval. One
might suspect that simple cuts or rules could yield a similarly good
result. The investigation of the 2-dimensional planes demonstrates that
this is not the case. In \reffi{fig:2022_C2N1W_scanM1} we present
the normalized determinant $\matr{D}^{\OS}_l $, 
see \refeq{eq:matrAOS},
or the best RS (left) and the corresponding RS (right) in: 
the $\mu - M_2$ plane for $M_1=200 \gev$ (top),
the $\mu - M_1$ plane for $M_2=300 \gev$ (middle), 
and in the $M_2 - M_1$ plane for $\mu=500 \gev$ (bottom), all for $\tb = 10$.
In all scenarios $\tb=10$. The color coding indicates the size of the
maximum normalized determinant (left) and the choice of the chosen RS
(right). In the left column one can observe that even in the most
``critical'' situations when all three mass parameters are of similar
size, the largest determinant does not go below $\sim 0.23$.
For two mass parameters close in range, the determinant only goes down
to about $\sim 0.4$ for $\mu \sim M_2$ and about $\sim 0.3$ for
$M_1 \sim M_2$. In the right column of
\reffi{fig:2022_C2N1W_scanM1}, where we indicate the selected RS, one
can observe that a multitude of schemes is chosen all over the planes. No
simple rules or cuts can reproduce the pattern of the selected RS. This
demonstrates the power of our new proposed algorithm, which cannot be
reproduced by simple selection cuts.


\newpage

\section{Conclusions}
\label{sec:conclusions}

The explorations of models beyond the Standard Model (BSM) naturally
involve scans over the BSM parameters. On the other hand, high precision
predictions require calculations at the loop-level and thus a
renormalization of (some of) the BSM parameters. Often many choices 
for the renormalization scheme (RS) are possible. This concerns the choice
of the set of to-be-renormalized parameters out of a larger set of BSM
parameters, but can also concern the type of renormalization condition
which is chosen for a specific parameter. A given RS can be well suited
to yield ``stable'' and ``well behaved'' higher-order corrections in one
part of the BSM parameter space, but can fail completely in other
parts. Such a failure may not even be noticed numerically if an isolated
parameter point (e.g.\ in a large scan) is investigated. Consequently,
the exploration of BSM models requires a choice of 
a good RS {\em before} the higher-order calculation of the physical
  observable is performed (counterterm calculations in various schemes,
  however, will be required).

In this paper we proposed 
a new method with which 
such a situation can be
avoided, i.e.\ how a ``good'' RS can be chosen {\em before} performing
the calculation. This new method is based on
the properties of the transformation matrix that connects the various
mass counterterms with the parameter counterterms.
The ``best RS'' is chosen as the one that (for the parameter point under
investigation) possesses the largest (normalized) determinant of the
transformation matrix, a quantity that can be evaluated for all RS under
consideration before the actual higher-order calculation is performed. 
This allows a
point-by-point test of all ``available'' or ``possible'' RS, and the
``best'' one can be chosen to perform the calculation.

Our idea is designed to work in all cases of RS choices (in BSM
models). In order to demonstrate its feasibility and power, we
concentrated on a 
more specific question: in many BSM models one can be faced with the
situation that one has $m$ underlying Lagrangian parameters
and $n > m$ particles or particle masses that can be renormalized
on-shell (OS). The calculation of the production and/or decay of BSM
particles naturally requires OS renormalizations of the particles
involved. Each choice of $m$ particles renormalized OS defines an
\rs{l}, of which we have $N$ in total. We have demonstrated how out of
these $N$ \rs{l} one can choose the ``best'' \rs{L}.
Starting from \DRbar\ input
parameters we provided a detailed and ``model independent'' 
description of how the
OS parameters can be derived. This can be done in one step, where the
OS masses and parameters are evaluated from the original
\DRbar\ parameters, the ``semi-OS'' scheme. This can also be done in two
steps, where in the second step after the semi-OS scheme the OS masses
are derived from OS parameters, the ``full-OS'' scheme. In our numerical
evaluation we have concentrated on the latter.

The numerical examples have been performed within the MSSM, concretely in
the sector of charginos and neutralinos, the supersymmetric (SUSY)
partners of the SM gauge bosons and the 2HDM-like Higgs sector. 
The sector is controlled by the three mass parameters $M_1$, $M_2$ and
$\mu$, as well as $\tb$. Three out of the six chargino/neutralino masses
can be renormalized OS, whereas the other three masses receive an extra
one-loop shift to yield the correct OS mass. While concentrating on the
chargino/neutralino sector of the MSSM
this constitutes a very specific example, we would like to stress 
our expectation for the general
applicability of our method to many types of BSM models and
types of RS choices. It is clear that a firm statement about the
applicability of our idea in another model and/or sets of RS can only be
made after its concrete implementation, which goes far beyond the scope
of this article. However, we find it conceivable that the method of a vanishing
determinant {\em always} indicates the failure of an RS, thus giving
rise to our expectation of much more general applications.

We have investigated eight mass hierarchies, in which two of the three
mass parameters were fixed,
and the third one changed continuously from very small to very large
values (with $\tb = 10$). This ensures that all mass hierarchies the
model offers are covered. We have concentrated on the RS renormalizing
either one chargino (\cnn{i}{j}{k}) or two charginos (\ccn{i}) OS and
left out the possibility of normalizing three neutralinos OS
(\nnn{i}{j}{k}). This yielded in total 16 RS that can potentially be
chosen. As physical process, we mostly concentrated on the decay width
$\Ga(\chap2 \to \neu1 W^+)$. 

For each scenario we presented five plots.
The first one showed the analysis of the determinants of the
transformation matrices, leading to the choice of ``the best RS'' before
performing any actual calculation.
Next we presented the tree results for all RS that are chosen in at
least one parameter point in the scan, resulting in four to five RS for
each benchmark scenario. The results for all RS are shown for the full
parameter region for comparison (with one fixed color for each chosen
RS), and we indicated which RS is best for each parameter point. 
One can observe that where a scheme is chosen, the tree level width
behaves ``well'' and smoothly.
On the other hand, outside the selected
interval the tree-level result behaves highly irregularly, induced by
the shifts in the mass matrices to obtain OS masses. 
Next, we presented a plot showing the ``loop plus real photon emission''
results with the same color coding as for the tree result.
One can observe that where a scheme is chosen the
loop corrections behave smoothly and the overall size stays at the level
of $\sim 10\%$ or less compared to the tree-level result (exceptions are
found only where the tree result goes to zero due to an accidental
zero crossing of the involved tree-level coupling). As before, 
outside the chosen interval the loop corrections take irregular values,
which sometimes even diverge, owing to a vanishing determinant.
The next plot presented for each benchmark scenario showed the sum of
tree and higher-order corrections, i.e.\ of the two previous plots. The
same pattern of numerical behavior can be observed. The chosen scheme
yields a reliable higher-order corrected result, whereas other schemes
result in highly irregular and clearly unreliable results. This is
summarized in the final plot, where we show the selected tree-level
result as dashed line, the loop result as dotted (multiplied
by~10 for better visibility), and the full result as  
solid line. The overall behavior is completely well-behaved and
smooth.
A remarkable feature can be observed in some cases, where
the selected tree-level result has a kink, because of a
change in the shift in the OS values of the involved chargino/neutralino
masses, caused by the change from switching one selected RS to another.
In this case, the loop corrections contain also a
corresponding kink, leading to a completely smooth full one-loop result.

In a final step we have 
presented an analysis of the automated RS choice with the results for
2-dimensional planes. Looking only at the 1-dimensional analysis, one
might suspect that simple cuts or rules could yield a similarly good
result. The investigation of the 2-dimensional planes demonstrated that
this is not the case. For three planes ($\mu - M_2$, $\mu - M_1$ and
$M_2 - M_1$) we show the value of the determinant of the chosen RS,
which in effectively all cases analyzed remains above $\sim 0.3$. We
furthermore showed in a separate plot the RS chosen for each point in
the parameter plane. One 
can observe that a multitude of schemes is chosen all over the planes. No
simple rules or cuts can reproduce the pattern of the selected RS. This
demonstrates the power of our new proposed algorithm, which cannot be
reproduced by simple selection cuts. 

\medskip
As next steps, we plan to apply our method to more involved calculations
such as $2 \to 2$ processes, as well as to Dark Matter related
calculations in parameter regions with small mass splittings between
various charginos/neutralinos.
We will also apply the new method to the other sectors of the
MSSM, where the renormalization can be even more involved than in
the chargino/neutralino sector (see
\citeres{Gluino-decay,Stau-decay,Hdec-sferm,Hdec-chaneu,eeHiggs,eeHpm,eeSlep}).
Furthermore, 
we intend to make public our modified versions of \FA/\FC\ 
which have been used to obtain the results presented here.
Moreover, we plan to apply our
method to different types of renormalization schemes in other (types of)
models. On the other hand, extending our idea to the two-loop level
  or beyond, we leave for now as an open question to be investigated in
  the future.
Only a method to select a reliable RS before the calculation is performed
will allow a fully automated set-up of BSM higher-order calculations.
We hope that our newly proposed method will enable such fully 
automated set-ups for BSM higher-order calculations.


\subsection*{Acknowledgments}

We thank C.~Schappacher for helpful discussions and
in particular for invaluable help with the chargino mass shifts
concerning the IR divergences.
The work of S.H.\ has received financial support from the
grant PID2019-110058GB-C21 funded by
MCIN/AEI/10.13039/501100011033 and by ``ERDF A way of making Europe".
MEINCOP Spain under contract PID2019-110058GB-C21 and in part by
by the grant IFT Centro de Excelencia Severo Ochoa CEX2020-001007-S
funded by MCIN/AEI/10.13039/ 501100011033.


\pagebreak
\clearpage


\begin{appendix}
\noindent{\Large\bf Appendix}
\setcounter{equation}{0}
\renewcommand{\thesubsection}{\Alph{section}.\arabic{subsection}}
\renewcommand{\theequation}{\Alph{section}.\arabic{equation}}

\setcounter{equation}{0}


\section{Transformation matrices for the counterterm}
\label{sec:matrA}

In order to give an explicit form for the transformation matrix 
$\matr{A}_l^{\DRbar}$, \refeq{ADRbar}, 
we order the model parameters as
\begin{align}
& P_1^{\DRbar} = M_1^{\DRbar},\;
P_2^{\DRbar} = M_2^{\DRbar},\;
P_3^{\DRbar} = \mu^{\DRbar}\,.
\end{align}
For a \ccn{n} RS the transformation matrix $ \matr{A}_l^{\DRbar}$
is given by 
\begin{align}
  \matr{A}^{\DRbar}_{l} &= 
\begin{pmatrix}
  0 & V^*_{11} U^*_{11} & V^*_{12} U^*_{12} \\ 
  0 & V^*_{21} U^*_{21} & V^*_{22} U^*_{22} \\
  (N^*_{n1})^2 & (N^*_{n2})^2 & -2 N^*_{n3} N^*_{n4}
\end{pmatrix}~,
\label{ACCN}
\end{align}
with $l = 1,\ldots,4$ for all possible values of $n$.
For a \cnn{c}{n}{n'} RS the matrix $ \matr{A}_l^{\DRbar}$ is given by
\begin{align}
  \matr{A}^{\DRbar}_{l} &= 
\begin{pmatrix}
  0 & V^*_{c1} U^*_{c1} & V^*_{c2} U^*_{c2} \\ 
  (N^*_{n1})^2 & (N^*_{n2})^2 & -2 N^*_{n3} N^*_{n4}\\
  (N^*_{n'1})^2 & (N^*_{n'2})^2 & -2 N^*_{n'3} N^*_{n'4}
\end{pmatrix}~,
\label{ACNN}
\end{align}
with $l=5,\ldots,16$, corresponding to all possible combinations
of $c$, $n$ and $n'$. 
It is worth noticing that in \refeqs{ACCN} and (\ref{ACNN}) the unitary
matrices $\matr{U}$, $\matr{V}$ and $\matr{N}$ have been obtained
diagonalizing the chargino and neutralino mass matrices given in terms
of the $\DRbar$ parameters. 
In order to obtain $\matr{A}_l^\Os$ the diagonalization matrices
$\matr{U}$, $\matr{V}$ and $\matr{N}$ have to be computed from 
chargino and neutralino mass matrices given in terms of $\Os$ parameters
of the corresponding\,RS.



\section{Additional plots}
\label{sec:plots}

\begin{figure}[h!]
\begin{center}
  \includegraphics[width=0.45\textwidth]{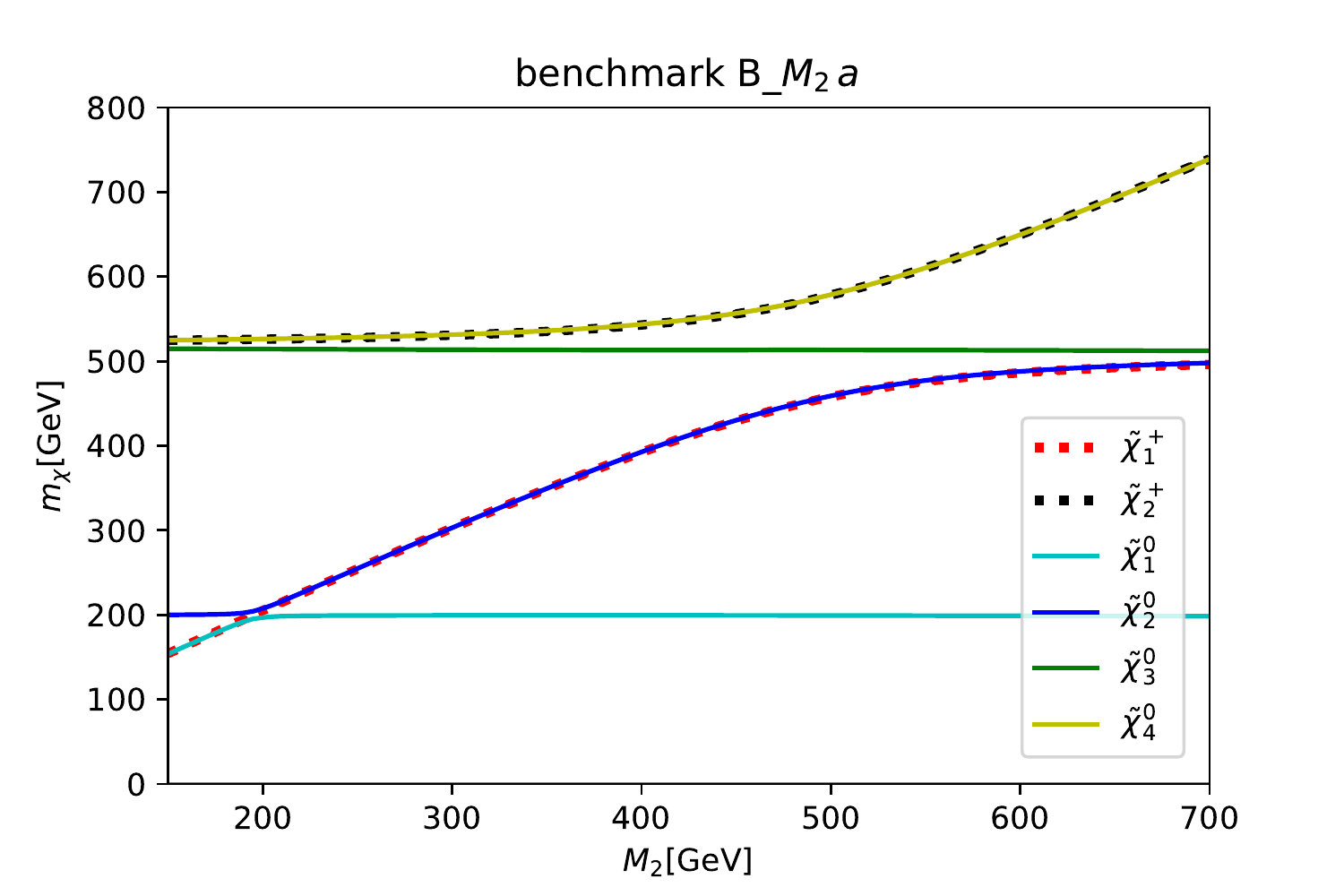}
  \includegraphics[width=0.45\textwidth]{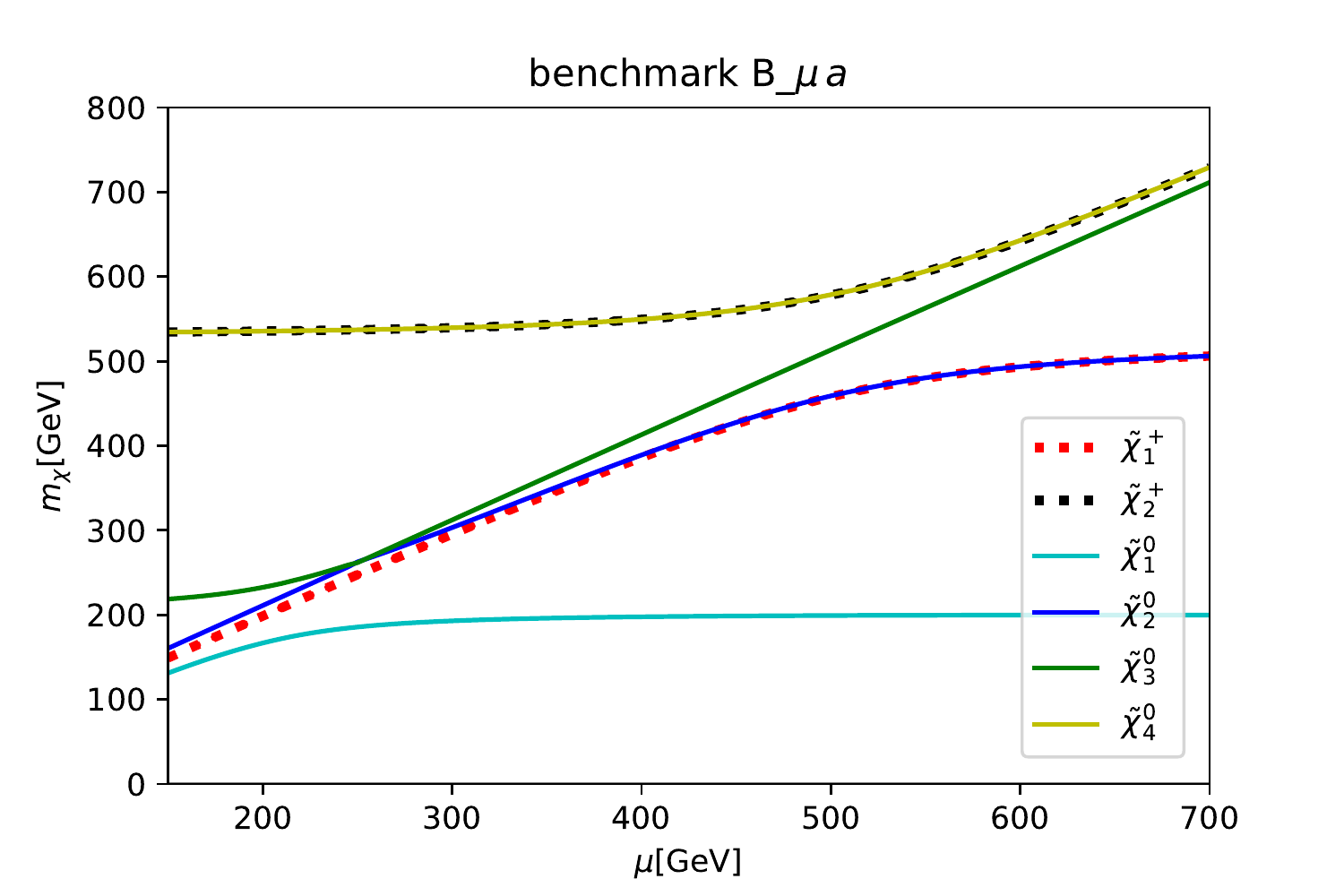}\\[0.5em]
  \includegraphics[width=0.45\textwidth]{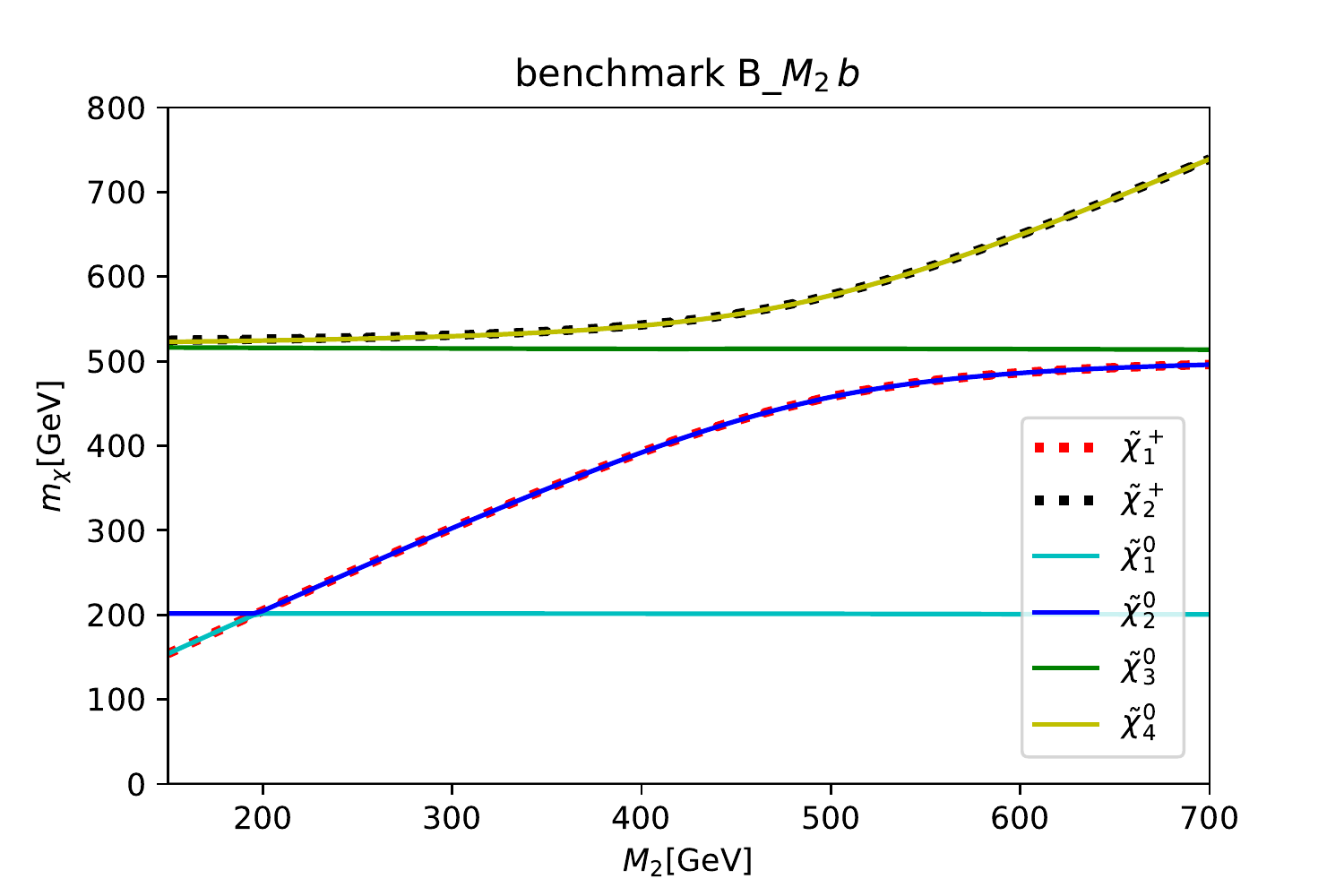}
  \includegraphics[width=0.45\textwidth]{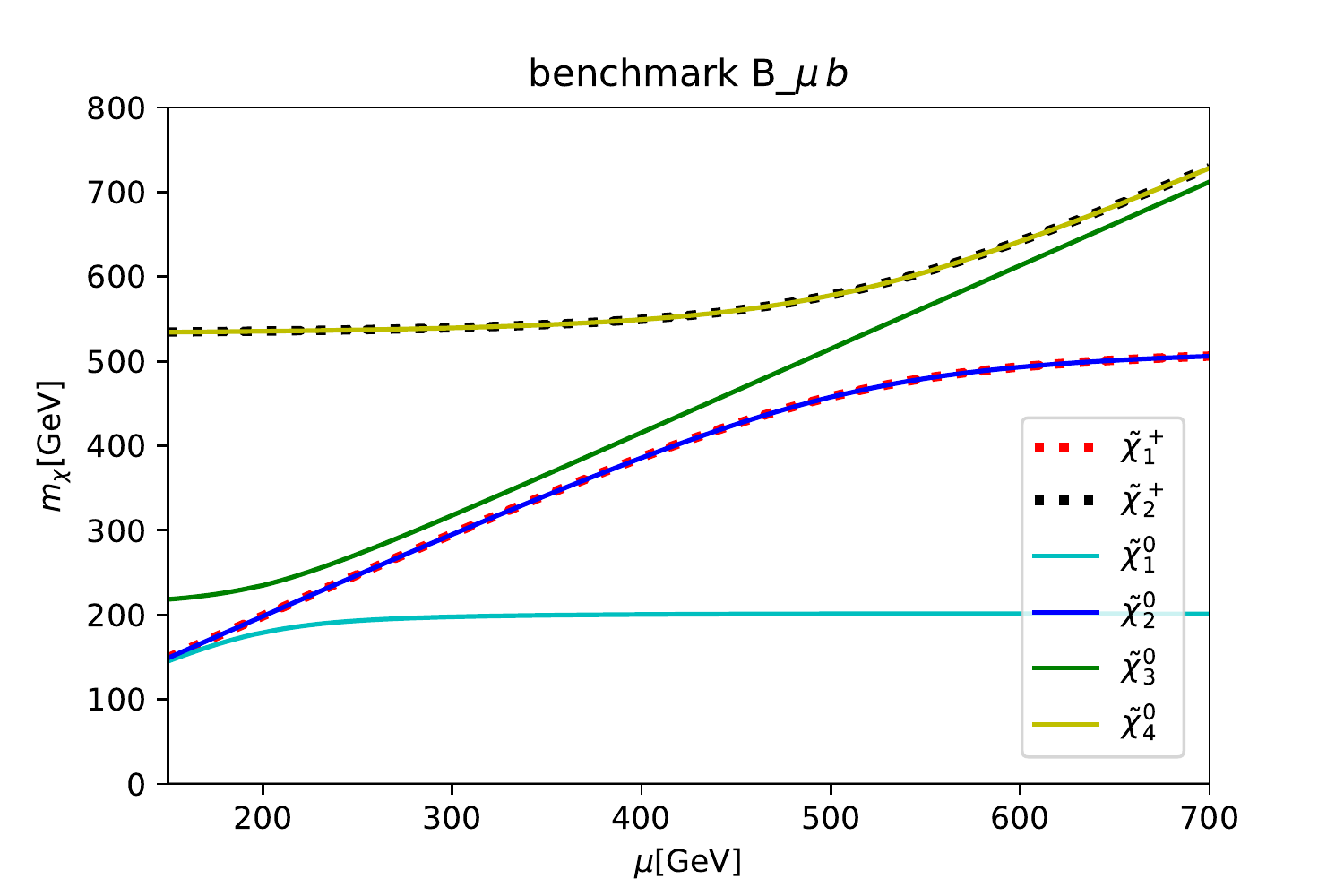}\\[0.5em]
  \includegraphics[width=0.45\textwidth]{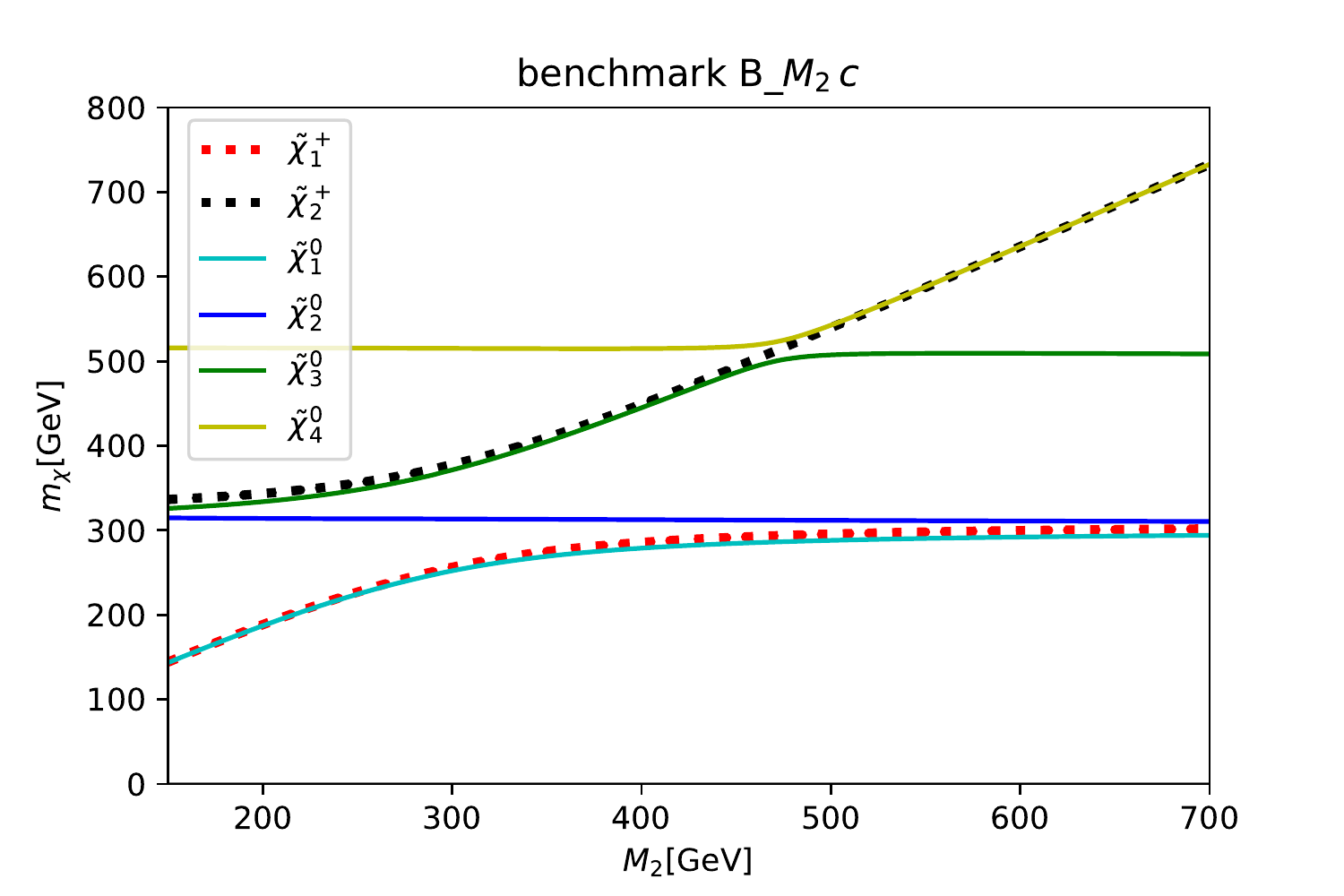}
  \includegraphics[width=0.45\textwidth]{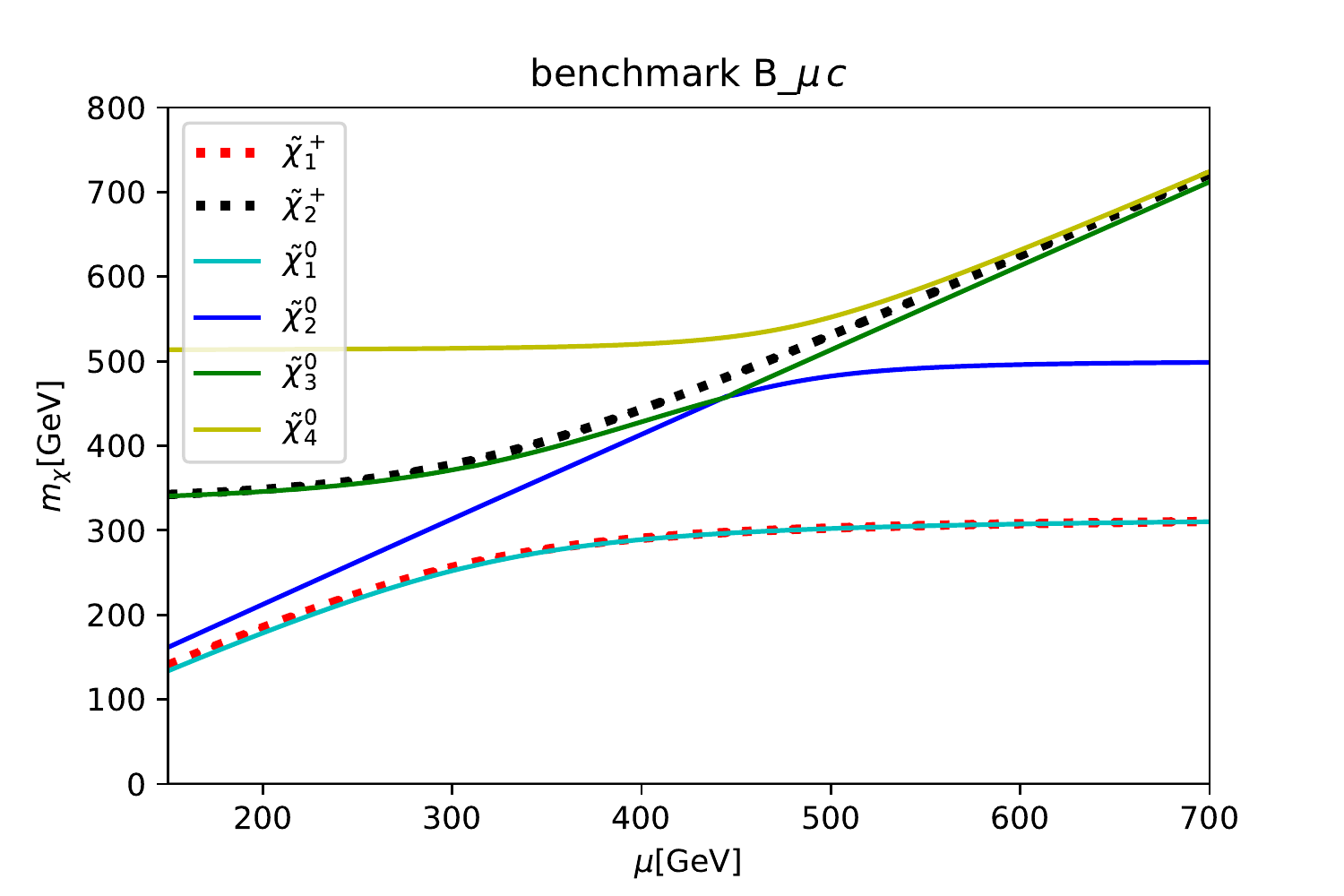}\\[0.5em]
  \includegraphics[width=0.45\textwidth]{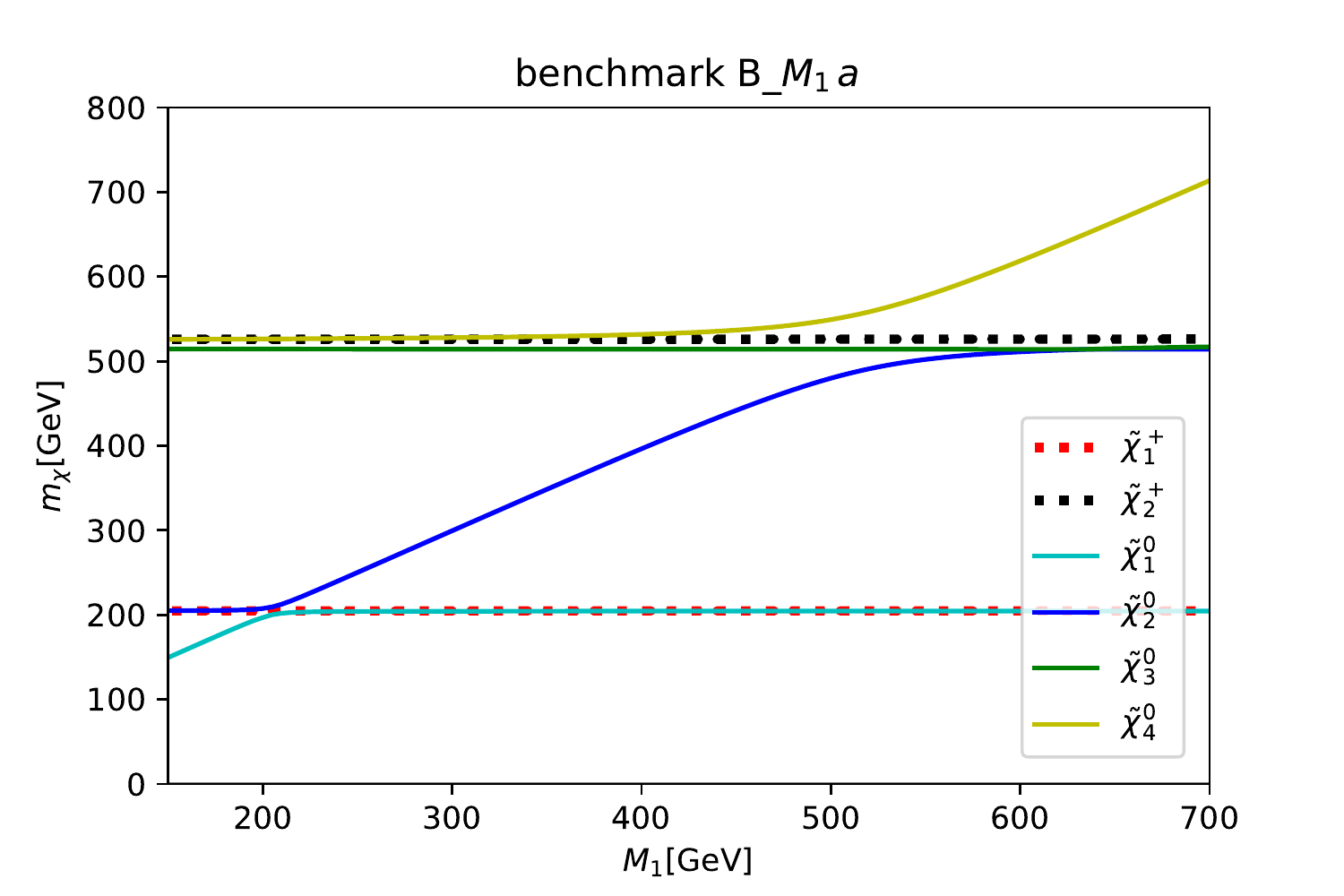}
  \includegraphics[width=0.45\textwidth]{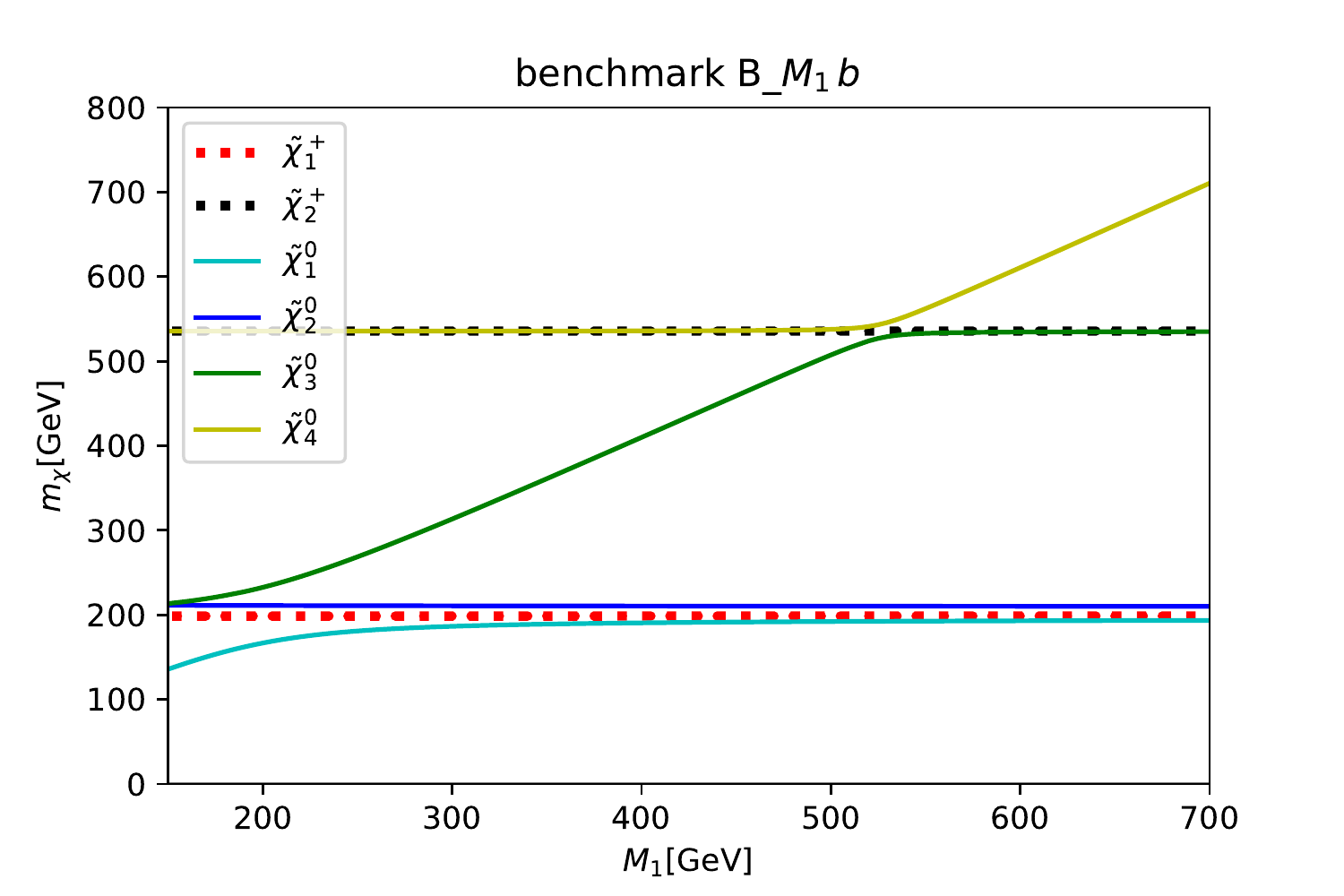}
  \caption{
Masses of charginos and neutralinos in all benchmark scenarios, see
\refta{tab:bench1}. 
  }
\label{fig:chimasses}
\end{center}
\end{figure}

In this appendix we collect additional plots that are not necessary to
follow the main line of argument in the article, but may give some
interesting additional information. In \reffi{fig:chimasses} we present
the masses of the two charginos and the four neutralinos in all eight
benchmark scenarios, as defined in \refta{tab:bench1}. The mass plots,
besides showing the masses themselves, illustrate the two different
types of ``level crossing''. As an example, in the upper left plot,
scenario B\_$M_2\,a$ at $M_2 \sim 200 \gev$ the $\neu1$ and $\neu2$ have
the same intrinsic $\cp$ parities, therefore mix with each other,
and thus experience a ``smooth'' level crossing, as can be observed by
the smooth dependence of the masses. On the other hand, in the second
row left, scenario B\_$M_2\,b$ the two neutralinos have different $\cp$
quantum numbers and do not mix. Consequently, they exhibit a ``sharp''
crossing of the two masses at $M_2 \sim 200 \gev$, see also the
discussion in \refse{sec:m2-var}.

\begin{figure}[h!]
\vspace{2em}
\begin{center}
  \includegraphics[width=0.55\textwidth]{PlotsFinal/Det_M2a.pdf}\\[1em]
  \includegraphics[width=0.45\textwidth]{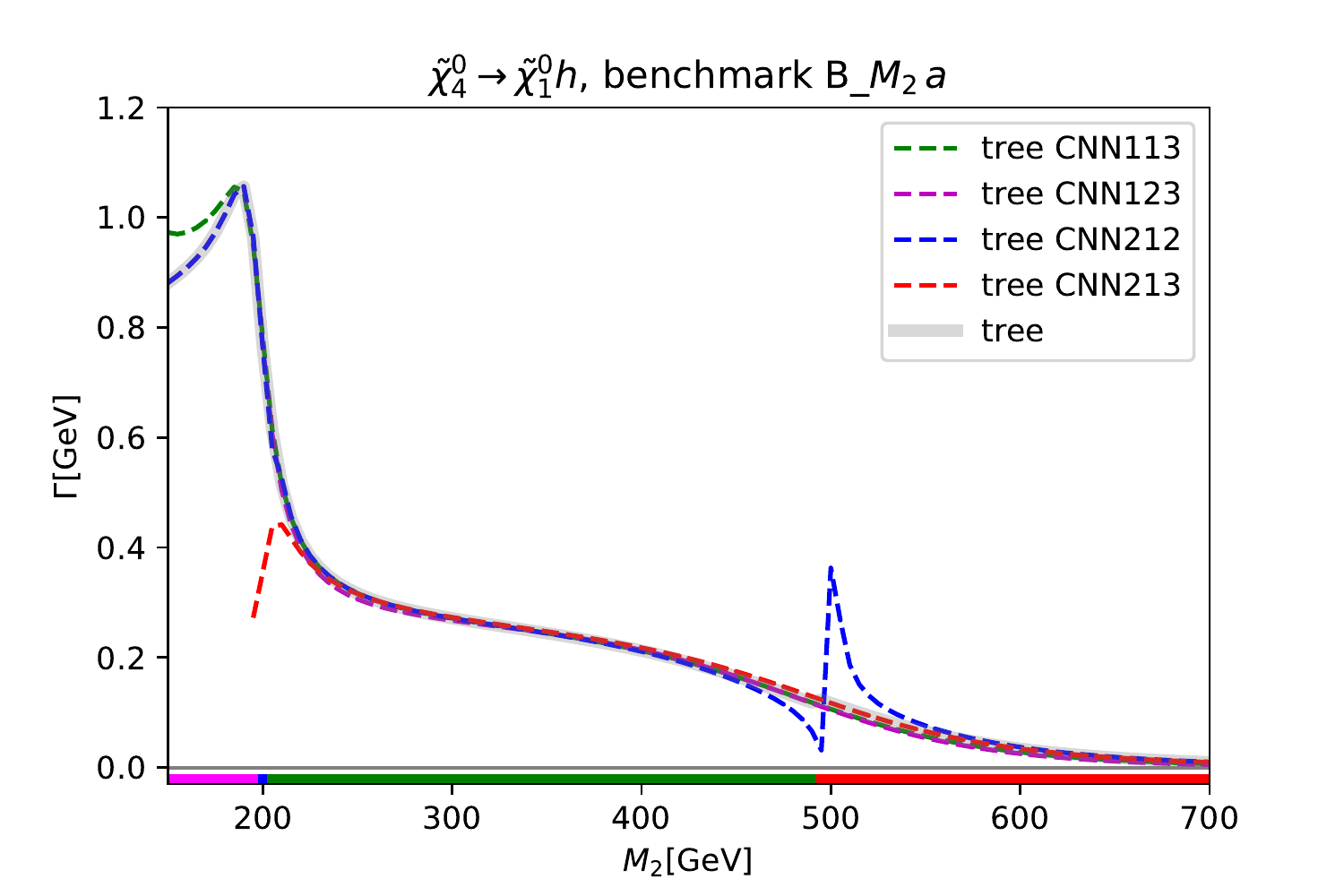}
  \includegraphics[width=0.45\textwidth]{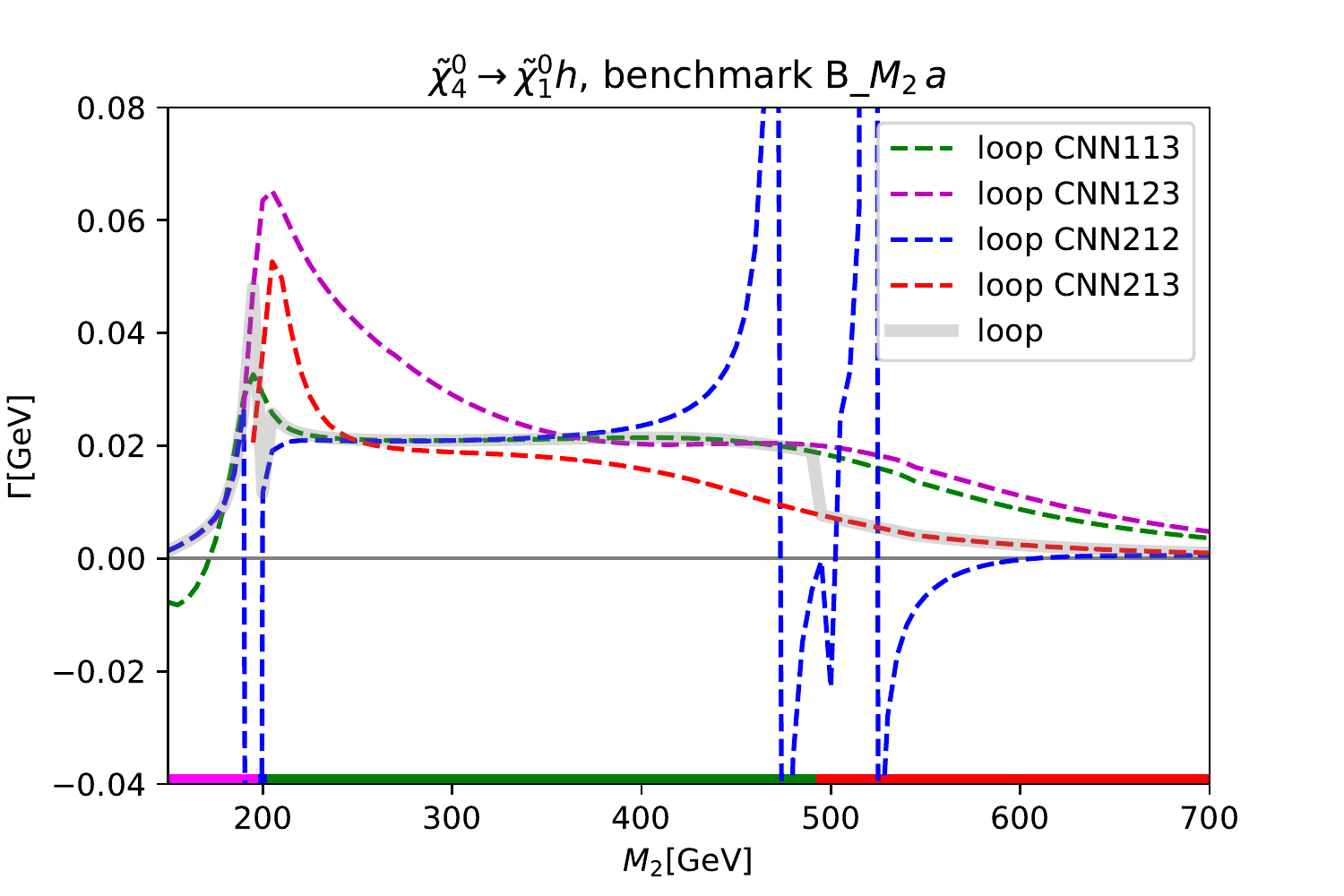}\\[1em]
  \includegraphics[width=0.45\textwidth]{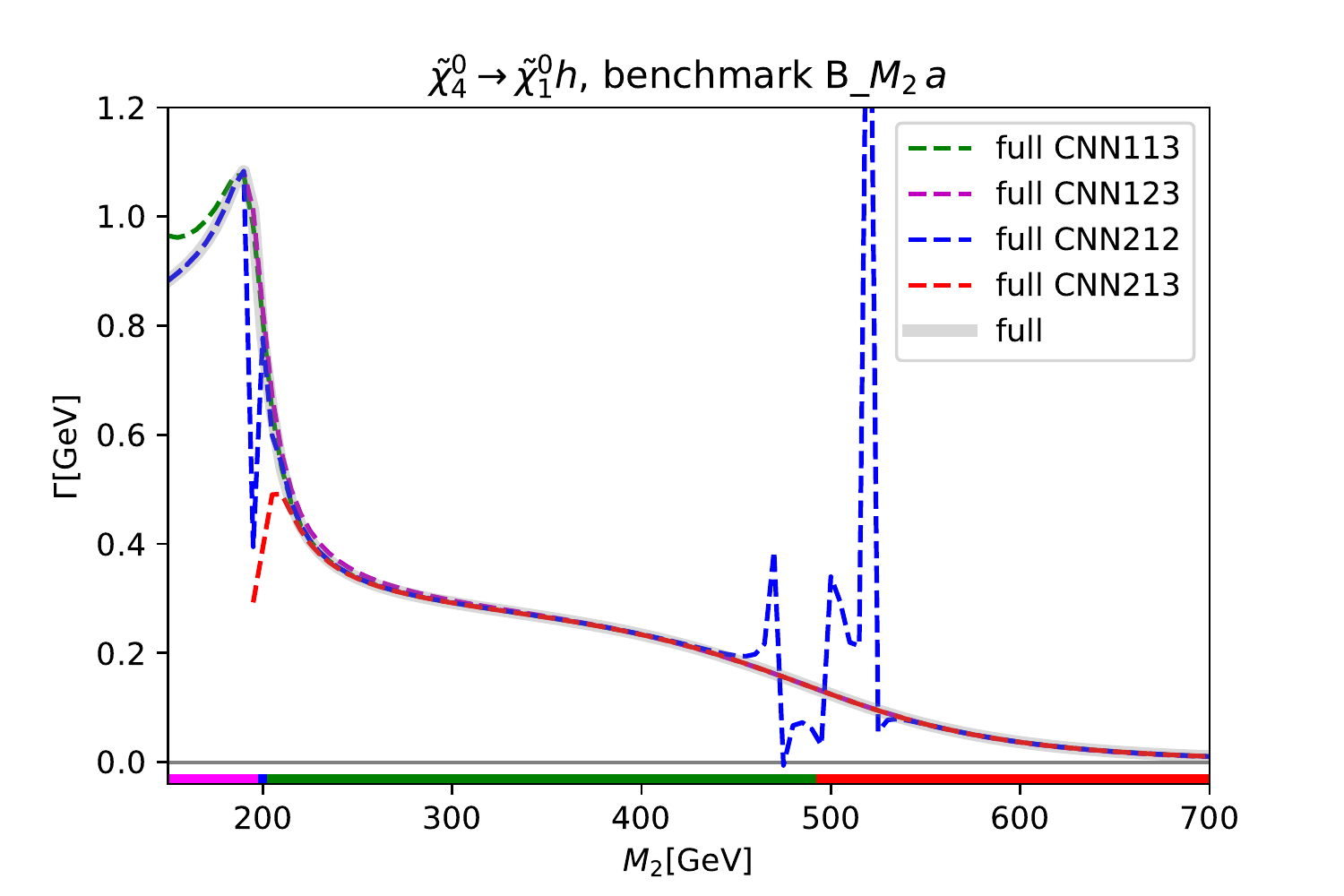}
  \includegraphics[width=0.45\textwidth]{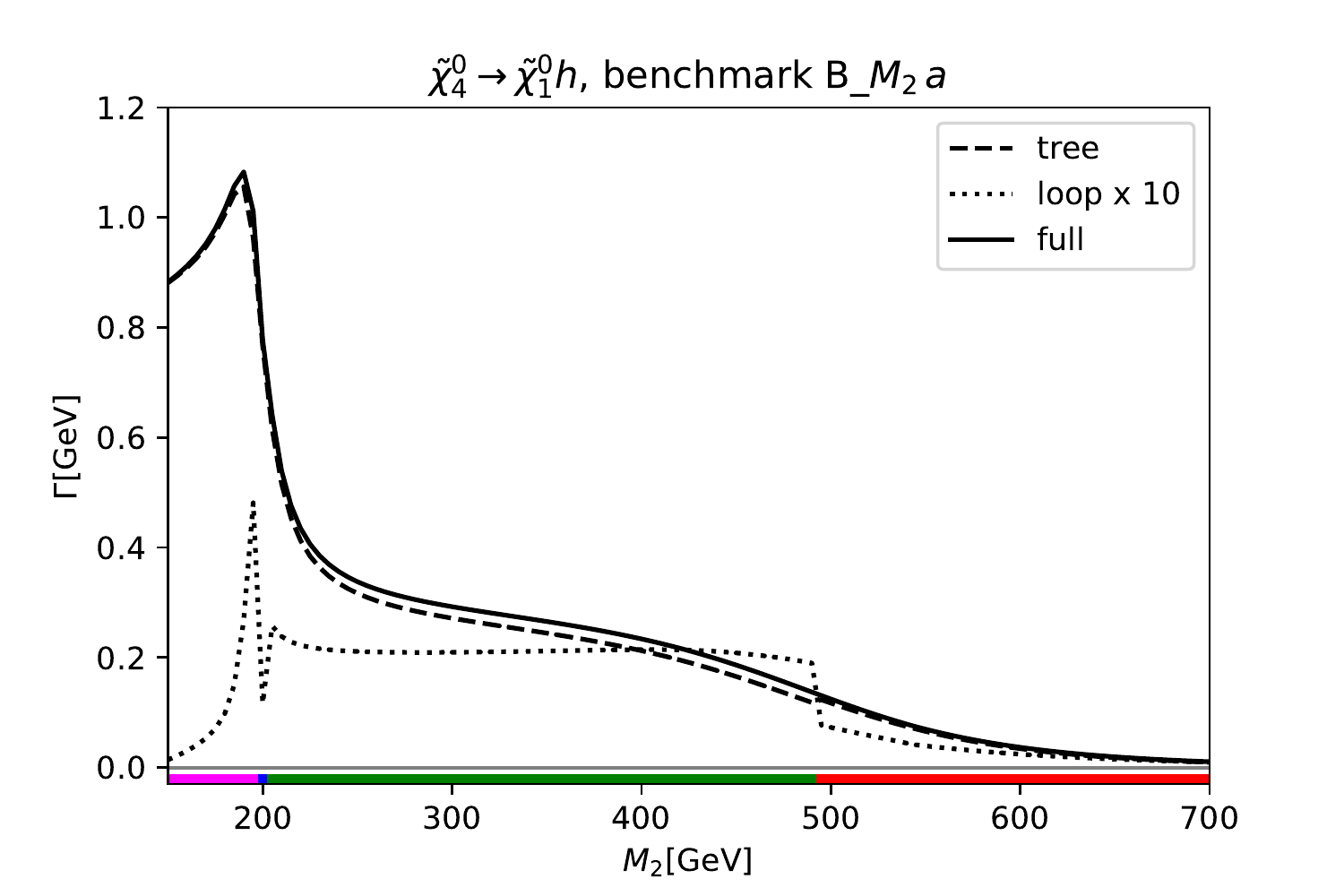}\\[1em]
  \caption{
    Decay width for  $\neu{4} \to \neu{1} h$
    as a function of $M_2$ in benchmark scenario B\_$M_2\,a$, 
    see \refta{tab:bench1},
    with $M_1= 200\gev$, $\mu=500\gev$, $\tb= 10$. 
    Shown are the four ``best RS'' in this range of parameters (see text).
    The plots show the same quantities as in
    \protect\reffi{fig:2022_C2N1W_mu_a}. 
   The horizontal colored bar shows the best RS for the corresponding
   value of  $M_2$,  
   following  the same color coding as the curves: 
   CNN$_{123}$ for $M_2\le 195\gev$, 
   CNN$_{212}$ for $200 \gev \le M_2\le 200\gev$, 
   CNN$_{113}$ for $205 \gev \le M_2\le 490\gev$, 
   CNN$_{213}$ for $495\gev\le M_2$. 
}
\label{fig:2022_N4N1H1_M2a}
\end{center}
\vspace{2em}
\end{figure}

\begin{figure}[h!]
\vspace{2em}
\begin{center}
  \includegraphics[width=0.55\textwidth]{PlotsFinal/Det_M2a.pdf}\\[1em]
  \includegraphics[width=0.45\textwidth]{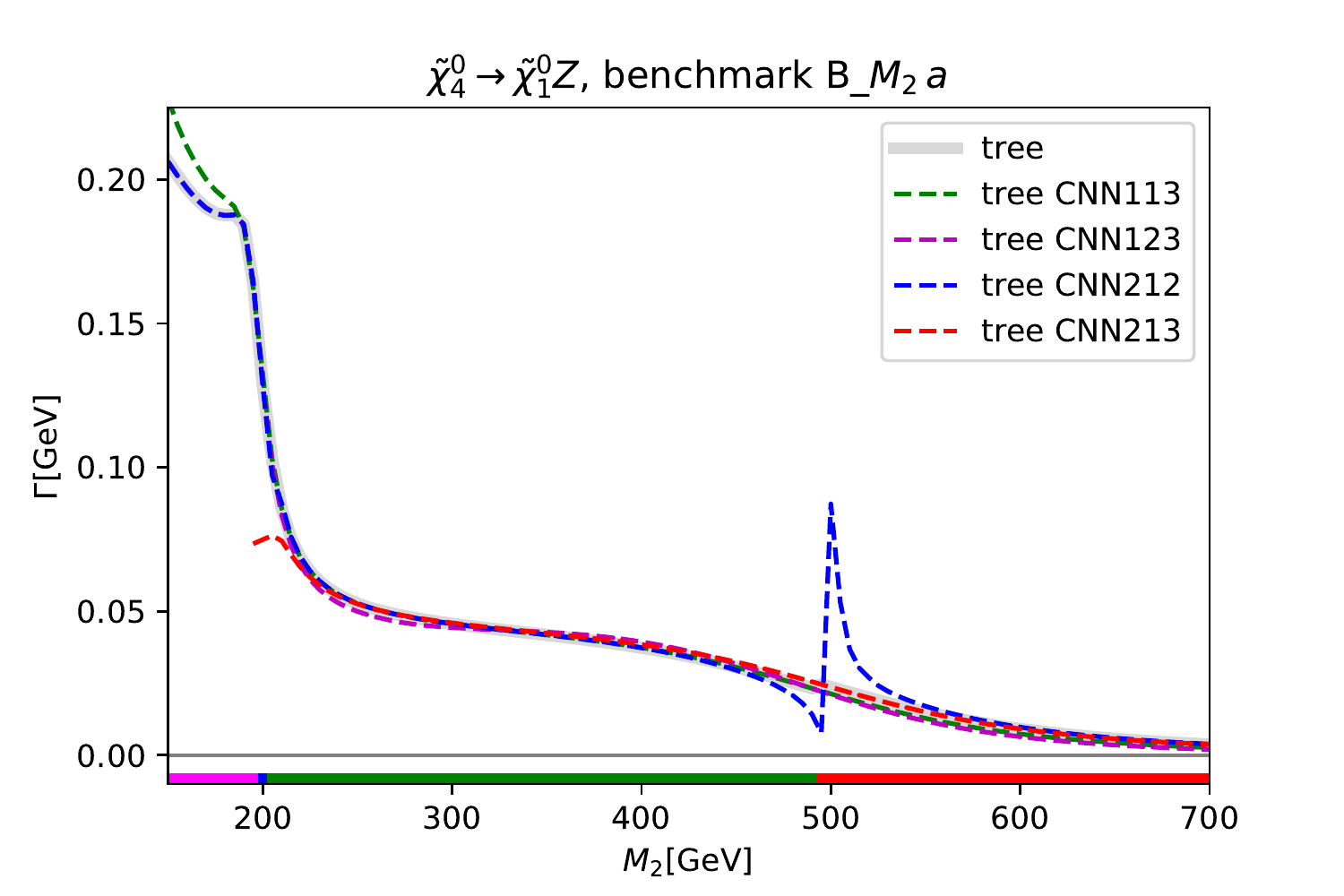}
  \includegraphics[width=0.45\textwidth]{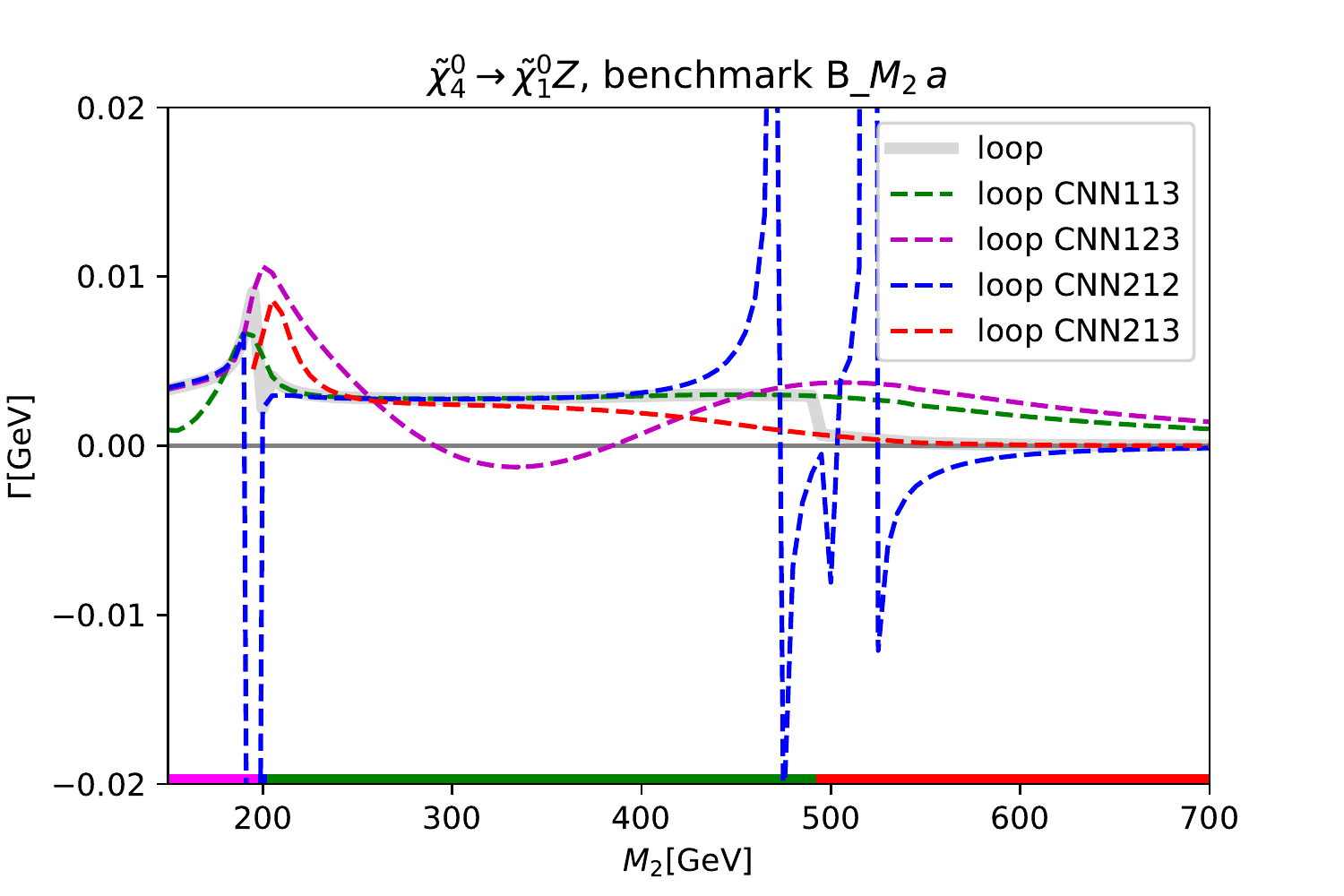}\\[1em]
  \includegraphics[width=0.45\textwidth]{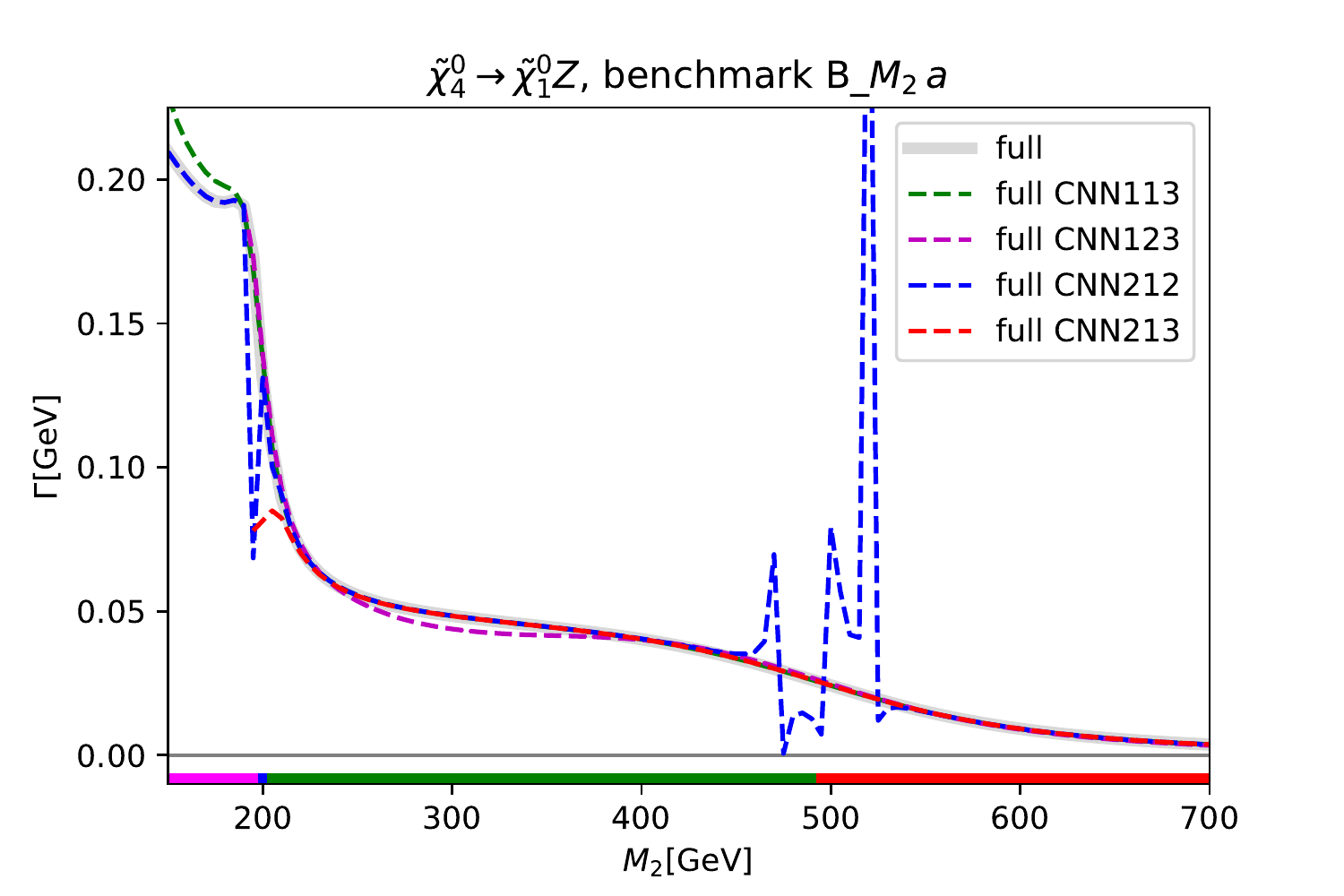}
  \includegraphics[width=0.45\textwidth]{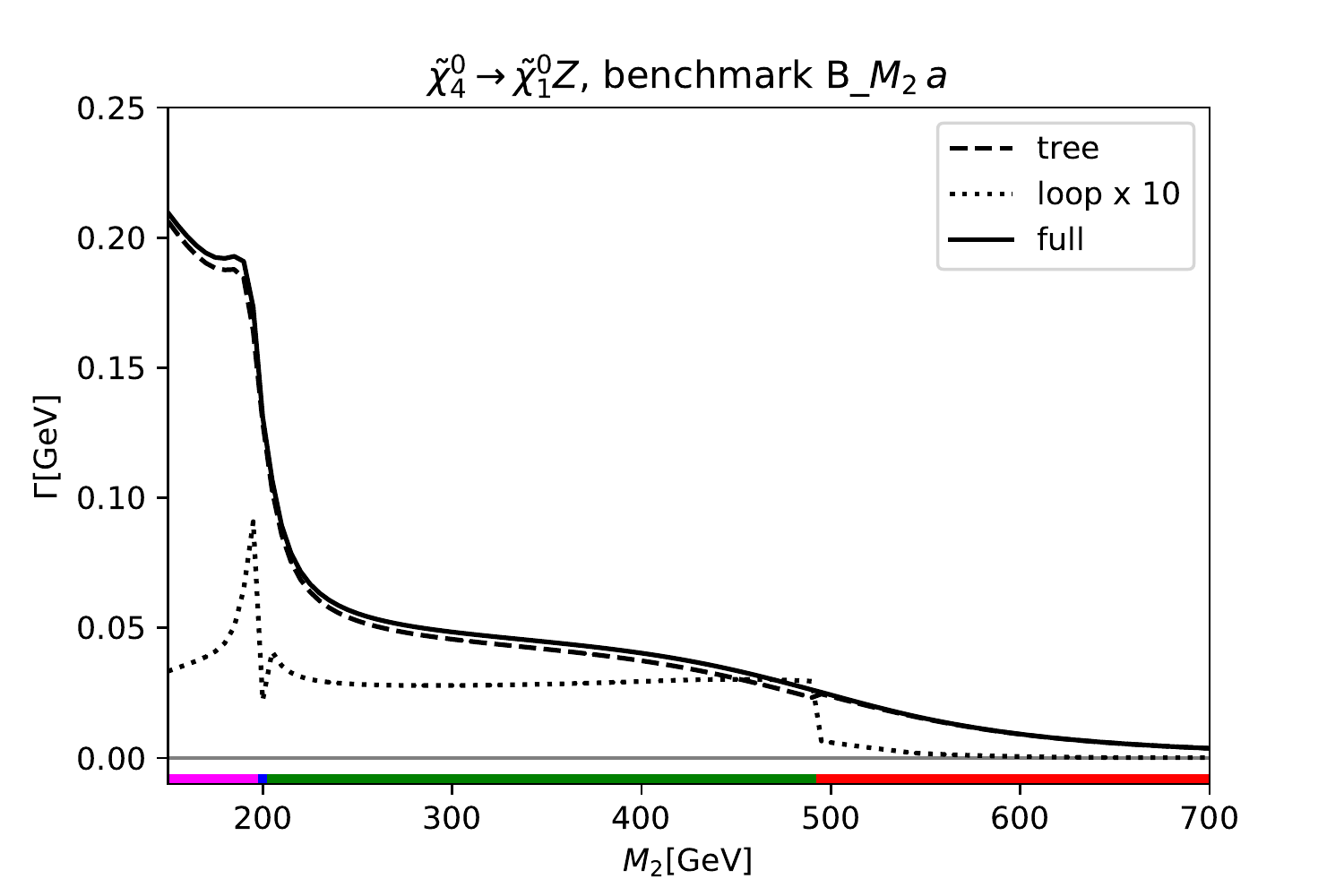}\\[1em]
  \caption{
    Decay width for  $\neu{4} \to \neu{1} Z$   
    as a function of $M_2$ in benchmark scenario B\_$M_2\,a$, 
    see \refta{tab:bench1},
    with $M_1= 200\gev$, $\mu=500\gev$, $\tb= 10$. 
    Shown are the four ``best RS'' in this range of parameters (see text).
    The plots show the same quantities as in
    \protect\reffi{fig:2022_C2N1W_mu_a}. 
   The horizontal colored bar shows the best RS for the corresponding
   value of  $M_2$,  
   following  the same color coding as the curves: 
   CNN$_{123}$ for $M_2\le 195\gev$, 
   CNN$_{212}$ for $200 \gev \le M_2\le 200\gev$, 
   CNN$_{113}$ for $205 \gev \le M_2\le 490\gev$, 
   CNN$_{213}$ for $495\gev\le M_2$.
  }
\label{fig:2022_N4N1Z_M2a}
\end{center}
\vspace{2em}
\end{figure}

In \reffis{fig:2022_N4N1H1_M2a} and \ref{fig:2022_N4N1Z_M2a} we show
``neutral decays'' involving either the $\sim 125 \gev$ Higgs boson,
$h$, or the $Z$~boson, $\neu4 \to \neu1 h/Z$. For these figures the
benchmark scenario B\_$M_2\,a$ has been chosen (and thus the top plots are
identical to the of \reffi{fig:2022_C2N1W_M2a}, but shown for
completeness). The arrangement of
plots is as in the main text, see the description in
\reffi{fig:2022_C2N1W_mu_a}. These two sets of plots simply demonstrate
the effectiveness of our newly proposed approach in two new decay
channels, with either the $h$ or the $Z$ in the final states. The
overall conclusions are as in the main text.

\end{appendix}


\newpage

\newcommand\jnl[1]{\textit{\frenchspacing #1}}
\newcommand\vol[1]{\textbf{#1}}

\end{document}